\documentclass[a4paper,11pt]{article}
\pdfoutput=1 
\usepackage{jcappub} 
\usepackage{xspace}
\usepackage{latexsym, amsmath, amssymb,amsthm,amsopn,amsfonts, graphicx, array, cleveref}
\usepackage{color}
\usepackage{relsize}
\usepackage[letterpaper]{geometry}
\usepackage{rotating}
\usepackage{tabularx}
\usepackage{dcolumn}
\usepackage{bm}
\usepackage{natbib}
\usepackage{multirow}

\makeatletter
\newcommand{\thickhline}{%
    \noalign {\ifnum 0=`}\fi \hrule height 1pt
    \futurelet \reserved@a \@xhline
}
\newcolumntype{"}{@{\hskip\tabcolsep\vrule width 1pt\hskip\tabcolsep}}
\makeatother

\def\simlt{\lower.5ex\hbox{$\; \buildrel < \over \sim \;$}}
\def\simgt{\lower.5ex\hbox{$\; \buildrel > \over \sim \;$}}
\def\simgtalt{\lower.5ex\hbox{$\buildrel > \over \sim \;$}}

\def\bd#1{{\bf #1}}
\def\l#1{\left #1}
\def\r#1{\right #1}

\def\ref#1{(\ref{#1})}

\def\core{COrE+}
\def\lb{LiteBIRD-ext}
\def\siv{Stage IV}
\def\pix{PIXIE}
\def\planck{\textit{Planck}}

\newcommand{\ns}{\ensuremath{n_{\rm s}}}
\newcommand{\as}{\ensuremath{A_{\rm s}}}
\newcommand{\h}{\ensuremath{H_0}}
\newcommand{\ombh}{\ensuremath{\Omega_{\rm b}h^2}}
\newcommand{\omch}{\ensuremath{\Omega_{\rm c}h^2}}
\newcommand{\nt}{\ensuremath{n_{\rm t}}}
\newcommand{\alphas}{\ensuremath{\alpha_{\rm s}}}
\newcommand{\omk}{\ensuremath{\Omega_{\rm k}}}
\newcommand{\mnu}{\ensuremath{M_\nu}}
\newcommand{\neff}{\ensuremath{N_{\rm eff}}}

\newcommand{\w}{\ensuremath{w}}
\newcommand{\wo}{\ensuremath{w_0}}
\newcommand{\wa}{\ensuremath{w_a}}
\newcommand{\fsky}{\ensuremath{f_{\rm sky}}}
\newcommand{\ukarc}{\ensuremath{\mu}K-arcmin}
\newcommand{\lmin}{\ensuremath{\ell_{\rm min}}}
\newcommand{\lmax}{\ensuremath{\ell_{\rm max}}}
\newcommand{\betad}{\ensuremath{\beta_{\rm d}}}
\newcommand{\Td}{\ensuremath{T_{\rm d}}}
\newcommand{\betas}{\ensuremath{\beta_{\rm s}}}

\def\lcdm{$\Lambda$CDM}
\def\lcdmr{\lcdm$+r$}
\def\lcdmrnt{\lcdm$+r+\nt$}
\def\lcdmi{\lcdm+inf}
\def\lcdmmnu{\lcdm$+\mnu$}
\def\lcdmneff{\lcdm$+\neff$}
\def\wcdm{$w$CDM}
\def\wwacdm{$w_0w_a$CDM}
\def\lcdmk{\lcdm$+\omk$}

\newcommand{\clfgr}{\ensuremath{C_\ell^{\rm fg\,res}}}
\newcommand{\reff}{\ensuremath{r_{\rm eff}}}
\newcommand{\clbbp}{\ensuremath{C_\ell^{BB,\,{\rm prim}}}}
\newcommand{\clbbl}{\ensuremath{C_\ell^{BB,\,{\rm lens}}}}
\newcommand{\clbbe}{\ensuremath{C_\ell^{BB,\,{\rm est}}}}
\newcommand{\clbbd}{\ensuremath{C_\ell^{BB,\,{\rm del}}}}

\usepackage{amsfonts}

\title{Robust forecasts on fundamental physics from the foreground-obscured, gravitationally-lensed CMB polarization}

\author[*,a,b,c,d]{Josquin Errard,}
\author[*,e]{Stephen M. Feeney,}
\author[f]{Hiranya V. Peiris,}
\author[e]{and Andrew H. Jaffe}

\affiliation[a]{Sorbonne Universit\'es, Institut Lagrange de Paris (ILP),\\ 98 bis Boulevard Arago, 75014 Paris, France }
\affiliation[b]{LPNHE, CNRS-IN2P3 and Universit\'es Paris 6 \& 7,\\ 4 place Jussieu, F-75252 Paris Cedex 05, France}
\affiliation[c]{Computational Cosmology Center, Physics Division, Lawrence Berkeley National Lab,\\ 1 Cyclotron Road, Berkeley CA 94720, USA}
\affiliation[d]{Space Sciences Laboratory, University of California,\\ 7 Gauss Way, Berkeley CA 94720, USA}
\affiliation[e]{Department of Physics, Blackett Laboratory, Imperial College London,\\ Prince Consort Road, London SW7 2AZ, United Kingdom}
\affiliation[f]{Department of Physics and Astronomy, University College London,\\ Gower Street, London WC1E 6BT, United Kingdom}
\affiliation[*]{{\bf these authors contributed equally to this work}}

\emailAdd{josquin.errard@lpnhe.in2p3.fr}
\emailAdd{s.feeney@imperial.ac.uk}
\emailAdd{h.peiris@ucl.ac.uk}
\emailAdd{a.jaffe@imperial.ac.uk}

\abstract{ Recent results from the BICEP, Keck Array and Planck Collaborations demonstrate that Galactic foregrounds are an unavoidable obstacle in the search for evidence of inflationary gravitational waves in the cosmic microwave background (CMB) polarization. Beyond the foregrounds, the effect of lensing by intervening large-scale structure further obscures all but the strongest inflationary signals permitted by current data. With a plethora of ongoing and upcoming experiments aiming to measure these signatures, careful and self-consistent consideration of experiments' foreground- and lensing-removal capabilities is critical in obtaining credible forecasts of their performance. We investigate the capabilities of instruments such as Advanced ACTPol, BICEP3 and Keck Array, CLASS, EBEX10K, PIPER, Simons Array, SPT-3G and SPIDER, and projects as \core, \lb, \pix\ and \siv, to clean contamination due to polarized synchrotron and dust from raw multi-frequency data, and remove lensing from the resulting co-added CMB maps (either using iterative CMB-only techniques or through cross-correlation with external data). Incorporating these effects, we present forecasts for the constraining power of these experiments in terms of inflationary physics, the neutrino sector, and dark energy parameters. Made publicly available through an online interface, this tool enables the next generation of CMB experiments to foreground-proof their designs, optimize their frequency coverage to maximize scientific output, and determine where cross-experimental collaboration would be most beneficial. We find that analyzing data from ground, balloon and space instruments in complementary combinations can significantly improve component separation performance, delensing, and cosmological constraints over individual datasets. In particular, we find that a combination of post-2020 ground- and space-based experiments could achieve constraints such as $\sigma(r)\,\sim1.3\times10^{-4}$, $\sigma(\nt)\,\sim\,0.03$, $\sigma( \ns )\,\sim\,1.8\times10^{-3}$, $\sigma(\alphas)\,\sim\,1.7\times10^{-3}$, $\sigma( \mnu )\,\sim\,31$ meV, $\sigma( \w )\,\sim\,0.09$, $\sigma( \wo )\,\sim\, 0.25$, $\sigma( \wa )\,\sim\, 0.50$, $\sigma( \neff )\,\sim\,0.024$ and $\sigma( \omk )\,\sim\,1.5\times10^{-3}$, after component separation and iterative delensing.}

\begin{document}
\maketitle
\flushbottom

\section{Introduction}
\label{sec:introduction}

Cosmic microwave background (CMB) $B$-mode polarization, generated by gravitational lensing of $E$-mode polarization at arcminute scales~\cite{1998PhRvD..58b3003Z} and, potentially, by inflationary gravitational waves on degree scales~\cite{1997PhRvL..78.2058K,1997PhRvL..78.2054S}, provides a unique window into the physics of both the early and evolved universe. As an integrated measure of the geometry and growth of structure in the Universe, the measurement of lensing $B$ modes provides both insight into late-time physics, such as the damping of structure formation by massive neutrinos~\cite{2003PhRvL..91x1301K,Lesgourgues} and the influence of dark energy~\cite{2001PhR...340..291B,2006PhRvD..74j3510A,2006ApJ...650L..13H}, and an opportunity to break the CMB geometric degeneracy~\cite{1997ApJ...488....1Z,1997MNRAS.291L..33B} and hence constrain the curvature of the Universe without resorting to external datasets~\cite{2011PhRvL.107b1302S}. The detection of degree-scale $B$ modes would provide convincing evidence that the Universe entered an inflationary phase at very early times~\cite{1981PhRvD..23..347G}, with measurement of their amplitude constraining the energy scale at which inflation took place~\cite{1997PhRvL..78.1861L, 2004PhRvD..70j3505S}. The lure of such scientific potential has driven considerable effort into the development of dedicated $B$-mode observatories, and this effort has now begun to come to fruition. The past year has yielded the first measurements of the lensed $B$-mode polarization~\cite{2014PhRvL.112x1101A, 2015PhRvL.114j1301B, 2014ApJ...794..171T, 2015arXiv150302315K, 2015arXiv150201582P, 2014JCAP...10..007N}, though primordial $B$ modes remain elusive. The drive for ever more sensitive $B$-mode measurements is therefore ongoing, with numerous ground-based, balloon and satellite missions funded and in various stages of taking data, and many more proposed.\footnote{See, for example, \url{http://lambda.gsfc.nasa.gov/product/expt/}.}

A natural consequence of the low amplitude of the $B$-mode polarization is that observatories designed to detect $B$ modes can also make exquisite measurements of the scalar-sourced $E$-mode polarization. Cosmic-variance limited $E$ modes, and indeed their cross-correlation with temperature, provide even more constraining power than their temperature counterparts \cite{2014PhRvD..90f3504G}. The next generation of CMB polarization experiments will therefore tighten our constraints on the standard cosmological parameters by factors of $\sim3$, as well as constrain additional parameters such as the number of relativistic degrees of freedom.

However, to fully exploit the constraining power of CMB polarization one must overcome several obstacles. Recent observations~\cite{2014arXiv1409.5738P,2015PhRvL.114j1301B,2015arXiv150201588P} have emphatically demonstrated that next-generation CMB experiments will have to characterize, control and remove significant astrophysical foregrounds. Polarized Galactic foreground contamination turns out to be important at all Galactic latitudes, either due to large total intensity or high polarization fraction. Beneath the astrophysical foregrounds, the lensing $B$ modes --- though highly important for cosmological constraints --- are a significant limiting factor in detecting low values of the tensor-to-scalar ratio~\cite{2002PhRvD..65b3505L,2002PhRvL..89a1303K,2002PhRvL..89a1304K}. Delensing the $B$ modes~\cite{2002PhRvL..89a1303K,2002PhRvL..89a1304K,2002ApJ...574..566H,2003PhRvD..67h3002O,2004PhRvD..69d3005S,2007PhRvD..76l3009M,2012JCAP...06..014S, 2014arXiv1410.0691S, 2015arXiv150205356S} is therefore a further necessity for maximizing the scientific output of the most sensitive polarization observatories. Significant prior effort has gone into CMB experimental design including the mitigation of foreground contamination and delensing~\cite{2006JCAP...01..019V, 2009AIPC.1141..222D, 2009AIPC.1141...10B, 2009arXiv0906.1188B, 2011arXiv1102.2181T, 2014JCAP...02..006A,2015arXiv150800017W,2015arXiv150902676H,2015arXiv150904714R,2015arXiv150905419E}. The Planck Collaboration has recently released maps of polarized dust and synchrotron~\cite{2015arXiv150201588P}, strongly motivating an updated evaluation of the abilities of current and planned CMB observatories to address these challenges. In this paper, we employ a parametric maximum-likelihood approach (Ref.~\cite{2011PhRvD..84f3005E} and references therein) to estimate the ability of a range of pre- and post-2020 CMB experiments to remove polarized foreground contamination, delens the resulting $B$ modes, and constrain the inflationary, neutrino, and dark energy sectors. The tool developed for this work allows the specification of instrumental configurations (observed frequencies, noise levels, beam sizes and sky fractions; it does not currently model the effect of further instrumental systematic errors) and uses a Fisher-matrix framework to forecast errors on the desired parameters. It has been made publicly available on a Lawrence Berkeley National Laboratory (LBNL) machine,\footnote{\url{turkey.lbl.gov}} with a user-friendly web interface allowing any specific instrumental configuration to be studied.

The paper is organized as follows. Section~\ref{sec:methodology} introduces the component separation approach, the delensing technique, and the Fisher framework that we use to evaluate the scientific performance of different experimental configurations. Section~\ref{sec:results} summarizes our results, and Sect.~\ref{sec:web_interface} describes the web interface which is publicly released with this work. Finally, we discuss the implications of our results in Sect.~\ref{sec:conclusions}. In addition, Appendix~\ref{app:instruments_specifications} details the specifications of each instrument considered in this work, and Appendix~\ref{app:fisher_tables} provides a comprehensive set of their forecasted constraints.

\begin{figure*}
	\centering
		\includegraphics[width=12cm]{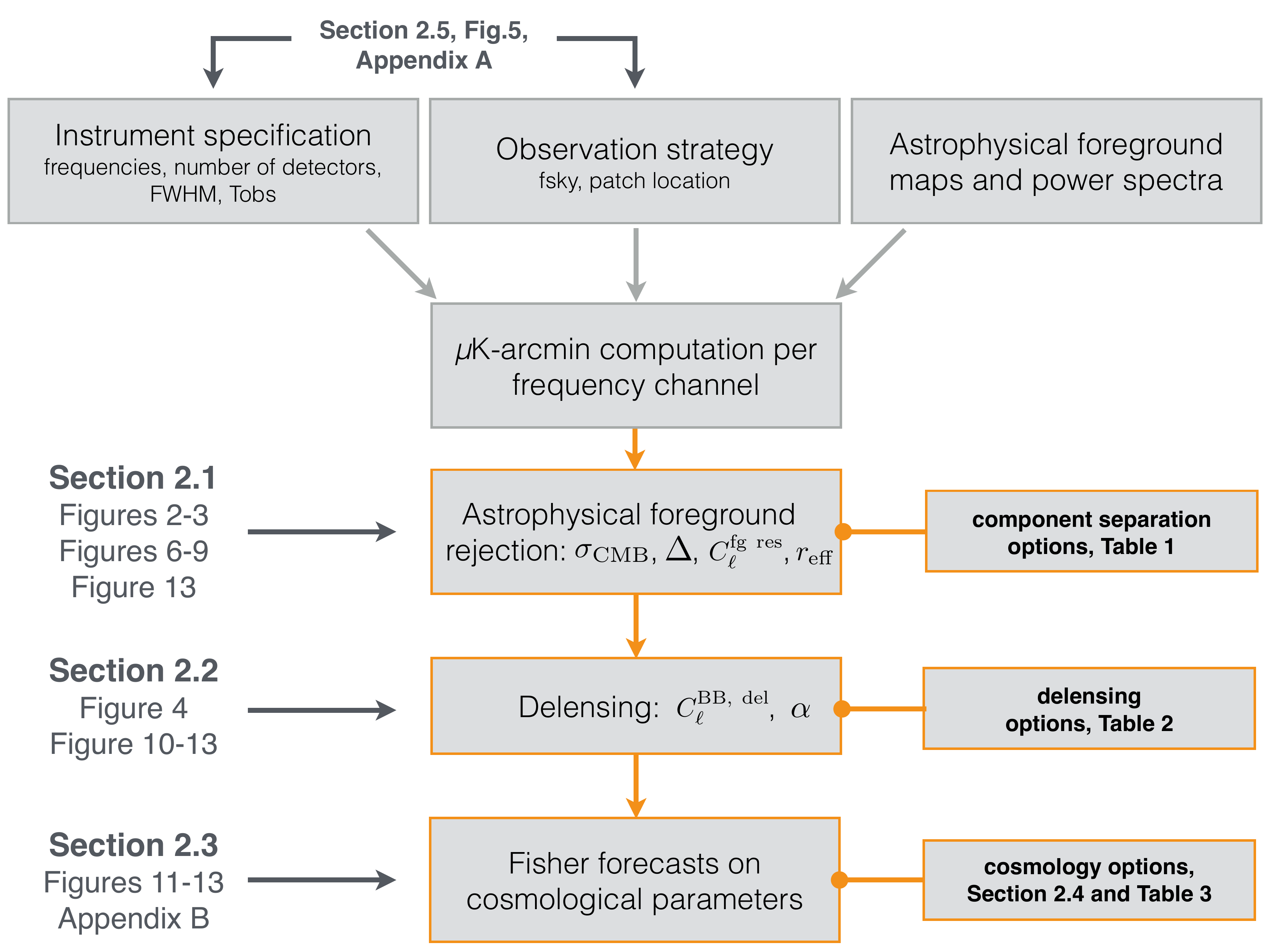}
	\caption{Schematic of the adopted methodology described in Sect.~\ref{sec:methodology}.}
	\label{fig:methodology}
\end{figure*}

\section{Methodology}
\label{sec:methodology}

We first describe the method we use to evaluate the scientific performance of a range of CMB experimental setups in the presence of Galactic foregrounds and gravitational lensing of the CMB. The precise specifications assumed for the experiments are described in Appendix~\ref{app:instruments_specifications}. Based on these specifications, we estimate each experiment's ability to first separate Galactic foreground emissions and then gravitational lensing from the primordial CMB, and hence constrain the cosmological parameters of interest. Figure~\ref{fig:methodology} summarizes the methodology; in the subsections below, we describe the main steps in detail.

\subsection{Foreground removal}
\label{ssec:foregrounds}
\begin{figure*}
\begin{center}
\includegraphics[width=7.5cm]{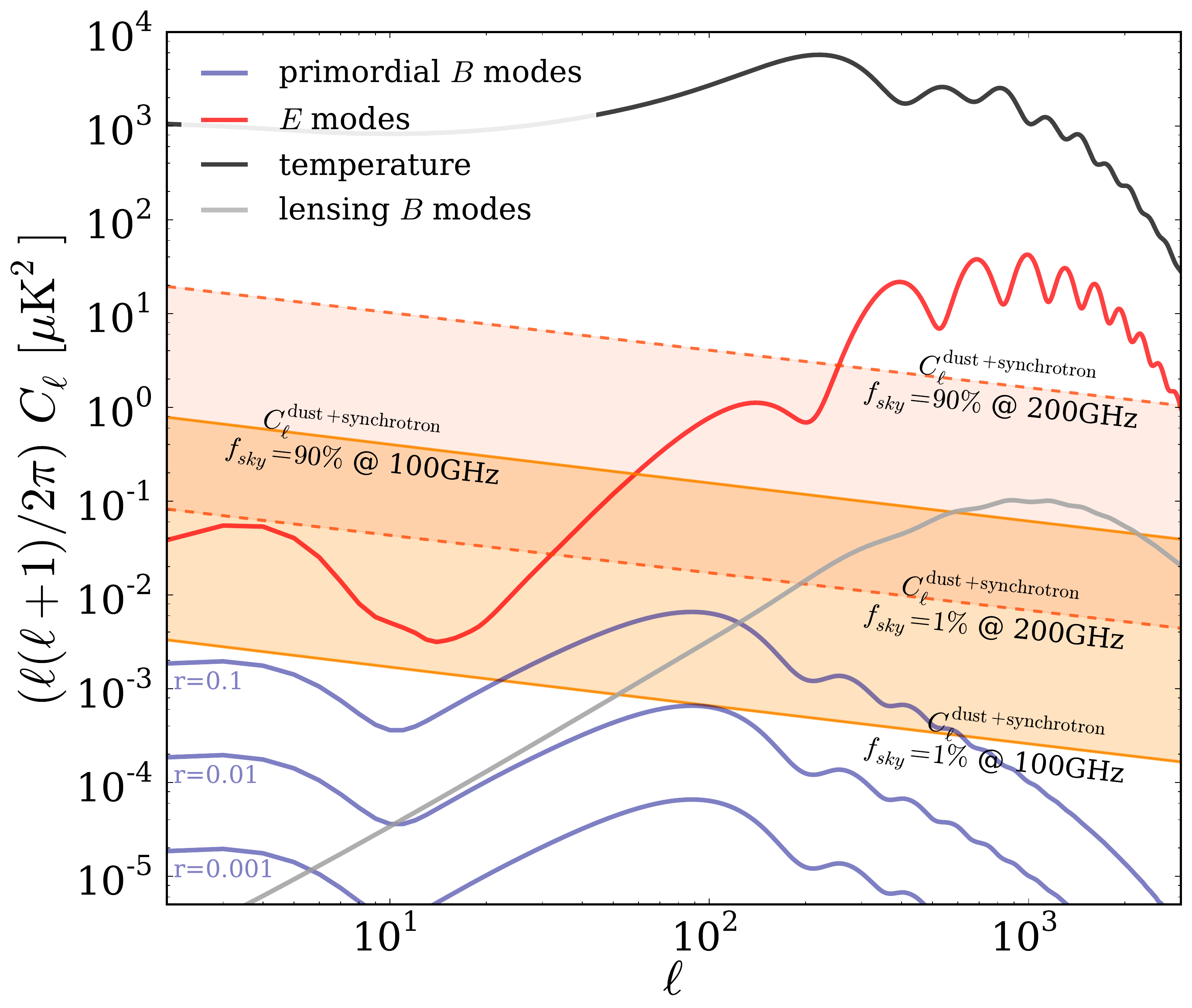}~\includegraphics[width=7.5cm]{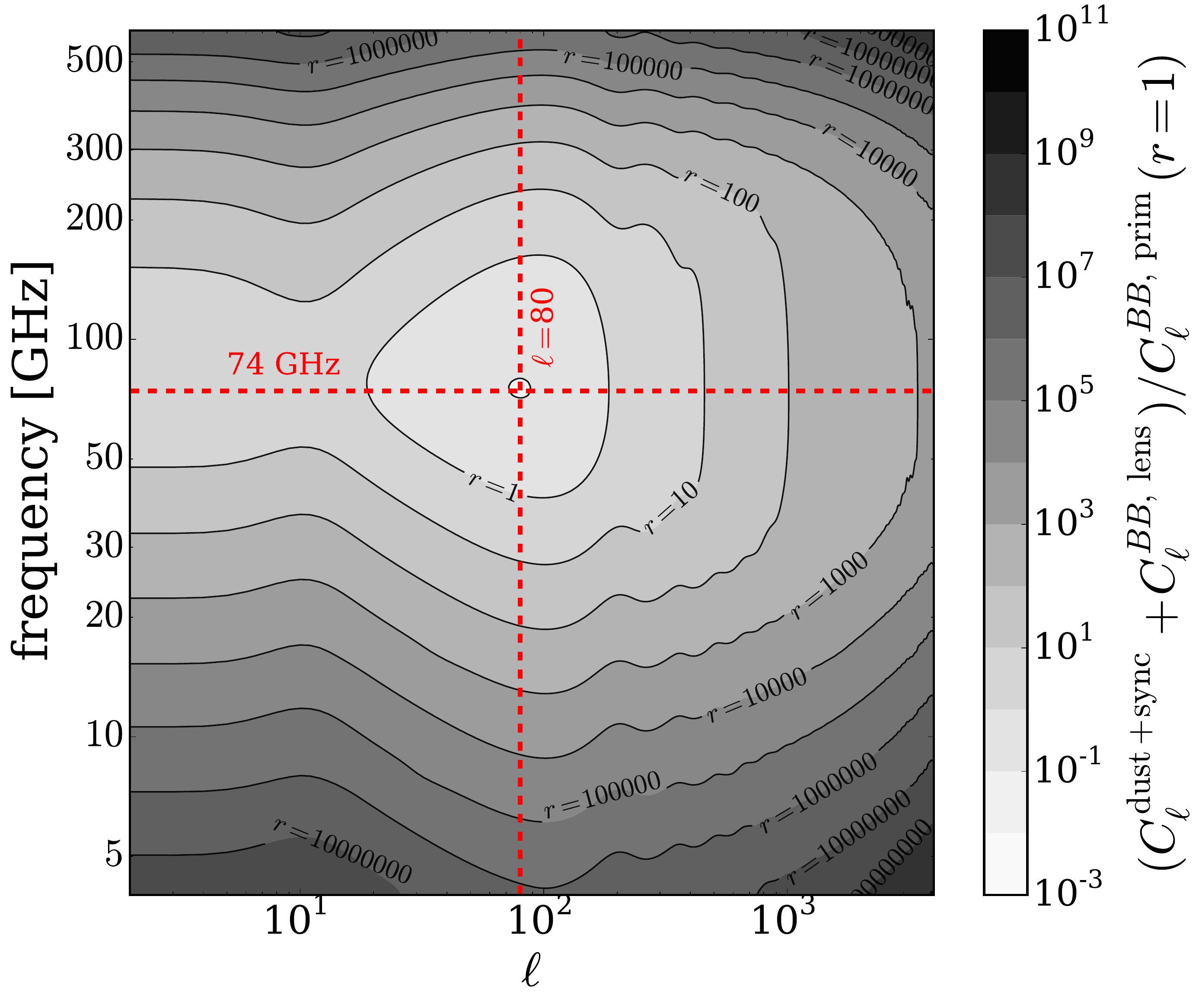}
\end{center}
\caption{\textit{Left panel:} Angular power spectra showing primordial $B$ modes, lensing $B$ modes, total intensity, and $E$ modes, as well as the total contribution of polarized $B$-mode foregrounds (dust plus synchrotron), expected on the cleanest 1--90\% of the sky, at $100$ and $200$\;GHz. Note that, as these results are derived from \planck's Galactic masks and are not therefore optimized for high-resolution, ground-based instruments, there is potential for discovery of small patches of sky (e.g., $\fsky \lesssim 5\%$) cleaner than those indicated here. \textit{Right panel:} The ratio of power spectra of foreground and lensing $B$ modes to primordial $B$ modes, assuming a tensor-to-scalar ratio $r=1$. The contours indicate, in effective values of $r$, the contamination due to foregrounds and lensing on primordial $B$-mode measurements. The $x$- and $y$-axes correspond to the multipole $\ell$ and frequency of observation, in GHz, respectively. The level of input foregrounds are estimated on a 50\% patch of the sky.}
\label{fig:input_foregrounds_plot}
\end{figure*}

As illustrated in Fig.~\ref{fig:input_foregrounds_plot}, polarized contamination from astrophysical foregrounds is an unavoidable challenge in the quest for primordial $B$-mode measurements. The left panel shows CMB temperature, $E$-mode and $B$-mode angular spectra for different values of the tensor-to-scalar ratio $r$, as well as the expected amplitude of dust and synchrotron $B$ modes, estimated using various portions of the \planck\ data~\cite{2015arXiv150201588P}. The right panel shows, as a function of frequency of observation and multipole $\ell$, the ratio of foreground plus lensing $B$ modes to primordial $B$ modes, assuming a tensor-to-scalar ratio $r=1$. Outside the half-sky Galactic mask that was considered, the minimum contamination is reached at multipole $\ell \approx 80$ and frequency $\nu \approx 74$\;GHz (as confirmed by Ref.~\cite{2015arXiv150905934C,2015arXiv151100532K}). This foreground-minimum region has an effective foreground amplitude of $r \sim 0.1$. Reaching values of $r$ less than 0.1 (or, indeed, improving the significance of any future detection of $r$ of this magnitude) therefore requires highly effective cleaning of both foregrounds and lensing $B$ modes.

\subsubsection{Formalism}
\label{ssec:formalism_fgs}

Our approach is based on the parametrization of the emission laws for two important astrophysical contaminants, namely dust and synchrotron \cite{2009MNRAS.392..216S,2010MNRAS.408.2319S,2011PhRvD..84f3005E,PhysRevD.85.083006}.\footnote{We restrict ourselves here to the polarized foregrounds known to be significant at CMB frequencies. As measurements of the polarized CMB sky improve, they may reveal further complexity in both the form and number of foreground components, with potential additional sources including spinning and magnetic dust~\cite{1998ApJ...508..157D,2013ApJ...765..159D} and carbon monoxide~\cite{1999ApJ...512L.139G}. We leave the treatment of these potential foregrounds to future investigation.} To estimate the impact of performing component separation with a given instrument, we use a parametric maximum-likelihood component-separation approach, as implemented in Ref.~\cite{2009MNRAS.392..216S}.\footnote{Note there exists a complementary approach to component separation that removes foregrounds by combining multi-frequency data to minimize the variance of the CMB component (see, e.g., Ref.~\cite{2015arXiv150205956P} for the various methods applied to \planck\ data). While we do not explicitly treat such methods in this analysis, we can expect broadly the same performance as for parametric models.} We assume the following linear data model, in which the signal measured in each frequency channel $i$ at each pixel $p$ is given by
\begin{equation}
	d_i(p) =  A_{ij} \, s_j(p) +  n_i(p) \, ,
	\label{eqn:dataModelFull}
\end{equation}
where we assume summation over the repeated index $j$ that runs over the underlying components (i.e., CMB, dust and synchrotron), and
\begin{itemize}
\item $\bd{d}(p)$ is a multi-frequency data vector, with each entry corresponding to a different frequency channel;
\item $\bd{s}(p)$ is a multi-component sky signal vector, each entry of which corresponds to a different polarized sky component to be estimated from the data;
\item $\bd{A}$ is a mixing matrix defining how the components need to be combined to give a signal for each of the considered frequency channels; and
\item $\bd{n}(p)$ is a vector containing the instrumental noise in each frequency channel, which is assumed to be Gaussian and uncorrelated with dispersion $\bd{N}$.
\end{itemize}
We will switch between index ($A_{ij}$) and matrix (\bd{A}) notation as convenient. For simplicity, both $\bd{A}$ and $\bd{N}$ are assumed to be pixel independent. The mixing matrix $\bd{A}$ takes into account each experiment's bandpasses
\begin{equation}
A_{ij} = \int{ \delta_i(\nu) A^{\rm raw}_j(\nu, \nu_{\rm ref}) \, d\nu},
\end{equation}
where $\delta_i(\nu)$ is the fractional bandpass of the $i^{\rm th}$ frequency channel, defined such that $\int{\delta_i(\nu) \,  d\nu }=1$, and $\bd{A}^{\rm raw}(\nu, \nu_{\rm ref})$ is a vector describing the frequency dependence of the components before bandpasses are taken into account.

We assume that component separation comprises two independent steps. First, we estimate the mixing matrix, $\mathbf{A}$, from the various frequency maps produced by a given instrument. In the parametric framework assumed here, estimating the mixing matrix corresponds to determining the parameters describing the frequency dependence of the dust and synchrotron emission. Second, we linearly combine the different frequency maps to disentangle dust and synchrotron from the CMB --- inverting Eq.\eqref{eqn:dataModelFull} to obtain the sky signals $\bd{s}(p)$ --- using the estimated mixing matrix. This process leads to a noise level in the reconstructed map which is higher than the simple quadratic combination of sensitivities from all frequency channels. Although the first step is specific to a considered component separation method, the second step is general to any approach --- though it can be performed either in pixel or Fourier space.

\textbf{Estimation of the mixing matrix $\mathbf{A}$.} As supported by the latest \planck\ results \cite{2015arXiv150201588P}, we assume for the synchrotron frequency dependence a simple power-law form with spectral index $\betas$,
\begin{equation}
A^{\rm raw}_{\rm sync}(\nu, \nu_{\rm ref}) \equiv \left(\frac{\nu}{\nu_{\rm ref}}\right)^{\betas},
\label{eq:As_def}
\end{equation}
where the reference frequency $\nu_{\rm ref}=150$\;GHz. We consider a modified grey-body emission law for the dust,
\begin{equation}
A^{\rm raw}_{\rm dust}(\nu, \nu_{\rm ref}) \equiv \left( \frac{\nu}{\nu_{\rm ref}}\right)^{\betad+1}\frac{e^{\frac{h\nu_{\rm ref}}{k\,\Td}}-1}{e^{\frac{h\nu}{k\Td}} -1 },
\label{eq:Ad_def}
\end{equation}
where $\betad$ is the dust spectral index and $\Td$ the dust temperature. As the dust temperature and spectral index are strongly anti-correlated in the frequency ranges targeted by CMB experiments, with higher frequencies required to constrain the temperature, we choose to fix the temperature to a reasonable value and only fit for the spectral index in the following. The component-separation scenarios studied in this work are summarized in Table~\ref{table:fgs_options}. We consider the most realistic scenario of synchrotron, dust and CMB to derive our main results, though in several intermediate figures we consider simplified foreground models for pedagogical purposes.

\begin{table*}
\caption{Summary of the component-separation scenarios and approaches to spatially-varying spectral indices described in Sect.~\ref{ssec:foregrounds}.}
	\begin{center}
		\begin{tabular}{|c|c|c|c|c|}
			\hline
			label &  no comp. sep. & d & s & s+d   \\
			\hline
			 & CMB & CMB & CMB & CMB \\
			components & only & + dust & + synchrotron & + synchrotron \\
			&   &  only   & only & + dust \\
			\hline
		\end{tabular}
		\newline
		\vspace*{0.5cm}
		\newline
		\begin{tabular}{|c|l|l|}
			\hline
			label & \multicolumn{1}{c|}{$n_p$ approach} & \multicolumn{1}{c|}{$\mathbf{A}$-expansion approach} \\
			\hline
			description &Assume $n_{\rm patch} \equiv 12\times \fsky \times 4^2$& Extend mixing matrix $\mathbf{A}$ and sky  \\ 
			& independent patches within & signal $\mathbf{s}(p)$ by expanding emission \\
			& observed sky fraction $\fsky$. & laws with respect to spectral \\
			&  Assume this hypothesis & indices $\betad$ and $\betas$ around \\
			& artificially boosts foreground & their mean values. Noise after  \\
			& residuals as $\clfgr \propto n_{\rm patch}$. &component separation is boosted.\\
			\hline
		\end{tabular}
	\end{center}
\label{table:fgs_options}
\end{table*}

Instead of performing simulations and running a full likelihood analysis, we use a Fisher-matrix approach to estimate the uncertainty on each foreground component's spectral index. As derived in Ref.~\cite{2011PhRvD..84f3005E}, the second-order derivatives of the $\beta$ spectral log-likelihood,  $\mathcal{L}(\beta)$, taken at the true values of the parameters, can then be written as
{\begin{eqnarray}
\label{eqn:secDervSat}
	\centering
 \bd{\Sigma} ^{-1} &\simeq& - \left.\left\langle\frac{\partial^2\mathcal{L}}{\partial \beta\partial \beta'}\right\rangle_{\rm noise}\,\right|_{\rm true\ \beta}\\
		&& \hspace{-0.5cm} = -\, {\rm tr}\,\l\{\l[ \frac{\partial \bd{A}}{\partial \beta}^T\, \bd{N}^{-1}\,  \bd{A}\, \l(\bd{A}^T\bd{N}^{-1}\bd{A}\r)^{-1} \,  \bd{A}^T\bd{N}^{-1} \frac{\partial \bd{A}}{\partial \beta'} - \, \frac{\partial \bd{A}}{\partial \beta}^T \, \bd{N}^{-1} \, \frac{\partial \bd{A}}{\partial \beta'}\,\r] \sum_p\, \bd{s}(p)\,\bd{s}^T(p) \r\}  \nonumber
\end{eqnarray}
where $\bd{s}(p)$ are template maps for each desired component (typically CMB, dust and synchrotron); see Sect.~\ref{ssec:input_fgs}.
$\bd{\Sigma}$ is a matrix of size $n_{\rm par}^2$, where $n_{\rm par}$ is the number of free parameters in the mixing matrix $\mathbf{A}$. In our case, $\mathbf{A}=\mathbf{A}(\betad, \betas)$ and $n_{\rm par} = 2$.
The uncertainty on the estimation of the foreground spectral indices is then given by \cite{2010MNRAS.408.2319S} 
\begin{equation}
\sigma( \beta ) \simeq \sqrt{ \l[\bd{\Sigma}\r]_{\beta\beta} }.
\label{eq:fg_beta_error}
\end{equation}
The residual foregrounds left in the CMB map after component separation can be derived from this uncertainty; their power spectrum is given by
\begin{equation}
\clfgr  \equiv \sum_{k,k'}\sum_{j,j'}\,\Sigma_{kk'}\,\bd{\kappa}^{jj'}_{kk'}  \, C^{jj'}_{\ell}, 
\label{eq:Clres_def}
\end{equation}
where $C^{jj'}_{\ell}$ are the input foreground spectra with $j,\,j'\ \in\, \{$cmb, dust, synchrotron$\}$\footnote{Note that cross-spectra, and in particular $C^{\rm dust\,\times\,sync}_{\ell}$, are included in the sum of Eq.~\eqref{eq:Clres_def}.}. The element $\bd{\kappa}^{jj'}_{kk'}$ is as defined in~\cite{2011PhRvD..84f3005E}
\begin{eqnarray}
	\centering
		\bd{\kappa}^{jj'}_{kk'} &\equiv& \alpha_k^{0j}\alpha_{k'}^{0j'}\\
		{\rm with}\  \alpha_k^{0j} &\equiv& -\left[\left(\mathbf{A}^T\mathbf{N^{-1}}\mathbf{A}\right)^{-1} \mathbf{A}^T\mathbf{N^{-1}}  \frac{\partial \mathbf{A}}{\partial \beta_{k}} \right]_{0j}.
	\label{eq:kappa_def}
\end{eqnarray}
The foreground residuals can ultimately bias the estimation of primordial $B$ modes in the reconstructed CMB map, see Ref.~\cite{1475-7516-2011-08-001}. To study the potential biases due to imperfect component separation, we therefore define the effective amplitude of foreground residuals, $\reff$, such that
\begin{equation}
\sum_{\ell=20}^{200}\clbbp ( \reff ) = \sum_{\ell=20}^{200} \clfgr,
\end{equation}
yielding
\begin{equation}
\reff \equiv \frac{ \sum_{\ell=20}^{200} \clfgr }{ \sum_{\ell=20}^{200} \clbbp (r=1) }.
\label{eq:reff_def}
\end{equation}

Equation~\eqref{eqn:secDervSat}, through the term $\sum_p\, \bd{s}(p)\,\bd{s}^T(p)$, contains the {\em a priori} assumption that the foreground spectral indices are constant over the sky. This is, however, not necessarily the case~\cite{2015arXiv150201588P}, and we account for this assumption in two ways. Firstly, following Ref.~\cite{PhysRevD.85.083006}, we break each experiment's observable patch into $n_{\rm patch}$ independent regions, within which the foreground spectral indices are assumed to be constant. Apportioning the data this way na\"ively leads to the approximation $\sigma(\beta) \propto \sqrt{n_{\rm patch}}$, or equivalently $\clfgr \propto n_{\rm patch}$, degrading the performance of the component separation and boosting the impact of the foreground residuals. 
Following Refs.~\cite{2014arXiv1409.5738P,2015arXiv150201588P}, we take the size of these independent patches to match $N_{\rm side} = 4$ {\tt HEALPix} pixels~\cite{2005ApJ...622..759G}, requiring 
\begin{equation}
n_{\rm patch} \equiv \lfloor12\times \fsky \times 4^2 \rfloor
\label{eq:np_definition}
\end{equation}
unique spectral indices to be fitted by an experiment covering a sky fraction $\fsky$ (rounding down to the next lowest integer). This approach, referred to as the ``$n_p$ approach" in the following, is the method used to derive the main results of this work.

Following Ref.~\cite{2005MNRAS.357..145S}, we also consider a second approach utilizing the first-order expansion of the mixing matrix $\mathbf{A}$ around the mean values of each foreground parameter
\begin{equation}
\mathbf{A} (\beta) \approx \mathbf{A}( \hat{\beta}) + \delta \beta (p) \left. \frac{\partial  \mathbf{A}}{\partial \beta}\right|_{\hat{\beta} } + \mathcal{O}\left(\delta \beta (p) ^2\right),
\label{eq:A_taylor_expansion}
\end{equation}
where $ \hat{ \beta } \equiv \langle\, \beta\, \rangle_p$ is computed over the observed sky pixels $p$, $\delta \beta \equiv \beta - \hat{\beta}$ and, as before, the dust temperature $\Td$ is assumed constant across the sky.
By increasing the dimensionality of the mixing matrix $\mathbf{A}$, we can extend the parametric framework to account for spatially varying foreground parameters.
Within this extended framework, the sky signal used in Eq.~\eqref{eqn:dataModelFull} is
\begin{equation}
\bd{s}(p) \equiv \left[ s_{\rm cmb}(p),\ s_{\rm dust}(p),\ \delta \beta_{d}(p)\,s_{\rm dust}(p), s_{\rm sync}(p),\ \delta \beta_{s}(p)\,s_{\rm sync}(p) \right] \, ,
\label{eq:s_p_convention}
\end{equation}
where each $s_j(p) \,\in\,\{Q, U\}$, and the corresponding mixing matrix $\mathbf{A}$ reads
\begin{equation}
\mathbf{A} = \left[ \mathbf{A}_{\rm cmb},  \mathbf{A}_{\rm dust},  \frac{\partial \mathbf{A}_{\rm dust}}{\partial \betad}, \mathbf{A}_{\rm sync},  \frac{\partial \mathbf{A}_{\rm sync}}{\partial \betas} \right].
\label{eq:A_matrix_convention}
\end{equation}
We denote this method the ``$\mathbf{A}$-expansion approach''. As we shall see in the following section, the expansion of the mixing matrix necessarily results in higher noise levels after component separation; however, the addition of new degrees of freedom in the foreground modeling can potentially yield better control of foreground residuals, provided an experimental configuration contains enough frequency channels to fully characterize the components present. An additional source of bias not considered here could result from using the wrong foreground model to clean the CMB. Throughout this work we assume that the foreground model is correct; see Ref.~\cite{2015arXiv150904714R} for a low-multipole treatment of this systematic.

The question of which of the $n_p$ and $\mathbf{A}$-expansion approaches is most appropriate is difficult to answer currently, as the evidence for spatial variation of the foreground parameters~\cite{2015arXiv150201588P} is still inconclusive. As yet, there is no compelling evidence for a particular angular scale in the variation of the spectral indices, and each project will have to determine whether the $n_p$ approach is a good approximation for their data. This will depend on the chosen region of the sky and on the observational integration depth. The current generation of (relatively)  noisy small-patch experiments will likely be driven towards adopting the $n_p$ approach---and hence relying on priors derived from foreground observations---in order to retain constraining power. The increased sensitivity of future projects, coupled with their broader frequency and sky coverage, will make the $\mathbf{A}$-expansion approach feasible, allowing critical tests of the priors and parametrizations used in the $n_p$ approach.

\textbf{Linear combination of frequency maps given $\mathbf{A}$.} Once the parameters of the mixing matrix are estimated, and assuming the estimates are unbiased, the noise variance in each pixel of the reconstructed CMB map in (\ukarc)$^2$ is given by~\cite{2009MNRAS.392..216S}
\begin{equation}
	\sigma_{\rm CMB}^2 \equiv \gamma_p^2\, \left[ \left( \bd{A} ^T\, \bd{N}^{-1}\,  \bd{A} \right)^{-1}\right]_{\rm CMB \, CMB},
	\label{eq:sigma_CMB_def}
\end{equation}
where $\gamma_{p}$ is the angular size of a sky pixel in arcmin.
Due to the $\mathbf{A}^T\mathbf{N^{-1}}\mathbf{A}$ inversion, $\sigma_{\rm CMB}$ is necessarily larger than the quadratic combination of all the sensitivities per channel.\footnote{$\sigma_{\rm CMB}$ is equal to the quadratic combination of all sensitivities only in the trivial case of $\mathbf{A}$ being the identity matrix, i.e., each component only existing at a single frequency.} The extent of this noise degradation depends on the condition of the mixing matrix $\mathbf{A}$, and thus the larger mixing matrix employed in the $\mathbf{A}$-expansion approach, Eq.~\eqref{eq:A_taylor_expansion}, can lead to worse noise levels than the $n_p$ approach. Complementary to the effective amplitude of residuals \reff\ defined in Eq.~\eqref{eq:reff_def}, we introduce the dimensionless noise degradation $\Delta$ defined as
\begin{equation}
	\Delta \equiv \left(\frac{\sigma_{\rm CMB}}{\sigma_{\rm quad}}\right)^2 \geq 1,
	\label{eq:Delta_def}
\end{equation}
where $\sigma_{\rm quad}$ is the simple quadratic combination of sensitivities across all frequency channels.

Figure~\ref{fig:sigma_cmb_example} shows the behavior of $\sigma_{\rm CMB}$ and $\Delta$ in the simple case of an experiment with three frequency channels --- $150$, $220$, and $353$\;GHz --- observing the CMB obscured only by dust. The sensitivities of the two low-frequency channels are allowed to vary, but the highest frequency has a noise level of $200$\;\ukarc, assumed to come from \planck. The noise in the reconstructed CMB map (left panel) is limited by the $150$\;GHz sensitivity across most of the parameter space considered, and we conclude that $150$\;GHz sensitivity should therefore be prioritized over $220$\;GHz until $\sigma_{150} \lesssim 5$ \ukarc. Below this value, there is an optimal balance to be found between the sensitivities in both $150$ and $220$\;GHz channels.

Turning to the noise degradation plot (right panel), we see exactly how much sensitivity is wasted by investing in $220$\;GHz sensitivity at the expense of $150$\;GHz: the noise in the final CMB map is over 20 times larger than the raw quadratic combination in the top-left of the plot. The noise is degraded the least when the $220$\;GHz noise is roughly twice that at $150$\;GHz. This toy example shows that \planck's 353\;GHz band can be a good dust-monitoring channel, though it is important to note that the foreground residuals resulting from imperfect estimation of the mixing matrix have not been considered here. Furthermore, adding synchrotron to the problem or letting the spectral indices in $\mathbf{A}$ vary across the sky would significantly degrade $\sigma_\mathrm{CMB}$ (and therefore $\Delta$) if more frequency channels with reasonable sensitivities were not available.

\begin{figure*}
\centering
\includegraphics[width=7.5cm]{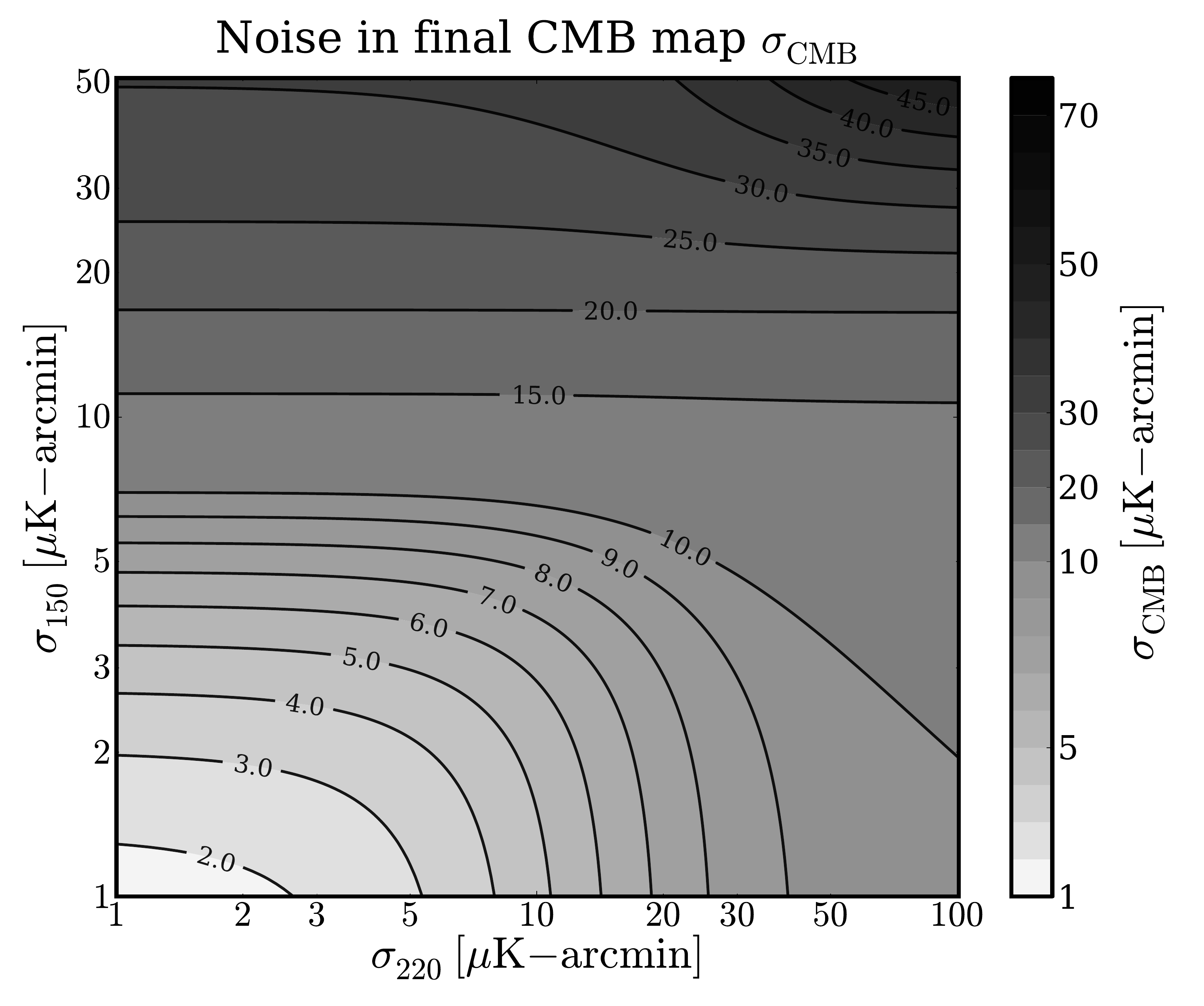}~\includegraphics[width=7.5cm]{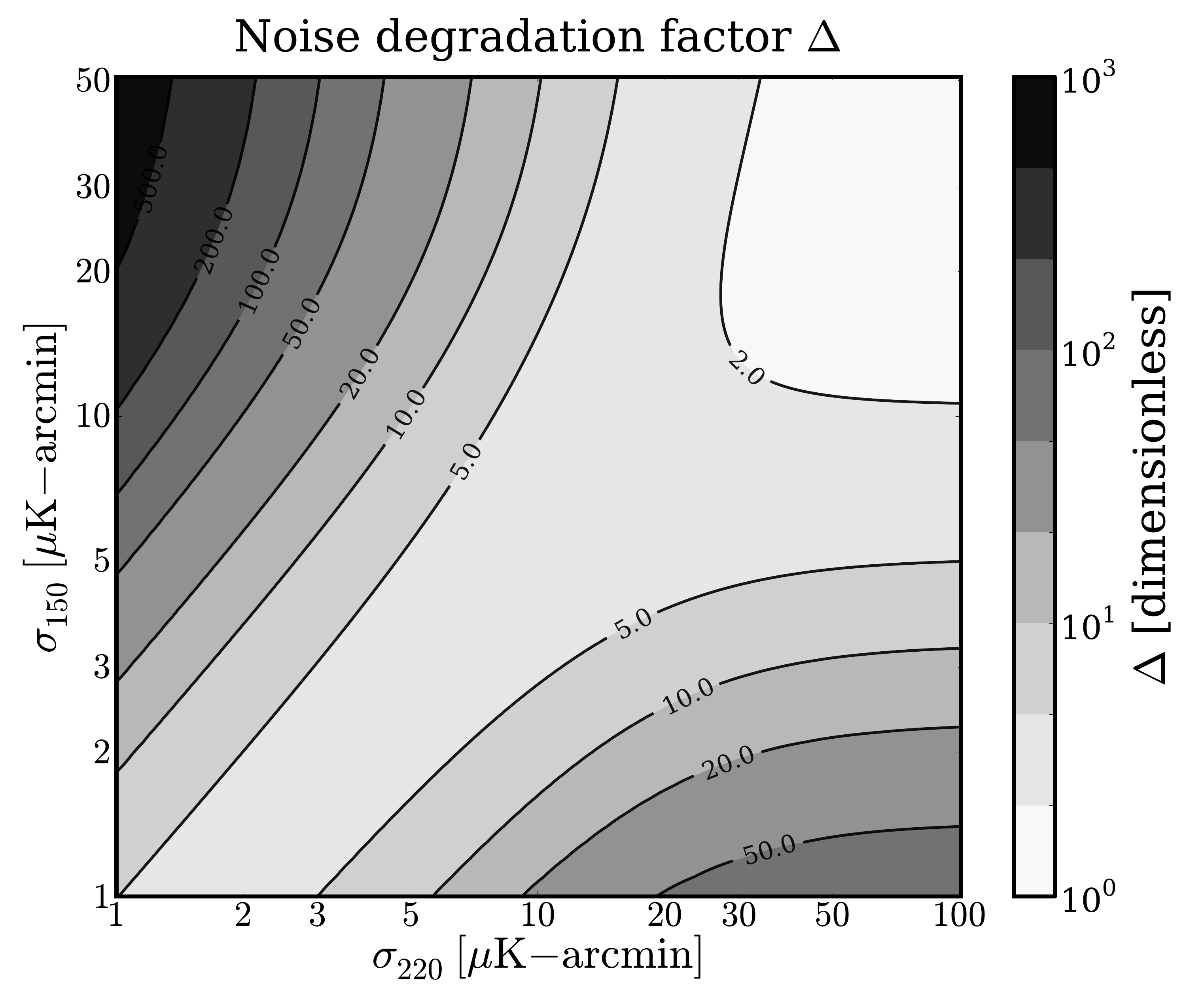}
\caption{\textit{Left panel:} the noise in the final CMB map for an experiment with two frequency channels of varying polarization sensitivity centered on 150 and 220\;GHz, combined with \planck's 353\;GHz channel. The sole foreground contaminant is dust. \textit{Right panel:} the noise degradation factor $\Delta$ between the final CMB map and the quadratic combination of all channels for the same experimental configuration. }
\label{fig:sigma_cmb_example}
\end{figure*}

\subsubsection{Input data}
\label{ssec:input_fgs}

The computation of the uncertainty on the foreground spectral indices, Eq.~\eqref{eqn:secDervSat}, requires $\bd{s}(p)$ --- polarized CMB, dust and synchrotron template maps --- as well as the spectral indices $\betad(p)$ and  $\betas(p)$ in the case of the $\mathbf{A}$-expansion approach. We use \planck\ component-separated maps~\cite{2015arXiv150205956P,2015arXiv150201588P} for this purpose, degrading all maps to their highest common resolution, {\tt HEALPix} $N_{\rm side} = 128$. Following Refs.~\cite{2015AA...576A.107P, 2015arXiv150201588P}, we further assume
\begin{align}
\betad&=1.59 \nonumber \\
\Td&=19.6\,{\rm K} \\
\betas&=-3.1 \nonumber
\end{align}
to build the elements of the input mixing matrix $\mathbf{A}$. Throughout this work, we set the dust temperature to its true value, and vary the two spectral indices $\betad$ and $\betas$. Where \planck\ is not directly included in the experimental configuration, we additionally adopt Gaussian priors from \planck's constraints on the foreground parameters~\cite{2015AA...576A.107P,2015arXiv150201588P}. We transform Eq.~\eqref{eqn:secDervSat} as $\bd{\Sigma} \rightarrow \bd{\Sigma} + \bd{\Sigma}_{Planck}$, assuming a diagonal $\bd{\Sigma}_{Planck}$ such that
\begin{align}
&\sigma_{Planck}( \betad = 1.59 ) = 0.04 \label{eq:priors_def}\\
&\sigma_{Planck}( \betas=-3.1 ) = 0.4. \nonumber
\end{align}
Regarding the foreground power spectra, as suggested by Ref.~\cite{2014arXiv1409.5738P}, we use the following power law forms for dust and synchrotron,
\begin{align}
C_\ell^{\rm dust} &\propto \left(\frac{\ell_0}{\ell}\right)^{2.4} \label{eq:Cl_fgs_ell_dependence}\\
C_\ell^{\rm sync} &\propto \left(\frac{\ell_0}{\ell}\right)^{2.6},\nonumber
\end{align}
where $\ell_0 = 80$. Several of these input spectra are shown in Fig.~\ref{fig:input_foregrounds_plot}. Note that the spatial and angular dependencies only enter into the calculation of the residual power spectrum, $\clfgr$, Eq.~\eqref{eq:Clres_def}. As we shall see, in many cases the experimental performance is dominated by the post-component-separation noise, $\sigma_{\rm CMB}$, which does not assume any information on the foregrounds beyond their frequency dependence.\footnote{Note that \planck\ has estimated the spectral indices in Eq.~\eqref{eq:Cl_fgs_ell_dependence} with uncertainties of $\sim\,5\,\%$, sourced by both genuine variation of the spectral indices across the sky and uncertainty at a given position. Varying the spectral indices within this range of uncertainty can---depending on the instrumental setup considered---lead to percent-level variations of the effective amplitude of foreground residuals, \reff, and sub-percent variations of constraints on cosmological parameters sensitive to large angular scales, such as $r$. In the remainder of this work we will neglect this uncertainty.} Finally, all the sky patches that we study throughout the paper use the Galactic masks created for the \planck\ data analysis~\cite{2015arXiv150702704P}.\footnote{These apodized Galactic masks and all other \planck\ data products can be downloaded from the \planck\ Legacy Archive at \url{http://pla.esac.esa.int/pla/}.} 

\subsection{Delensing}
\label{ssec:delensing}

For experiments with {\em foreground-cleaned} noise levels below 4-5 \ukarc, CMB lensing becomes the dominant obstacle to observing weak primordial $B$ modes~\cite{2002PhRvD..65b3505L}. Removing this contaminant through delensing~\cite{2002PhRvL..89a1303K,2002PhRvL..89a1304K,2002ApJ...574..566H} requires a measurement of the lensing potential, which one can use to estimate the lensed CMB $B$ modes for subtraction from the total observed signal. We base our delensing procedure on Ref.~\citep{2012JCAP...06..014S}, which provides the following analytical expression for the estimated lensing $B$ modes:
\begin{equation}
\label{eq:smith_delensing} 
\clbbe = \frac{1}{2\ell + 1}\sum_{\ell_1, \ell_2} \left|f_{\ell \ell_1 \ell_2}^{EB}\right|^2 \times \frac{(C_{\ell_1}^{EE})^2}{C_{\ell_1}^{EE}+N_{\ell_1}^{EE}} \frac{(C_{\ell_2}^{\phi \phi})^2}{C_{\ell_2}^{\phi \phi}+N_{\ell_2}^{\phi \phi} } \, , 
\end{equation}
where $f_{\ell \ell_1 \ell_2}^{EB}$ is a geometric coupling factor. The delensed $B$ mode is then given by
\begin{equation}
\clbbd \equiv \clbbl - \clbbe.
\label{eq:delensing_def}
\end{equation}
Note that the presence of the noise in Eq.~\eqref{eq:smith_delensing} guarantees that $\clbbl \geq \clbbe$.

In this work, we consider three sources for the lensing potential estimate: the CMB polarization itself (``CMB$\times$CMB'' delensing), the cross-correlation of the CMB and the cosmic {\em infrared} background (``CMB$\times$CIB''), and measurements of the large-scale structure using, for example, cosmic shear or 21cm radiation  (``CMB$\times$LSS''). In the CMB$\times$CMB case, the noise on this estimate is~\cite{2003PhRvD..67h3002O}
\begin{equation}
\label{eq:eb_estimator}
N_\ell^{\phi\phi} = \left[ \frac{1}{2\ell+1} \sum_{\ell_1 \ell_2} |f_{\ell_1 \ell_2 \ell}^{EB}|^2 \left( \frac{1}{C_{\ell_1}^{BB} + N_{\ell_1}^{BB}} \right) \times \left( \frac{(C_{\ell_2}^{EE})^2}{C_{\ell_2}^{EE} + N_{\ell_2}^{EE}} \right) \right]^{-1}.
\end{equation}
The crucial observation here is that the lensed $B$ mode appears as a source of noise in the lensing potential estimator. Its performance therefore improves if the {\em delensed} $B$ mode is substituted for the lensed $B$ mode, in turn allowing better estimates of the lensing $B$ mode, and so on. Iterating this estimator can greatly improve the ability of an experiment to delens the CMB~\cite{2004PhRvD..69d3005S}: indeed, in the noiseless limit perfect delensing is achievable. For realistic instrumental configurations, this process converges after a few steps, where we use the convergence criterion as below:
\begin{equation}
\left| \sum_\ell \frac{ N_{\ell}^{\phi \phi, i} - N_{\ell}^{\phi \phi, i-1} }{ N_{\ell}^{\phi \phi,i} } \right|\,\leq\,1\%.
\label{eq:iterative_delensing_def}
\end{equation}

CMB$\times$CIB and CMB$\times$LSS delensing rely on non-CMB datasets for their lensing potential estimates, a feature that is particularly useful for experiments with moderate polarization sensitivities. The CIB is known to be a good tracer of the lensing potential~\cite{2003ApJ...590..664S,2013ApJ...771L..16H,2014A&A...571A..18P,2015arXiv151105116B}, and Refs.~\citep{2015arXiv150205356S, 2014arXiv1410.0691S} have recently demonstrated its utility in delensing the CMB. Following their example, to simulate CMB$\times$CIB delensing we make the replacement
\begin{equation}
C_\ell^{\phi \phi}+N_\ell^{\phi \phi} \rightarrow C_\ell^{\phi \phi} / \rho_\ell^2
\label{eq:rho_cor_cib}
\end{equation}
in Eq.~\eqref{eq:smith_delensing}, where $\rho_\ell$ is the $\ell${\em -dependent} correlation coefficient between the CMB lensing and CIB, as presented in Fig. 1 of~Ref.~\cite{2015arXiv150205356S}.

The CMB$\times$LSS delensing case requires high signal-to-noise large-scale structure measurements extending to high redshift (e.g., cosmic shear~\citep{2007PhRvD..76l3009M} or 21cm radiation~\citep{2006ApJ...653..922Z, 2005PhRvL..95u1303S, 2006ApJ...647..719M} measurements). Following Ref.~\citep{2012JCAP...06..014S}, we assume a (crude but illustrative) sharp cutoff in LSS tracers at $z_{\rm max}$, providing a perfect estimate of the gravitational potential out to this redshift. In this setting, the estimated lensing $B$ mode takes the same form as Eq.~\eqref{eq:smith_delensing}, with the following replacement 
\begin{equation}
\frac{(C_{\ell}^{\phi \phi})^2}{C_{\ell}^{\phi \phi}+N_{\ell}^{\phi \phi} } \rightarrow C_\ell^{\phi \phi,\,z_{\rm max}},
\label{eq:lss_delens}
\end{equation}
where $C_\ell^{\phi \phi,\,z_{\rm max}}$ is the power spectrum of the lensing potential including contributions from $z \le z_{\rm max}$ only (calculated using a modified version of {\tt CAMB}~\cite{2000ApJ...538..473L}). In this case, delensing performance is limited by the noise in the $E$ measurements and the redshift range of the LSS measurements. In this work, we take $z_{\rm max}$ to be 3.5: the proposed limit of the capabilities of the Large Synoptic Survey Telescope (LSST)~\cite{LSST_Science_Book}.

To evaluate the quality of the various delensing techniques in each experimental context we introduce the delensing factor $\alpha$, adapted from Ref.~\cite{2015arXiv150205356S} and analogous to Eq.~\eqref{eq:reff_def}:
\begin{equation}
\alpha \equiv \frac{ \sum_{\ell=20}^{200} \clbbd }{ \sum_{\ell=20}^{200} \clbbl }.
\label{eq:alpha_def}
\end{equation}
A summary of the delensing options is presented in Table~\ref{table:delens_options}. 

\begin{table}
\caption{Delensing options employed in this work and references to implementation details.}
\begin{center}
\begin{tabular}{|c|c|c|c|c|}
\hline
label & no delensing & CMB $\times$ CMB  & CMB $\times$ CIB & CMB $\times$ LSS  \\
\hline
references & - & Ref.~\cite{2012JCAP...06..014S} & Ref.~\cite{2015arXiv150205356S} & Ref.~\cite{2012JCAP...06..014S} \\ 
\hline 
properties & no delensing & iterative delensing & delensing using & delensing with \\ 
& performed & using CMB &  correlation between & perfect LSS \\ 
& & polarization only & CMB and CIB & measurements \\ 
& & & & to $z_{\rm max}$ \\ 
\hline
\end{tabular}
\end{center}
\label{table:delens_options}
\end{table}

The left panel of Fig.~\ref{fig:sigma_r_mnu_vs_fsky_uKarcmin} shows the delensing factor $\alpha$ introduced above as a function of the {\em post-component-separation} sensitivity of a CMB experiment with a Gaussian beam with full-width at half-maximum (FWHM) of $3^\prime$. Plotted are all three delensing methods considered in this work: iterative delensing using the CMB polarization (Eqs.~\eqref{eq:eb_estimator} and \eqref{eq:iterative_delensing_def}), and via cross-correlation with the CIB (Eq.~\eqref{eq:rho_cor_cib}) and LSS (Eq.~\eqref{eq:lss_delens}). The first point to note is that delensing using the CMB$\times$LSS correlation only slightly outperforms CMB$\times$CIB delensing, even though CMB$\times$CIB delensing makes use of existing \planck\ data products~\cite{2015arXiv150205356S}. Moreover, for both cross-correlation methods the sensitivity of the CMB instrument does not matter in the regime $\leq 5-10$ \ukarc, as the $E$-mode observations are largely cosmic-variance limited for these sensitivities (see Eq.~\eqref{eq:smith_delensing}). Iterative delensing using CMB polarization becomes more effective than the cross-correlation techniques when the post-component separation noise level drops below $\sim\,3$ \ukarc.\footnote{Other quadratic combinations of CMB fields --- and, indeed, their minimum-variance combination~\cite{2002ApJ...574..566H} --- will outperform the ``$EBEB$'' estimator considered here in the high-noise regime.}
This is a consequence of $N_\ell^{\phi\phi}$, Eq.~\eqref{eq:eb_estimator}, which continues to decrease as the instrumental sensitivity increases, in the framework of iterative delensing~\cite{2004PhRvD..69d3005S}.

One final point is to note that the curves become thicker as the noise level decreases. This effect is due to varying the area of sky to which the experiment has access (sky fractions considered here span the range 1--100\%). Larger-scale experiments have access to more multipoles (we assume $\lmin \propto \fsky^{-1/2}$) which allows for improved iterative delensing at low noise levels.

\begin{figure}
\centering
\includegraphics[width=7.5cm]{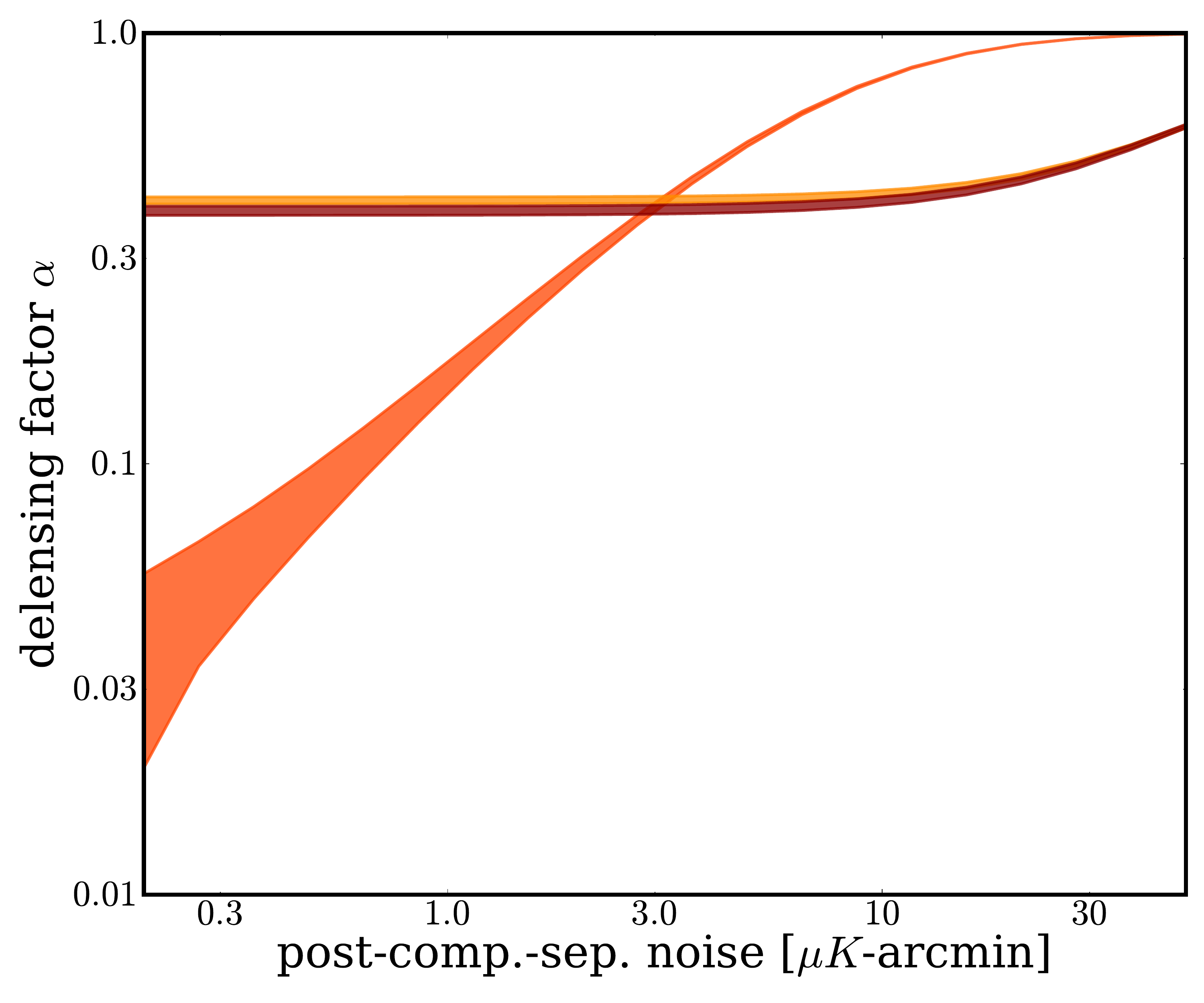}~\includegraphics[width=7.5cm]{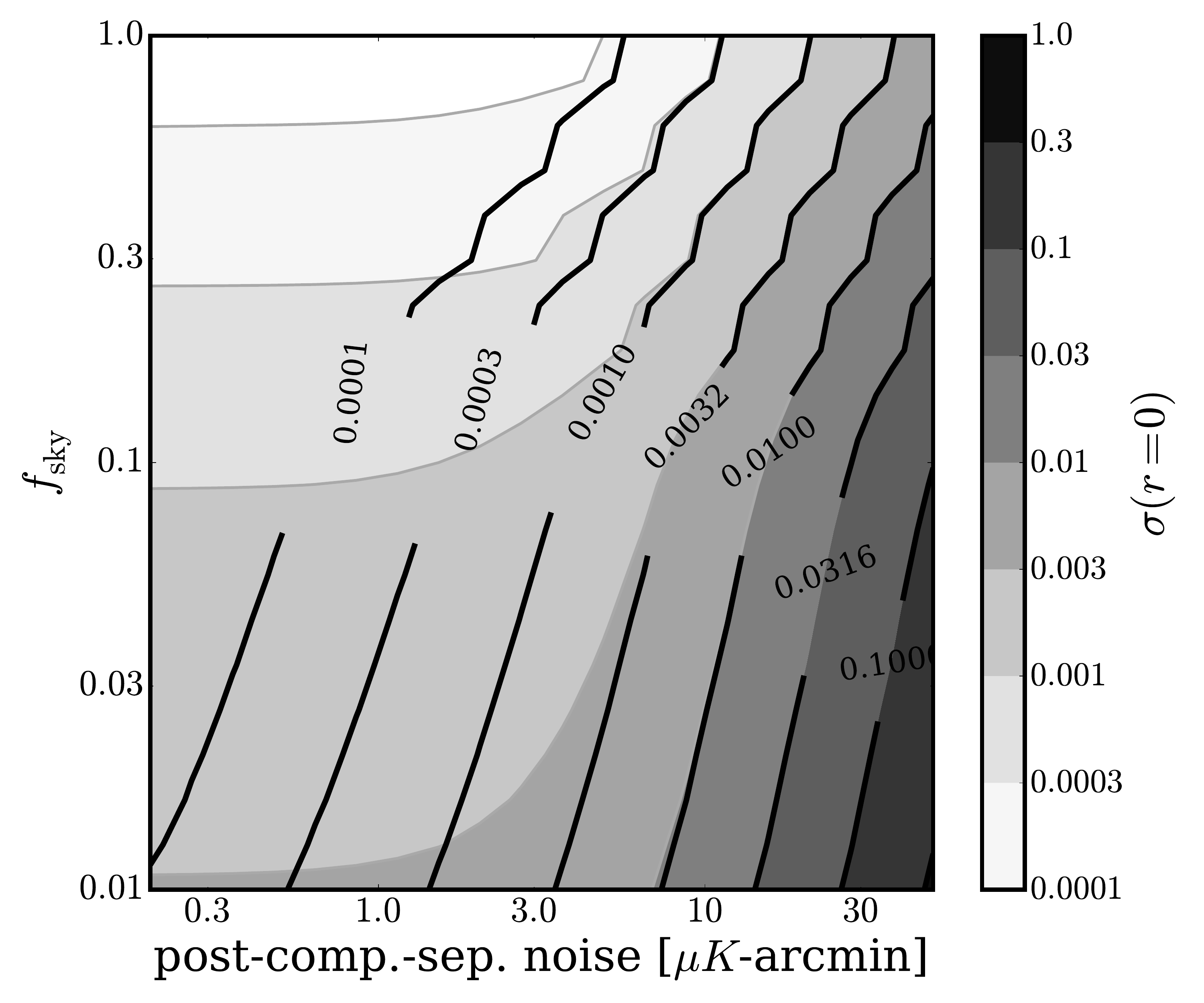}
\caption{\textit{Left panel:} the delensing improvement factor, $\alpha$ (defined in Eq.~\eqref{eq:alpha_def}), for iterative CMB$\times$CMB (dark orange), CMB$\times$CIB (light orange) and CMB$\times$LSS (dark red) delensing, as a function of the noise in the CMB map after component separation. We assume FWHM=$3'$ and a multipole range of $\sqrt{ 2\pi / \fsky} \le \ell \le 3000$. Figure~\ref{fig:sigma_cmb_example} and its accompanying discussion describe three-frequency experimental configurations producing comparable noise levels after the removal of dust. \textit{Right panel:} One-sigma limit on $r=0$, $\sigma(r=0)$, as a function of both the noise in the CMB map and the observed fraction of the sky. Grey (black) contours correspond to the no delensing (CMB $\times$ CMB  delensing) case. No foreground residuals are taken into account in this result.}
\label{fig:sigma_r_mnu_vs_fsky_uKarcmin}
\end{figure}

\subsection{Scientific performance forecast}
\label{sec:forecast_formalism}

We adopt a Fisher matrix approach to provide estimates of the scientific potential of each experimental configuration~\cite{1997astro.ph..7265T}. Following, e.g., Refs.~\cite{2009AIPC.1141..121S, 2014ApJ...788..138W}, we write the Fisher matrix element $F_{ij}$ for CMB spectra as
\begin{equation}
F_{ij} = \sum_{\ell=\lmin}^{\lmax} \frac{2\ell+1}{2} \fsky {\rm tr} \left(  \boldsymbol{C}^{-1}_\ell \frac{\partial \boldsymbol{C}_\ell}{\partial \theta_i} \boldsymbol{C}^{-1}_\ell \frac{\partial \boldsymbol{C}_\ell}{\partial \theta_j}  \right) \, ,
\label{eq:fisher_definition}
 \end{equation}
where $\theta_i$ and $\theta_j$ are two of the cosmological parameters of interest. The covariance matrix $\boldsymbol{C}_\ell$ is defined as
\begin{equation}
\label{eq:covariance_definition} 
 \boldsymbol{C}_\ell \equiv 
 \begin{bmatrix}
\bar{C}_\ell^{TT} + N_\ell^{TT} & \bar{C}_\ell^{TE} & 0 & C_\ell^{Td} \\
\bar{C}_\ell^{TE} & \bar{C}_\ell^{EE} + N_\ell^{EE} & 0 & C_\ell^{Ed} \\
0 & 0 & \bar{C}_\ell^{BB} + N_\ell^{BB} & 0  \\
C_\ell^{Td}  & C_\ell^{Ed} & 0 & C_\ell^{dd} + N_\ell^{dd}
\end{bmatrix},
\end{equation}
where $C_\ell$ are the various auto- and cross-power spectra of the CMB temperature ($T$), polarization ($E, B$) and deflection ($d$) components. So as to not double-count the lensing information encapsulated in the deflection field, we use only unlensed $T$, $E$ and $B$ information, as denoted by barred $C_\ell$s~\cite{Lesgourgues}. In Eq.~\eqref{eq:covariance_definition}, the diagonal elements of the covariance matrix contain ``catch-all'' Gaussian noise terms $N^{XX}_\ell$. For the components $X = \{T,E,B\}$, this noise power spectrum accounts for the effects of instrumental noise, imperfect foreground removal and, in the case $X = B$, delensing:
 \begin{equation}
N^{BB}_\ell = N^{BB,\,{\rm  inst}}_\ell + \clfgr + \clbbd;
\label{eq:ellnoise}
\end{equation}
\clfgr\ is defined in Eq.~\eqref{eq:Clres_def} and $\clbbd$ in Eq.~\eqref{eq:delensing_def}. Interpreting the foreground residuals as a bias with unknown shape, we parametrize \clfgr as 
\begin{eqnarray}
	\centering
		\clfgr \equiv A_{\rm fg\ res}\times C^{\rm fg\ res}_{\ell_0}\times\left( \frac{\ell}{\ell_0}\right)^{b_{\rm fg\ res}}
	\label{eq:clfgrs_parametrization}
\end{eqnarray}
and derive all Fisher constraints on cosmological parameters after marginalizing over $A_{\rm fg\ res}$ and $b_{\rm fg\ res}$ (when foregrounds are present, cf. Table~\ref{table:models}). The choice of parametrization in Eq.~\eqref{eq:clfgrs_parametrization} is similar to one chosen in Ref.~\cite{2011JCAP...08..001F}.
The instrumental noise power spectra,  $N^{XX,\,{\rm  inst}}_\ell$, are given by~\cite{1995PhRvD..52.4307K}
\begin{equation}
N^{XX,\,{\rm  inst}}_\ell  = \left[ \sum_\nu{ w_{X,\nu} \exp \left(- \ell(\ell+1) \frac{\theta^{\ 2}_{\textsc{fwhm},\nu}}{8\log2}\right)} \right]^{-1},
\label{eq:beamnoise}
\end{equation}
where $w_{X,\nu}^{-1/2}$ is the instrumental white noise level of a given frequency channel $\nu$ in $\mu$K$_{\rm CMB}$-rad and $\theta_{\textsc{fwhm},\nu}$ is the full-width at half-maximum beam size in radians. We assume fully polarized detectors, such that $w_E^{-1/2} = w_B^{-1/2} = \sqrt{2} w_T^{-1/2}$. Eq.~\eqref{eq:beamnoise} is only valid in its given format in the case of no component separation, i.e., the CMB-only case of Table~\ref{table:fgs_options}. For the realistic cases in which component separation is performed, we rescale $N_\ell^{XX,\,{\rm  inst}}$ using $\Delta$, the noise degradation due to component separation defined in Eq.~\eqref{eq:Delta_def}, such that
\begin{equation}
N^{\rm XX,\,{\rm inst}}_\ell \rightarrow \Delta N_\ell^{\rm XX,\,{\rm  inst}}.
\label{eq:post_comp_sep_Nl_rescaling}
\end{equation}
The uncertainty on the reconstruction of the deflection field, $N_\ell^{dd}$, depends on the delensing procedure adopted. When the lensing potential estimate is derived solely using the CMB (the ``no delensing'' and ``CMB$\times$CMB'' delensing cases), the noise is given by Eq.~\eqref{eq:eb_estimator}, appropriately reweighted in $\ell$. When external data are used, $N_\ell^{dd}$ can be derived from the form of the correlations between lensing tracers and the potential (Eqs.~\eqref{eq:rho_cor_cib} and~\eqref{eq:lss_delens}).

The Fisher matrix (Eq.~\eqref{eq:fisher_definition}) requires partial derivatives of the covariance matrix with respect to the parameters of interest. To calculate these derivatives numerically, we evaluate four additional sets of power spectra for each parameter. For parameters with non-zero values in the fiducial cosmology, the additional power spectra are calculated at 90\%, 95\%, 105\% and 110\% of the parameter's fiducial value; for parameters whose fiducial values are zero, we calculate the additional power spectra at $\pm0.1$ and $\pm0.05$.\footnote{The exceptions to these rules are the energy density in curvature, $\omk$, and the sum of the neutrino masses, $\mnu$. For the curvature we calculate power spectra at $\omk = \pm0.002$ and $\pm0.001$; for $\mnu$ we use 80\%, 90\%, 110\% and 120\% of the fiducial value.} We interpolate between these and the fiducial power spectra at each $\ell$, using a cubic spline to get a smooth curve, before finally forming a two-sided numerical derivative with very small step size ($10^{-3}$) from the interpolated power spectra. The step sizes used during the first stage are tuned so that derivative computations converge for all multipoles and numerical noise is minimal.

The Fisher formalism allows forecasting of uncertainties either conditional on the other parameters taking their fiducial values or marginalized over the parameters taking any value. Conditional errors are given simply by the inverse of individual entries in the Fisher matrix, $1/\sqrt{F_{ij}}$; marginal errors, which we employ throughout, are given by inverting the Fisher matrix:
\begin{equation}
\sigma_i \equiv \sigma (p_i) = \sqrt{[\mathbf{F}^{-1}]_{ii}}.
\end{equation}

As an illustration of the method described above, the right panel of Fig.~\ref{fig:sigma_r_mnu_vs_fsky_uKarcmin} plots the one-sigma limits on $r=0$, $\sigma(r=0)$, as a function of the $3^\prime$-FWHM example experiment's post-component-separation noise level and fraction of sky covered. Grey filled contours show the limits without delensing, black contours the limits with iterative CMB$\times$CMB delensing. We assume only $B$-mode information is used to derive the limits on $r$, and all other cosmological parameters are fixed; in this setting, the Fisher formalism reduces to~\cite{2000PhRvD..61h3501J}
\begin{equation}
\label{eq:bb_r_only_fisher}
\sigma(r=0) = \left[ \sum_\ell \frac{(2\ell+1) \fsky}{2} \left(\frac{C_\ell^{BB,\,{\rm prim}}(r=1)}{N_\ell^{BB}}\right)^2 \right]^{-1/2},
\end{equation}
where the total $B$-mode noise, $N_\ell^{BB}$, comprises the post-component-separation instrumental noise and the lensed $B$ mode or its post-delensing residual, as appropriate. Without delensing, the limits on $r$ achievable by low-noise experiments (with noise levels $\lesssim 3$ \ukarc) are fixed by the sky fraction, which sets the number of modes available for analysis.\footnote{The stepped appearance of the contours for large sky fractions is due to the conversion of the minimum multipole to an integer for the sum over modes.} Iterative CMB delensing completely removes this $\fsky$ performance floor, allowing arbitrary improvement in the limits on $r=0$ as the experimental noise approaches zero~\cite{2004PhRvD..69d3005S}. The impact of delensing is greatest for experiments with smaller sky fractions as the variance due to lensing $B$ modes becomes increasingly dominant over the putative signal as $\ell$ increases (cf. Eq.~\eqref{eq:bb_r_only_fisher} and Fig.~\ref{fig:input_foregrounds_plot}).

\subsection{Models and fiducial cosmologies}
\label{ssec:models_fid_cosmologies}

The aim of this work is to estimate the ability of upcoming experiments to constrain the parameters of extensions to the standard cosmological model, \lcdm, in the inflationary, neutrino and dark energy sectors. Our analysis therefore always includes the standard six \lcdm\ parameters: the spectral index, $\ns$, and amplitude, $\as$, of the power spectrum of scalar perturbations, the optical depth to reionization, $\tau$, the Hubble constant, $\h = 100h$ km/s/Mpc, and the current energy densities in baryons, $\ombh$, and cold dark matter, $\omch$. The extended models are described by the running of the scalar spectral index $\alphas$, the ratio of the primordial tensor and scalar power spectra $r$, the spectral index of the tensor power spectrum, $\nt$, the current energy density in curvature, $\omk$, the sum of the neutrino masses, $\mnu$, the number of relativistic degrees of freedom in the early Universe, $\neff$, and two parameterizations describing the dark energy equation of state: a constant equation of state parameter, $w$, and a time-dependent version, $\w\equiv w(a) = \wo + (1-a)\wa$, where $a$ is the scale factor~\cite{2001IJMPD..10..213C,2003PhRvL..90i1301L}. When deriving constraints on the additional parameters, we do so in the context of minimally extended models: for example, constraints on $r$ are derived assuming the true cosmological model can be described by the six \lcdm\ parameters and $r$ only; constraints on $\neff$ are derived assuming a six-parameter \lcdm\ cosmology with a variable number of relativistic degrees of freedom, etc.\footnote{The one exception to this rule are the constraints on $\ns$ and $\alphas$, which are derived from a model with ``full'' inflationary freedom: a running scalar spectral index, tensors with unknown amplitude and spectral index, and curvature.} Precise definitions of the models considered can be found in Table~\ref{table:models}, along with the parameter constraints for which the models are used. In practice, we calculate the Fisher matrix once for all parameters and derive constraints on individual parameters of interest by inverting the submatrices corresponding to the appropriate models.

\begin{table}
\begin{center}
\begin{tabular}{|l|c|c|c|c|c|c|c|c|c|c|c|}
\hline
model & $A_{\rm fg\ res}$ & $b_{\rm fg\ res}$ & $\alphas$ & $r$ & $\nt$ & $\omk$ & $\mnu$ & $\neff$ & $\w$ & $\wa$ & {\rm constraints}\\
\hline
\lcdmr\ & $\checkmark$ &$\checkmark$ && $\checkmark$ &  &  &  &  &  & & $\sigma( r )$ \\
\lcdmrnt & $\checkmark$ &$\checkmark$ && $\checkmark$ & $\checkmark$ &  &  &  &  & &  $\sigma( \nt )$\\
\lcdmi & $\checkmark$ &$\checkmark$ &$\checkmark$ & $\checkmark$ & $\checkmark$ & $\checkmark$ &  &  &  & & $\sigma(\ns)$, $\sigma(\alphas)$\\
\lcdmk &  $\checkmark$ &$\checkmark$ &&  &  & $\checkmark$ &  &  &  & &  $\sigma(\omk)$ \\
\lcdmmnu & $\checkmark$ &$\checkmark$ & &  &  &  & $\checkmark$ &  &  & & $\sigma(\mnu)$ \\
\lcdmneff & $\checkmark$ &$\checkmark$ & &  &  &  &  & $\checkmark$ &  & & $\sigma(\neff)$\\
\wcdm &  $\checkmark$ &$\checkmark$ & &  &  &  &  &  & $\checkmark$ & & $\sigma(\w)$ \\
\wwacdm &  $\checkmark$ &$\checkmark$ & &  &  &  &  &  & $\checkmark$ & $\checkmark$ & $\sigma(\wo)$, $\sigma(\wa)$ \\
\hline
\end{tabular}
\end{center}
\caption{Models considered and their parameterizations. Constraints reported on additional cosmological parameters are derived using specific cosmologies, as given in the right column. In addition, all constraints are obtained after marginalization over $A_{\rm fg\ res}$ and $b_{\rm fg\ res}$, the foreground residuals' amplitude and $\ell$-dependence.}
\label{table:models}
\end{table}

Our fiducial cosmology is a flat \lcdm\ universe with one massive and two massless neutrinos. We set the majority of our parameter values to the best-fit $\Lambda$CDM values found in \planck's analysis of a compilation of CMB, lensing, baryon acoustic oscillation, supernova and expansion data~\cite{2015arXiv150201589P}. Specifically, we use $\ns = 0.9667$, $\as = 2.142\times10^{-9}$, $\tau = 0.066$, $\h=67.74$ km/s/Mpc, $\ombh = 0.0223$, $\omch = 0.1188$, $\alphas = \omk = 0.0$, $\mnu = 0.06$ eV, $\neff = 3.046$, $\w = \wo = -1$ and $\wa=0$. For the majority of models we additionally assume a very small tensor component (satisfying the inflationary consistency relation~\cite{2014AA...571A..22P}, $\nt = -r/8$), setting $r = 10^{-3}$, $\nt = -1.25\times10^{-4}$. In order to derive interesting constraints on $\nt$ (using the \lcdm+$r$+$\nt$ model) we boost this tensor component to $r = 10^{-1}$, $\nt = -1.25\times10^{-2}$. In all cases, the scalar and tensor spectral indices are measured at pivot scales of 0.05 and 0.002 Mpc$^{-1}$, respectively.

\subsection{Instruments}
\label{ssec:missions}

\begin{figure}
\centering
\includegraphics[width=7.5cm]{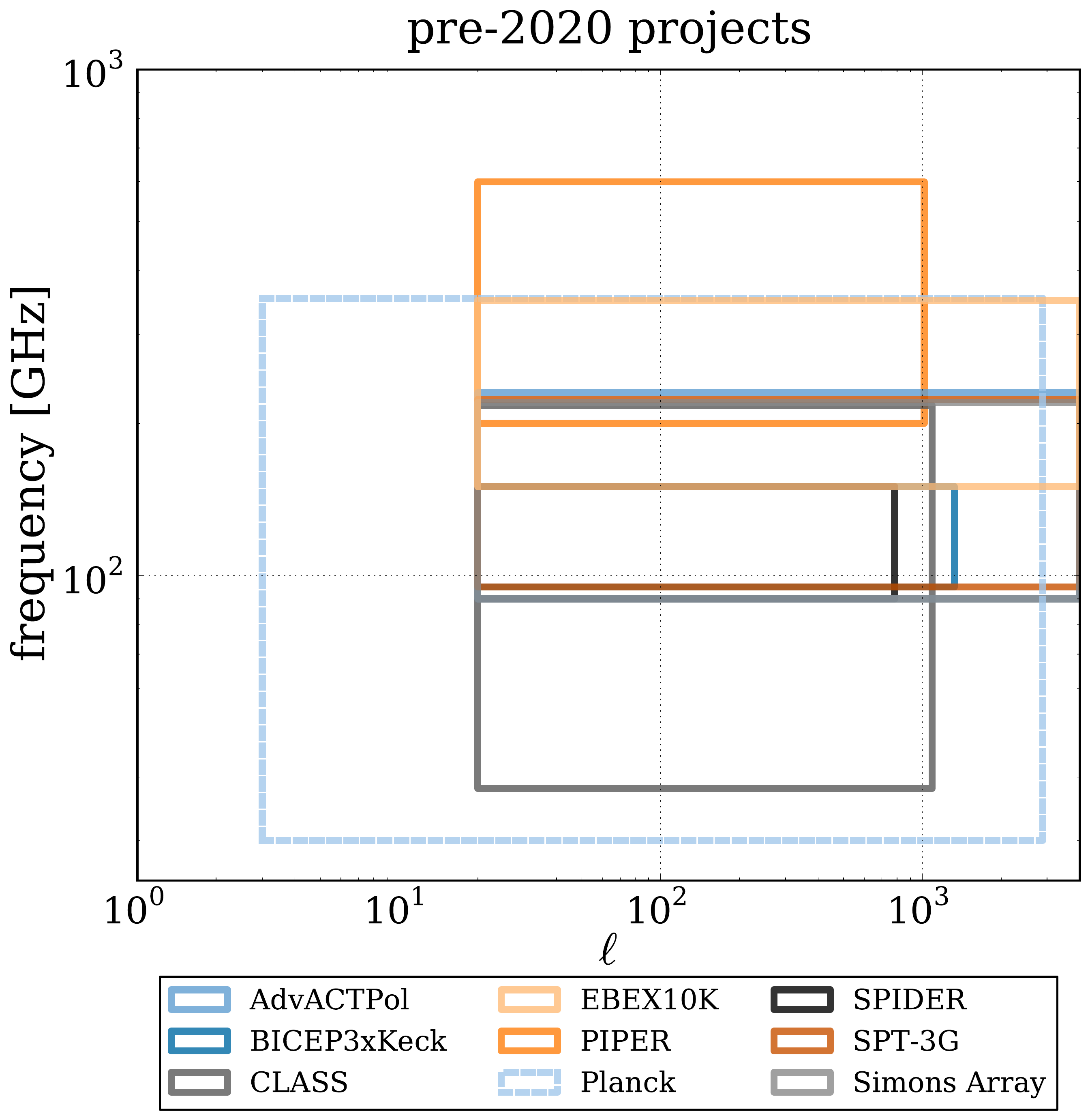}~\includegraphics[width=7.5cm]{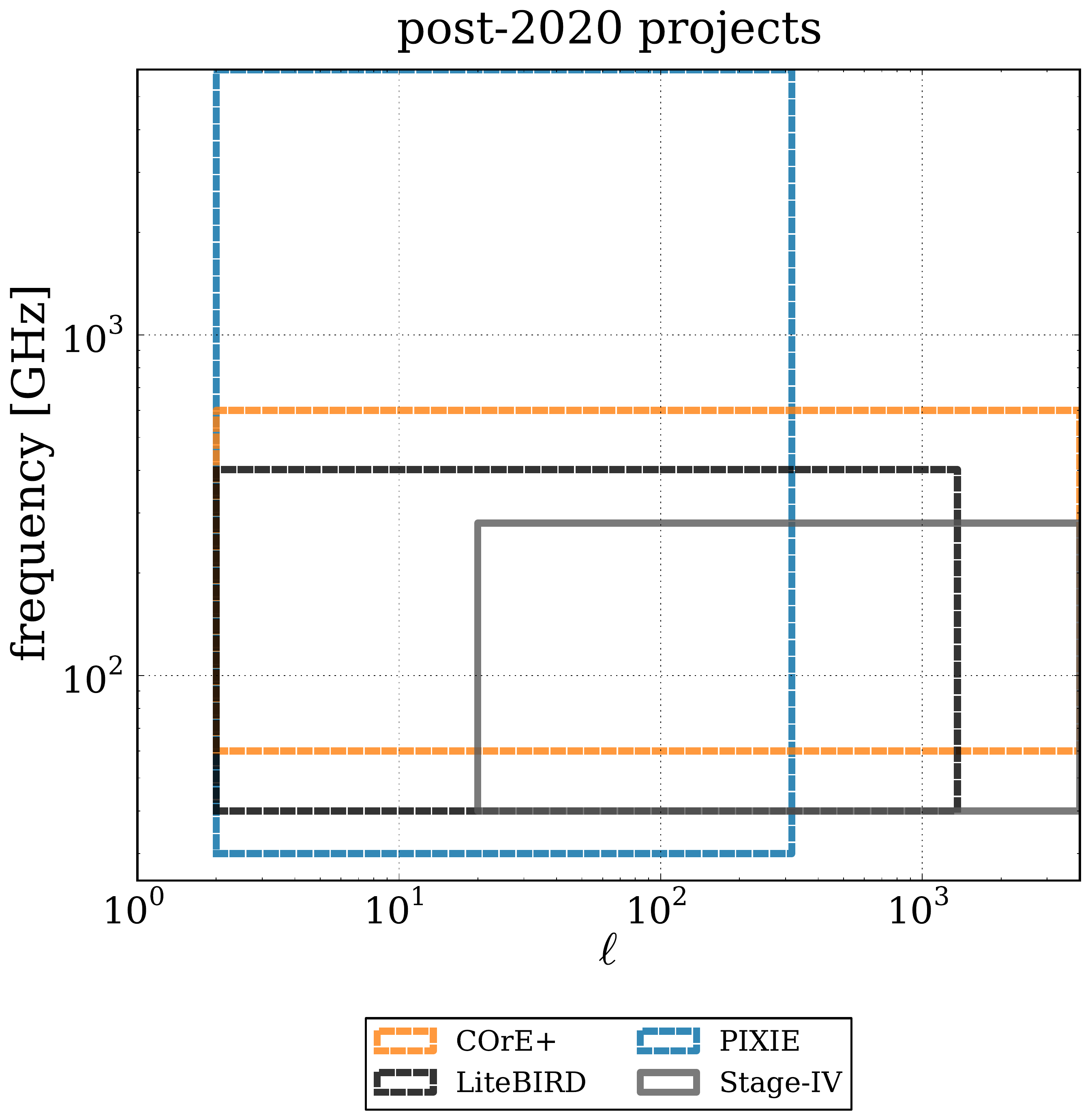}
\caption{Illustration of the frequency and multipole coverage of the instruments detailed in Sect.~\ref{ssec:missions} and Appendix~\ref{app:instruments_specifications}. The $\lmax$ values plotted indicate the multipole at which the noise and beam dominate all cosmological signal (where this is lower than the maximum multipole considered in this work, 4000).}
\label{fig:freq_ell_coverage}
\end{figure}

The full scope of possible instrumental configurations that could be tested by our framework is too large to be considered in this work, and the code is accordingly made available via a web interface, as described in Sect.~\ref{sec:web_interface}.
To illustrate the different options and characteristics of our algorithm, we focus on various experiments already in place or expected to deploy in two timeframes, 2015--2020 and 2020--2030, namely:
\begin{itemize}
\item {\bf 2015--2020.} The ground-based Stage-III experiments Advanced ACTPol\footnote{Ref.~\cite{2014JCAP...08..010C} and E. Calabrese (private comm.). Note that an update to the Advanced ACTPol specifications~\cite{2015arXiv151002809H} appeared after the completion of this work. The updated design includes additional channels at 28 and 41 GHz, which will certainly help Advanced ACTPol disentangle synchrotron from the CMB.}, BICEP3 and the Keck Array\footnote{Z. Ahmed (private comm.).}, CLASS\footnote{Ref.~\cite{2014SPIE.9153E..1IE} and T. Essinger-Hileman (private comm.).}, SPT-3G\footnote{Ref.~\cite{2014SPIE.9153E..1PB} and B. Benson (private comm.).}, and the Simons Array\footnote{Ref.~\cite{2014SPIE.9153E..1FA} and A. Lee and K. Arnold (private comm.).}, as well as the balloon-borne missions EBEX10K\footnote{Proposal to NASA in 2015. and S. Hanany (private comm.).}, PIPER\footnote{Ref.~\cite{2014SPIE.9153E..1LL} and J. Lazear (private comm.).} and SPIDER\footnote{Ref.~\cite{2008SPIE.7010E..2PC} and C. Contaldi (private comm.).}. We consider these experiments both alone and in combination with \planck\footnote{Ref.~\cite{2015arXiv150201588P}.}. The balloon observatories are designed at least partially as dust-monitoring observatories through the use of frequencies well above $150$\;GHz; Ref.~\cite{2015PhRvL.114j1301B} has unequivocally demonstrated that the presence of dust-monitoring channels with similar sensitivity to the CMB channels is crucial to the search for primordial B modes at any Galactic latitude.
\item {\bf 2020--2030.} A ground-based Stage-IV-like instrument\footnote{In the absence of a finalized design for Stage-IV, we choose its specifications so that its noise after component separation, $\sigma_{CMB}$, is around $1$ \ukarc, Ref~\cite{2014ApJ...788..138W,2015APh....63...66A}.} and the satellites \lb\footnote{\url{http://ltd16.grenoble.cnrs.fr/IMG/UserFiles/Images/09\_TMatsumura\_20150720\_LTD\_v18.pdf}}, COrE+\footnote{\url{http://conservancy.umn.edu/handle/11299/169642}} and PIXIE\footnote{Ref.~\cite{2011JCAP...07..025K} and A. Kogut (private comm.).}. The ground-based Stage-IV-like instrument is assumed to have increased sky and frequency coverage compared to its Stage-III counterparts; the satellites provide high-sensitivity, broad-frequency coverage of the entire sky. We forecast the performance of these instruments both in isolation and in combination.
\end{itemize}

Figure~\ref{fig:freq_ell_coverage} illustrates the frequency coverage and multipole range observed by each instrument (reproducing the format used in Ref.~\cite{2015arXiv150800017W}). We define the minimum observable multipole for each experiment using its sky patch to be $\lmin = \lceil\pi / (2\fsky^{1/2})\rceil$, rounding up to the next integer. We further assume that systematic effects such as atmospheric contamination~\cite{2015ApJ...809...63E} and $E$-$B$ leakage~\cite{2001PhRvD..64f3001T, 2015arXiv150606409F} will render the lowest multipoles of sub-orbital experiments unsuitable for science, and set an $\lmin$ floor of 20 for all such experiments (see Ref.~\cite{2015arXiv150800017W,2015arXiv150904714R} for more optimistic forecasts). We assume a maximum multipole $\lmax = 4000$ for all experiments, though the noise and beam for many of those considered becomes dominant at much lower multipoles. In Fig.~\ref{fig:freq_ell_coverage}, we plot an effective maximum multipole, chosen to be the lower of 4000 and the multipole at which the instrumental noise reaches $10^4 \, \mu{\rm K}^2$, where it dominates over even the temperature power spectrum. Note that the formalism does not presently account for any further systematic errors beyond those encoded in the choice of these parameters.

In addition to these experiments, we also consider C-BASS\footnote{\url{http://www.astro.caltech.edu/cbass}} and QUIJOTE-CMB\footnote{Ref.~\cite{2012SPIE.8444E..2YR}.} as dedicated synchrotron monitors when examining the more challenging $\mathbf{A}$-expansion approach to handling spatially varying foreground spectral indices. Due to QUIJOTE-CMB's location in the Northern Hemisphere, we include it only in conjunction with space-based missions guaranteeing considerable overlap in coverage. The specifications for all experiments can be found in Tables~\ref{table:quijote_cbass_planck_specs},~\ref{table:pre2020_specs} and~\ref{table:post2020_specs} of Appendix~\ref{app:instruments_specifications}. Note that a number of these experiments are still at the proposal stage, and their precise specifications are therefore subject to change.

When considering the combination of two instruments A and B, to avoid double-counting data we add three Fisher matrices, the first containing information from both instruments from the patch of sky they have in common, the second containing any remaining information from the instrument covering the larger sky fraction, and the third containing any low-$\ell$ information from the common sky patch that is only accessible to one experiment. The combined Fisher matrix is then
\begin{equation}
\mathbf{F}^{\rm A\times B} \equiv \mathbf{F}(\fsky^{\rm A\times B}) + \mathbf{F'}(\Delta \fsky) + \mathbf{F''}_{{\rm low}-\ell}(\fsky^{\rm A\times B}),
\end{equation}
where $\fsky^{\rm A\times B}$ is the common sky fraction and 
\begin{equation}
\Delta \fsky \equiv \max\left( \fsky^{\rm A}, \fsky^{\rm B} \right) - \fsky^{\rm A\times B}
\end{equation}
is the remainder (should there be any). To build $\mathbf{F}(\fsky^{\rm A\times B})$, we take the overlap in multipole and sky coverage of the two instruments, i.e.,
\begin{align}
\lmin^{\rm A\times B} &=  \max\left( \lmin^{\rm A}, \lmin^{\rm B} \right) \\
\fsky^{\rm A\times B} &=  \min\left( \fsky^{\rm A}, \fsky^{\rm B} \right); \nonumber\label{eq:combination_specs}
\end{align}
as stated previously, we consider $\lmax = 4000$ in all cases. With these equations satisfied, the frequency channels, sensitivities and beams of the two instruments are concatenated and fed into the foreground-cleaning, delensing and Fisher algorithms. 
$\mathbf{F'}(\Delta \fsky)$, if needed, is computed using the specifications of the instrument having the largest sky coverage, substituting $\Delta \fsky$ for the instrument's original sky fraction. Similarly, $\mathbf{F''}_{{\rm low}-\ell}(\fsky^{\rm A\times B})$, if needed, is computed using the specifications of the instrument observing the lowest multipoles, using only the multipoles between $\min\left( \lmin^{\rm A}, \lmin^{\rm B} \right)$ and $\lmin^{\rm A\times B}$.

As stated in Sect.~\ref{ssec:delensing}, we consider two further sources of experimental data for delensing purposes: \planck's 545\;GHz data~\cite{2015arXiv150205356S} are used in the CMB$\times$CIB delensing forecast (Eq.~\eqref{eq:rho_cor_cib}), and LSST-like LSS observations out to redshift $z_{\rm max}=3.5$~\cite{LSST_Science_Book} are used in the CMB$\times$LSS delensing forecast (Eq.~\eqref{eq:lss_delens}).

\section{Results}
\label{sec:results}

\subsection{Foreground cleaning}

In this section, we will first illustrate the detailed foreground cleaning performance of selected experiments from each timeframe considered, before providing comprehensive cleaning results for the full set of experiments.
Figure~\ref{fig:comp_sep} demonstrates the ability of two pre-2020 and two post-2020 experiments (and their cross-correlations) to clean a variety of foreground components, under the $n_p$-independent patches approach. Overlaid upon the power spectra of primordial $B$ modes with $r = \{ 10^{-3}, 10^{-2}, 10^{-1}\}$ (solid dark grey) and lensing $B$ modes (solid light grey), we plot the post-component-separation residuals (short-dashed) and CMB map noise (long-dashed) for the dust-only (dark orange), synchrotron-only (light orange) and synchrotron plus dust (dark red) cases. The raw quadratic combination of the sensitivities in all channels is plotted as a dotted black curve.

\begin{figure}
\centering
\includegraphics[width=15cm]{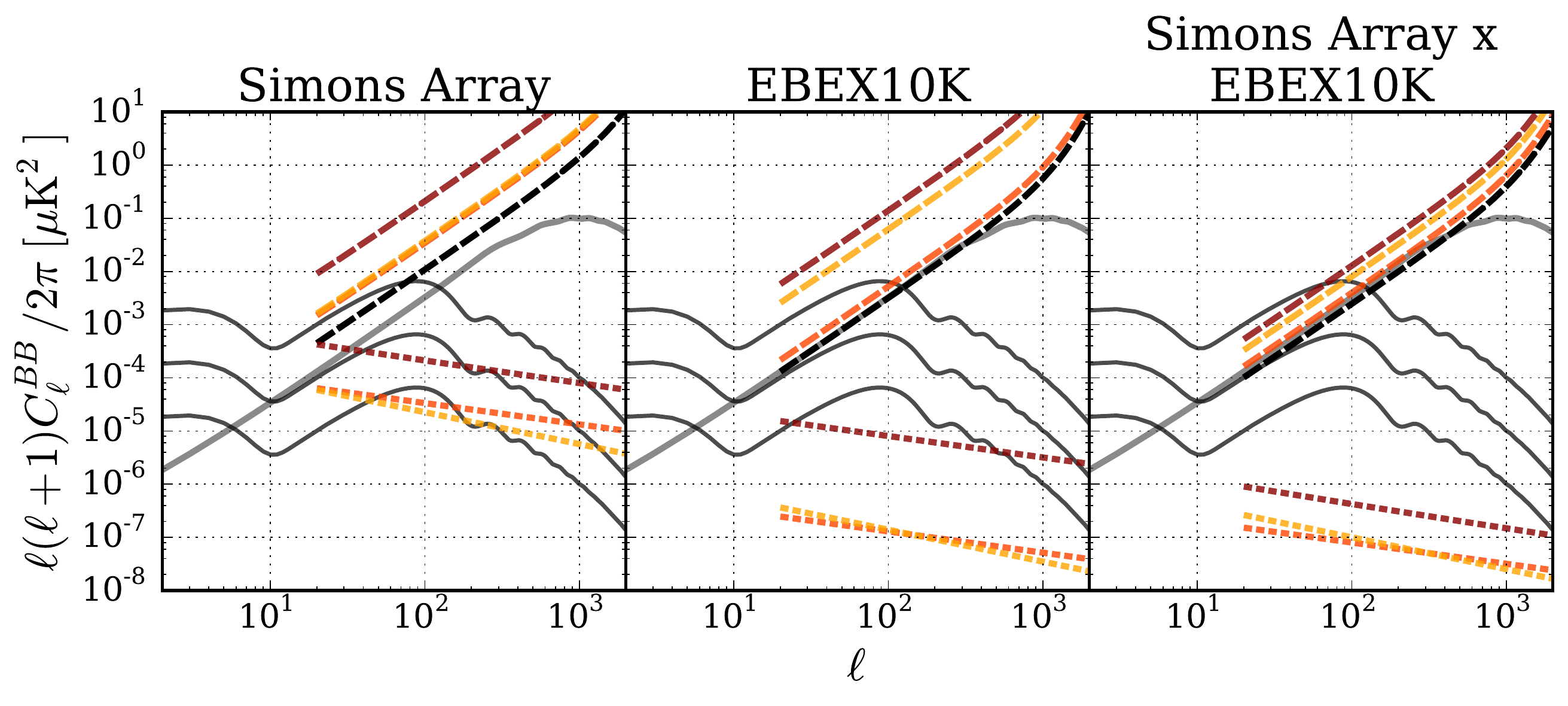}\\
\includegraphics[width=15cm]{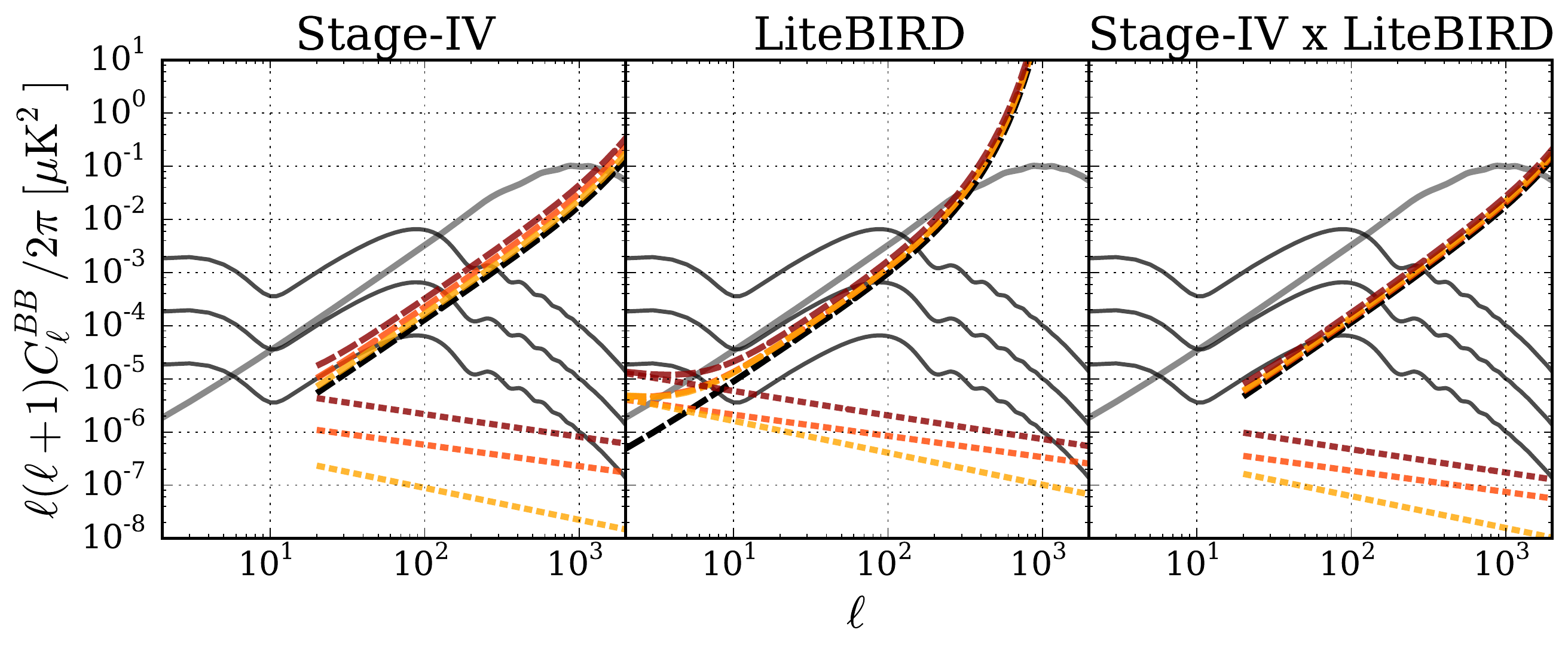}
\caption{Component-separation performance for selected experiments and their cross-correlation. Short-dashed colored curves correspond to the residuals (Eq.~\eqref{eq:Clres_def}) and long-dashed colored curves correspond to the noise (cf. Eq.~\eqref{eq:sigma_CMB_def}) after removing only synchrotron (light orange), only dust (dark orange), or synchrotron+dust (dark red). These curves are compared to the primordial $B$-mode spectra for $r = \{ 10^{-3}, 10^{-2}, 10^{-1}\}$ (solid dark grey), and lensing $B$ modes (solid light grey). The long-dashed black noise curve corresponds to the simple quadratic combination of sensitivities in all channels. Note that \planck\ data are not explicitly included in any of the instrumental configurations above; however, priors on \betad\ and \betas, Eqs~\eqref{eq:priors_def}, derived from \planck\ data, are employed.}
\label{fig:comp_sep}
\end{figure}

The figure is split into two panels, the top panel focusing on the pre-2020 Simons Array and EBEX10K instruments, the bottom on the post-2020 Stage-IV and \lb\ observatories. For the pre-2020 experiments we find noise degradation values in the ranges 3.2 -- 19, 1.6 -- 44 and 1.6 -- 5.2 for the Simons Array, EBEX10K and their combination, respectively, with the largest values corresponding to the synchrotron plus dust case. For these experiments, the foreground residuals are subdominant to the noise, with values in the range $3.8\times 10^{-6} \le \reff \le 10^{-2}$, comparable in the worst-case scenario to the smallest cosmological signal plotted. EBEX10K's design as a dust observatory, amply demonstrated by the similarity between the pre- and post-dust-removal noise curves ($\Delta = 1.6$), means it alone does not perform well in the cases with both dust and synchrotron. While Simons Array is not exceptional under any foreground assumption, its combination with EBEX10K is much more robust and sensitive, though the combination is still noise- rather than lensing-limited for measuring primordial $B$ modes. The increased frequency coverage and sensitivity of the post-2020 missions brings order of magnitude improvements in post-component-separation noise ($1.2 \le \Delta \le 2.5$), with \lb\ and PIXIE particularly robust to the foreground composition. Though the foreground residual level $\reff$ is at worst $1.1 \times 10^{-4}$, the residuals become non-negligible at the largest scales attainable by these experiments, and these scales should therefore be treated with care. In all cases, the post-component-separation noise is below the lensing, and we should therefore expect delensing to help considerably in constraining primordial $B$ modes.

Having considered a handful of experiments in detail, we now present results for the full suite of experiments. Figure~\ref{fig:comp_sep_performance} illustrates the foreground-cleaning performance of all experiments considered in terms of the noise degradation ($\Delta$) and effective level of foreground residuals ($\reff$), adopting the $n_p$ approach to spatially-varying foreground spectral indices. Recall that $\Delta$, which is obtained by assuming the true mixing matrix $\mathbf{A}$, quantifies the increase in variance imparted by the foreground cleaning, and the complementary $\reff$ measures the level of bias left in the $B$-mode power spectrum due to the mixing matrix's misestimation. For a given instrumental configuration, both quantities do not necessarily perform equally well, especially when there are too few frequency channels available.

\begin{figure}
\centering
\includegraphics[width=15cm]{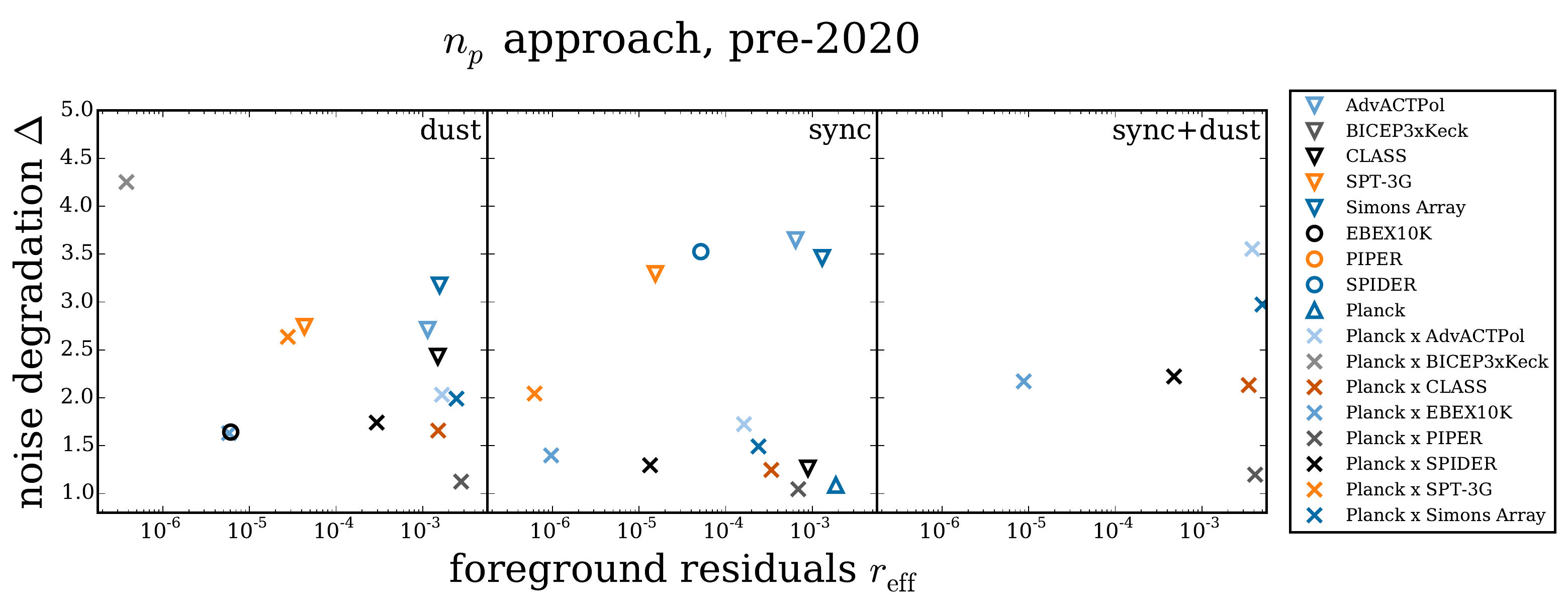}
\includegraphics[width=15cm]{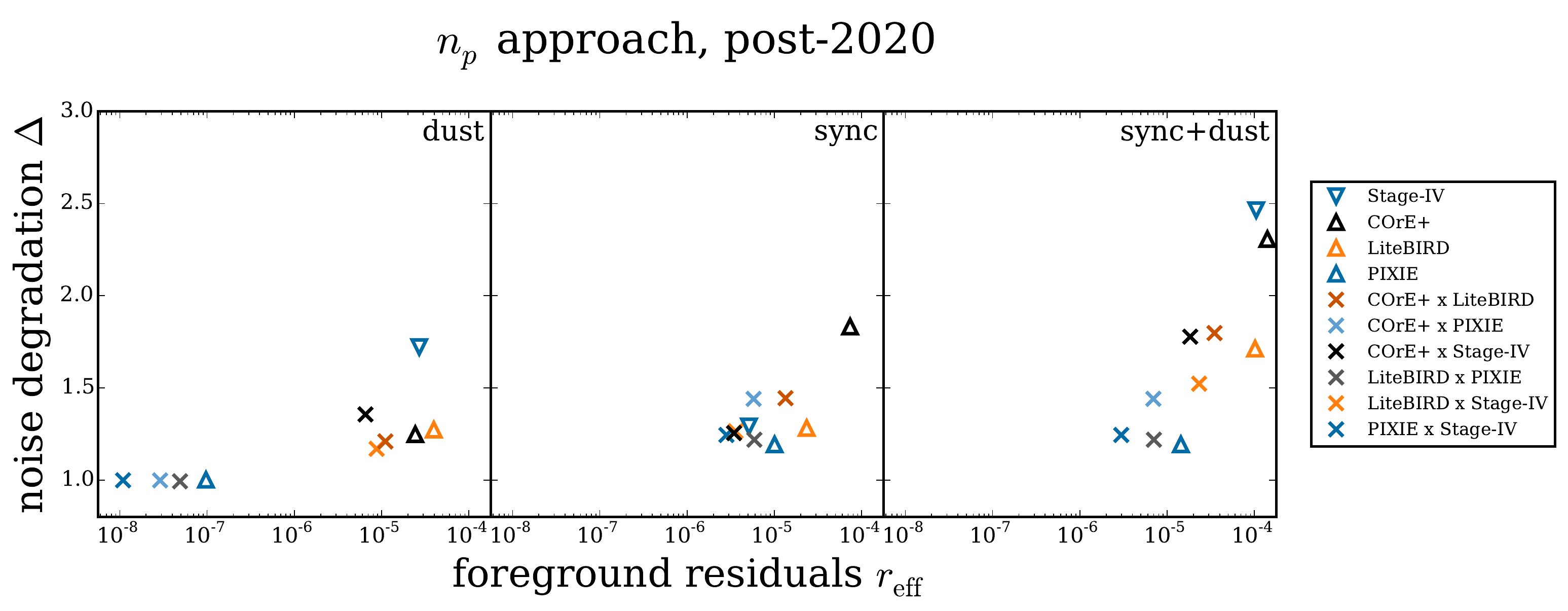}
\caption{\textit{Top panel:} Component-separation performance of selected pre-2020 experiments in the presence of CMB, dust and synchrotron. The $x$-axis corresponds to the effective amplitude of residual astrophysical foregrounds, \reff\  (Eq.~\eqref{eq:reff_def}), in the reconstructed CMB map. The $y$-axis is the degradation factor (Eq.~\eqref{eq:Delta_def}), or noise boost, driven by the linear combination of frequency maps during the foreground cleaning process. Upward triangles are for satellites, circles are for balloons, downward triangles are for ground-based instruments, and crosses indicate cross-correlations between experiments. \textit{Bottom panel:} Same as top panel but for experiments due to begin their observations post-2020. Note different axis ranges.}
\label{fig:comp_sep_performance}
\end{figure}

Concentrating first on the pre-2020 experiments, Fig.~\ref{fig:comp_sep_performance} indicates that, when considering a single foreground component, foreground residuals are under control in most cases, with effective amplitudes $\reff \lesssim 10^{-3}$. Similarly, the noise degradation is below 5 for most configurations considered. However, both the noise degradation and the residuals become considerably worse when including more sky components, and only the cross-correlations between Stage-III instruments and \planck\ remain below $\Delta\,\leq\,5$ in the presence of dust and synchrotron (top-right panel). The precise values of $\Delta$ and $\reff$ for the synchrotron plus dust case are tabulated for the cross-correlations with \planck\ in Appendix~\ref{app:fisher_tables}. Note that, even when considering both dust and synchrotron, dust-monitoring instruments such as EBEX10K and PIPER perform especially well in cross-correlation with \planck. As expected, the post-2020 instruments (bottom panel) perform better with respect to both criteria than the previous generation's observatories, with their cross-correlations yielding even more control over foregrounds. The experiments perform consistently well, with typical residual contaminations of $\mathcal{O}(10^{-4})$ for individual experiments and $\mathcal{O}(10^{-5})$ for combinations; we find noise degradation factors ranging from 2.5 for Stage-IV alone to 1.2 for PIXIE alone and Stage-IV's cross-correlations with PIXIE and \lb.

Figure~\ref{fig:comp_sep_performance_stolyarov} shows the component-separation performance obtained using the expansion of the $\mathbf{A}$ matrix, as opposed to the $n_p$ approach. In this case, the foreground residuals remain under control (for experiments with a number of frequency channels larger than the dimension of the expanded $\mathbf{A}$), but the noise degradation worsens considerably:  we find noise degradation values of $\Delta \gtrsim 3$ even for the cross-correlations of post-2020 experiments. These effects arise from the added complexity of the mixing matrix: adding extra dimensions to $\mathbf{A}$ means there are sufficient degrees of freedom to fit the foreground emission (yielding good control of residuals), but this also leads to a larger noise boost after component separation (due to the mixing inversion in the computation of $\sigma_{\rm CMB}$, Eq.~\eqref{eq:sigma_CMB_def}). In order to help control the additional foreground complexity, we add C-BASS to each experimental configuration as a dedicated synchrotron monitor. These results are plotted in Fig.~\ref{fig:comp_sep_performance_stolyarov_cbass}. Adding C-BASS yields factors of at least 2-3 improvement in the noise degradation and almost an order of magnitude in foreground residuals for the post-2020 experiments. Dedicated synchrotron monitoring is necessary to maximize the scientific output of Stage-IV-era CMB experiments if we are to fit for spatially-varying foreground spectral indices (a finding confirmed by Ref.~\cite{2015arXiv150905934C}, who fit for scale-dependent spectral indices). We note that the high-frequency coverage of PIXIE is particularly effective in combination with the synchrotron monitoring, with PIXIE and its cross-correlations suffering noise degradations around half those of the other post-2020 experiments.

As briefly mentioned in section~\ref{ssec:foregrounds}, selecting the optimal complexity for the foreground parametrization amounts to trading off the bias of foreground residuals against the variance in the reconstructed CMB maps. Though using the $\mathbf{A}$-expansion approach assumes less about the spatial variations of spectral indices, helping to minimize the bias, this inevitably leads to increased CMB noise after cleaning. Although combining data from dedicated foreground monitors will help reduce both bias {\em and} variance, determining whether the foregrounds are complex enough to warrant paying the price of constraining power will be difficult to check in practice, requiring careful testing using reliable sky models.

\begin{figure}
\centering
\includegraphics[width=15cm]{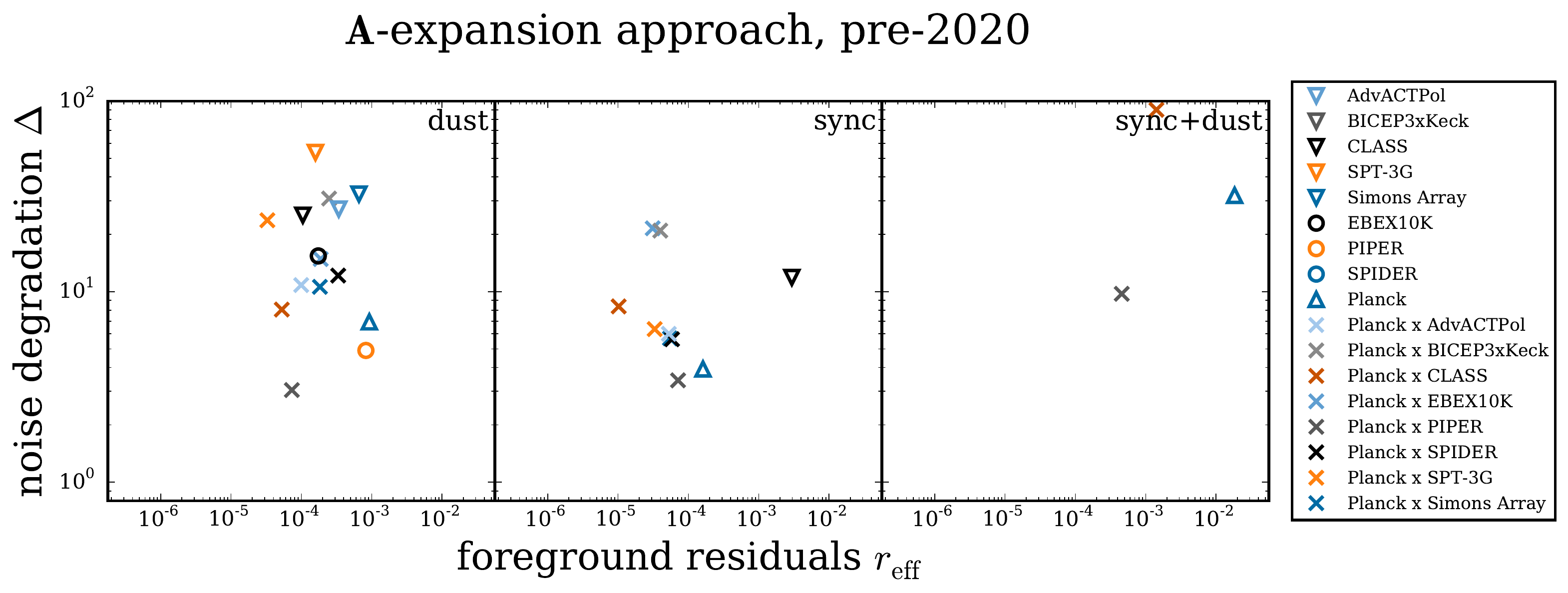}
\includegraphics[width=15cm]{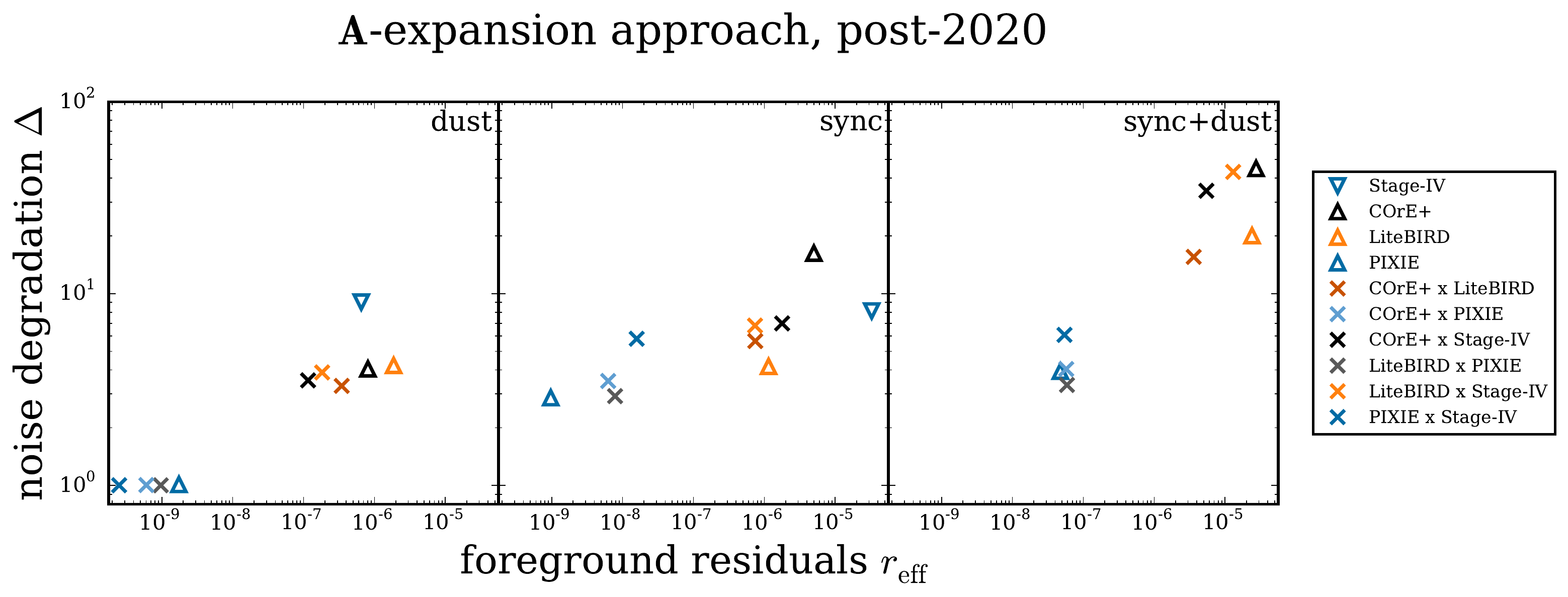}
\caption{Same as Fig.~\ref{fig:comp_sep_performance}, but using the expansion of the mixing matrix $\mathbf{A}$, as described in Eq.~\eqref{eq:A_taylor_expansion}. On one hand, this expansion leads to a significant increase of the noise degradation $\Delta$ due to the inversion of a larger matrix, cf. Eq.~\eqref{eq:sigma_CMB_def}. On the other hand, the foreground residuals can be slightly better than for the $n_p$ approach. This is due to the modeling of both dust and synchrotron emission encapsulating more degrees of freedom, trading increased noise variance in the final CMB map for reduced residual bias.}
\label{fig:comp_sep_performance_stolyarov}
\end{figure}

\begin{figure}
\centering
\includegraphics[width=15cm]{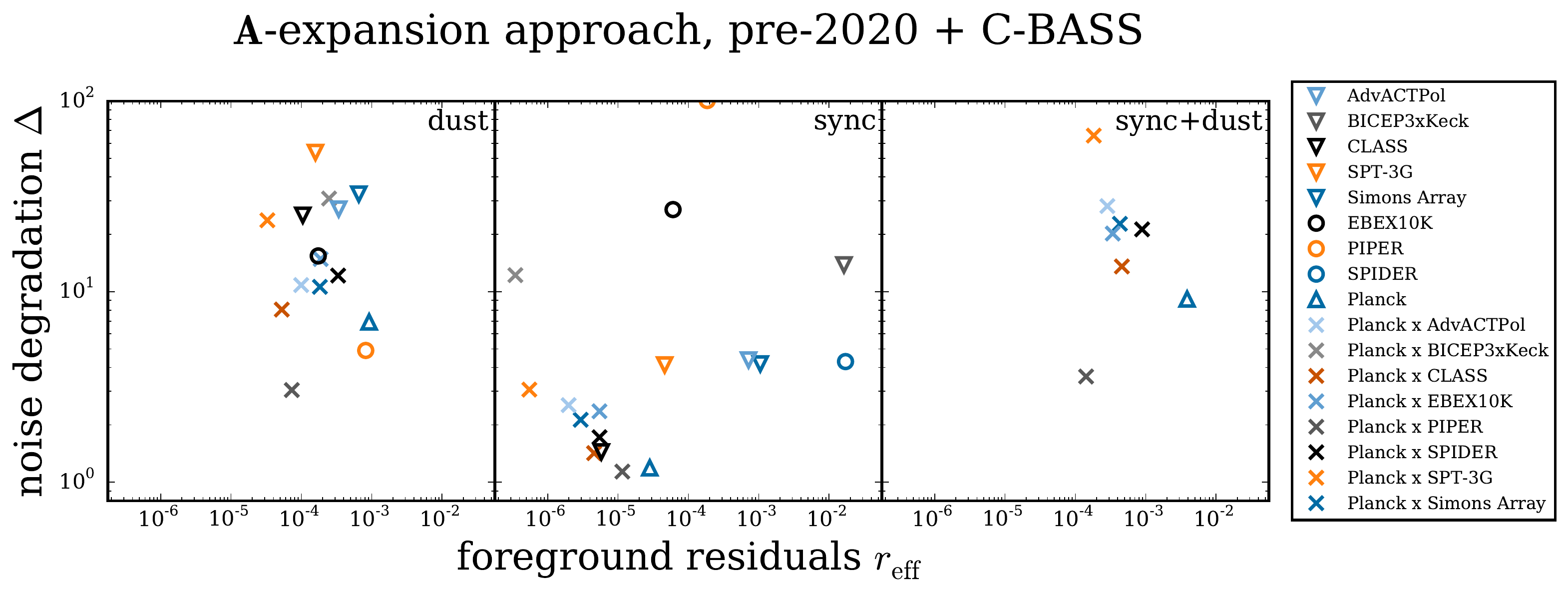}
\includegraphics[width=15cm]{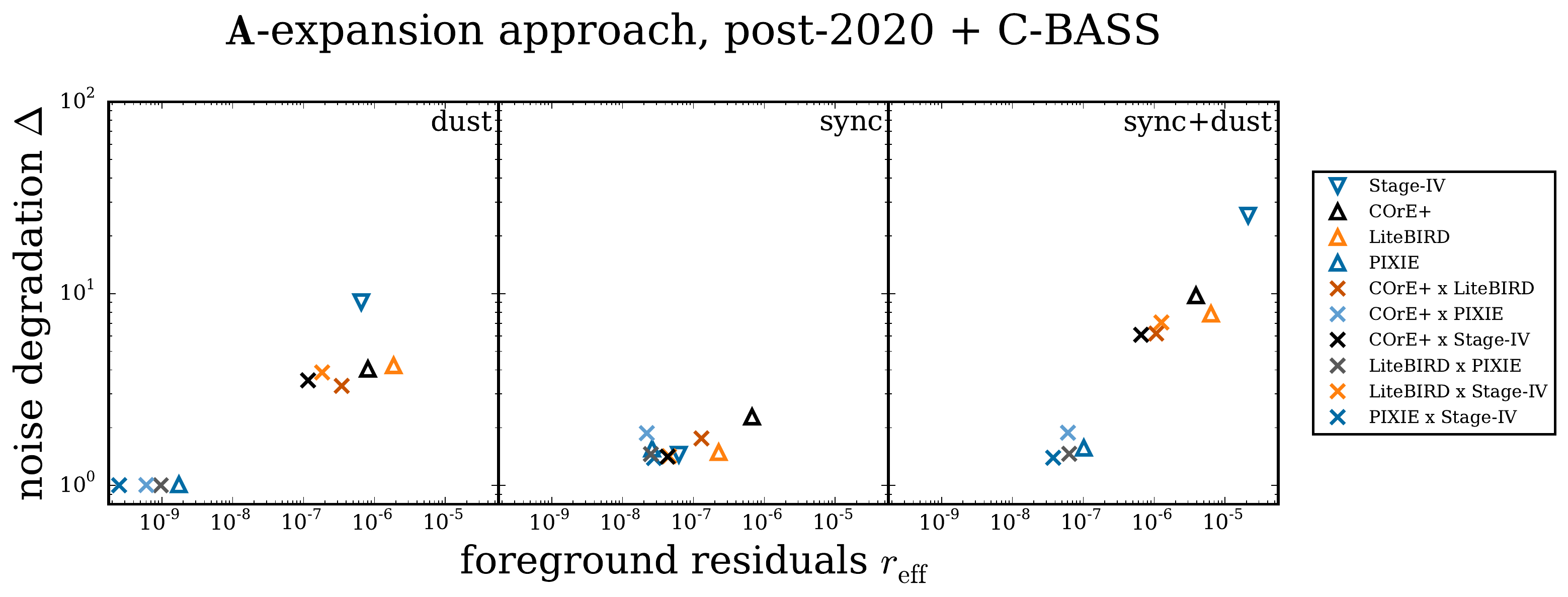}
\caption{Same as Fig.~\ref{fig:comp_sep_performance_stolyarov}, but including C-BASS in each experimental combination.}
\label{fig:comp_sep_performance_stolyarov_cbass}
\end{figure}

\subsection{Delensing and detectability limits on tensor-to-scalar ratio}

As with the foreground-cleaning performance, in this section we will first illustrate the detailed delensing performance using our two pedagogical pre- and post-2020 experiments, before plotting the comprehensive delensing performance for the full set of experiments. Figure~\ref{fig:BB_delensed_example} demonstrates the ability of each experimental configuration to delens their foreground-cleaned CMB maps. Plotted is the fraction of the lensing $B$ mode remaining after delensing, $\clbbd/\clbbl$, as a function of $\ell$, for CMB$\times$CMB (solid), CMB$\times$CIB (long-dashed) and CMB$\times$LSS (short-dashed) delensing after cleaning only dust (dark orange), only synchrotron (light orange) or dust and synchrotron (dark red), as well as without performing any cleaning (black).

\begin{figure}
\centering
\includegraphics[width=15cm]{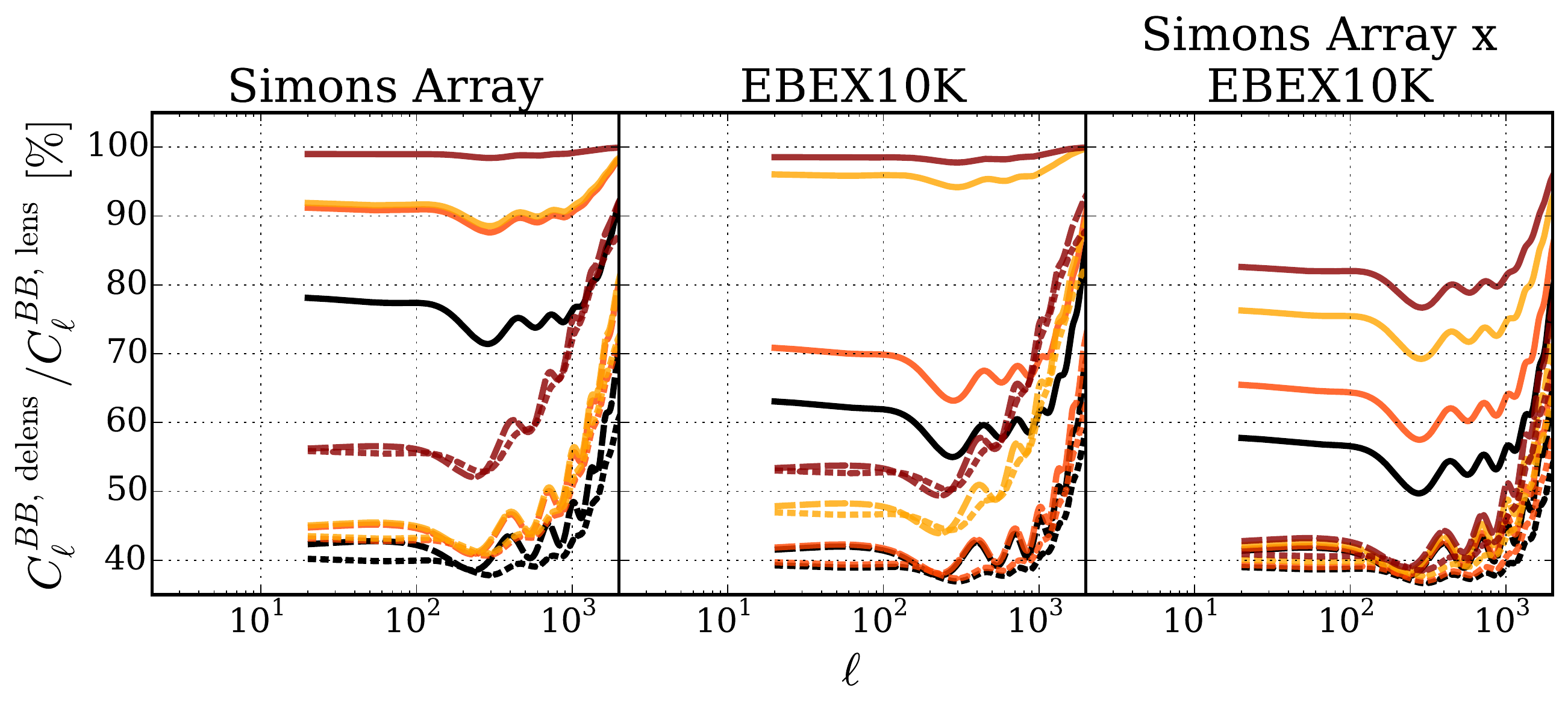}\\
\includegraphics[width=15cm]{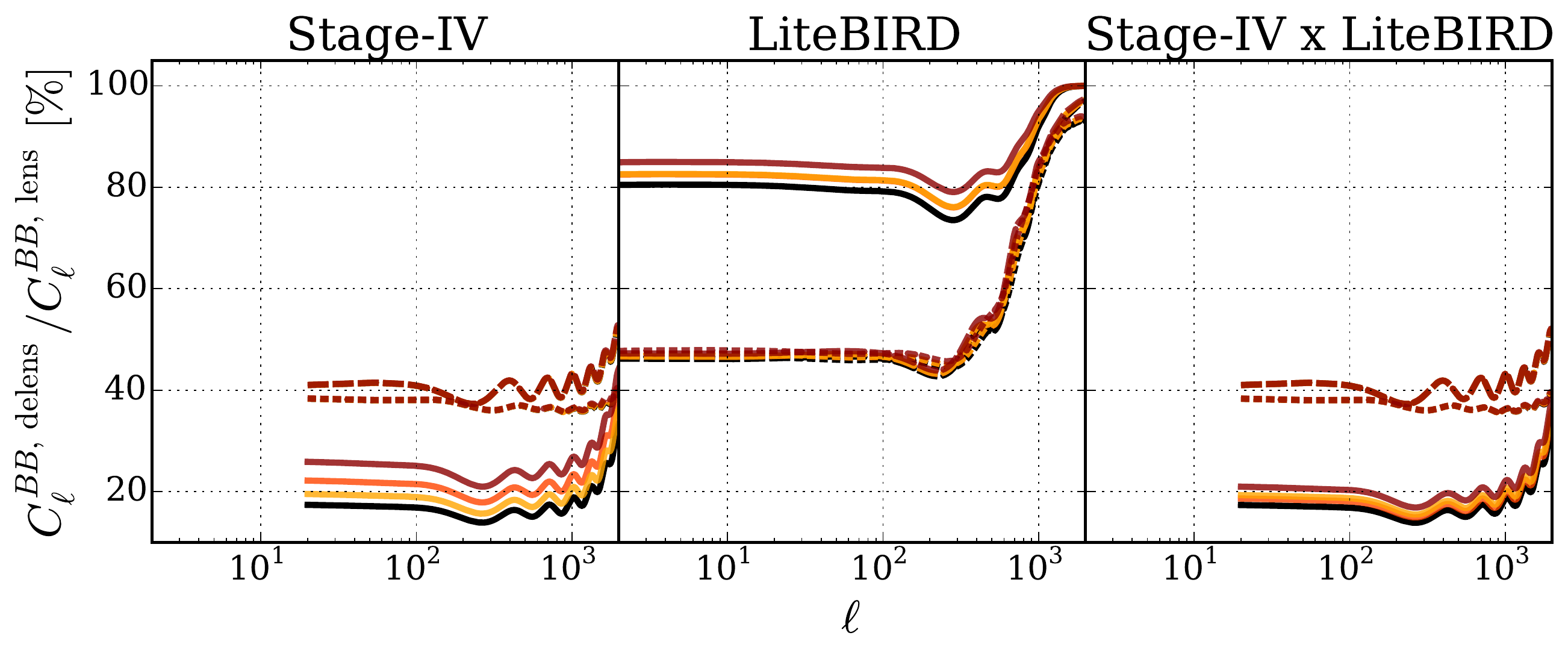}
\caption{Delensing performance of the different experiments. The curves show the fractional delensing residual after CMB$\times$CMB (solid), CMB$\times$CIB (long-dashed) and CMB$\times$LSS (short-dashed) delensing. We consider cases in which, prior to delensing, synchrotron (light orange), dust (dark orange), or synchrotron and dust (dark red) have been removed, as well as the CMB-only case in which no foregrounds are present (black). Please note the different y-axis ranges. Note that \planck\ data are not explicitly included in any of the instrumental configurations above; however, priors on \betad\ and \betas, Eqs~\eqref{eq:priors_def}, derived from \planck\ data, are employed.}
\label{fig:BB_delensed_example}
\end{figure}

From these plots, it is clear that CMB$\times$CMB delensing for experiments in the 2015--2020 period (and, indeed, \lb) can reduce the lensing contamination by factors of only a few tens of percent, with considerable variation in performance due to the composition of the foregrounds (note, however, that we only consider polarization delensing here: further improvement may well be possible using estimators based on the temperature or the minimum-variance combination of temperature and polarization~\cite{2002ApJ...574..566H}). Delensing through cross-correlation with the CIB or LSS --- which require only well-measured CMB $E$ modes --- look promising, yielding a factor of two reduction in lensing contamination, robust to the precise form of the foregrounds. Recall, however, that the $B$ modes will still be noise-dominated after delensing (cf. Fig.~\ref{fig:comp_sep}), and thus we should expect only small improvement in the attainable constraints on $r=0$. Iterative CMB$\times$CMB delensing is much more promising for the Stage-IV instrument, offering a factor of five reduction in lensing contamination, again, robust to the foreground composition.

Figure~\ref{fig:alpha_vs_sigma_r_results} summarizes the delensing performance of the full suite of experimental configurations, plotting the delensing factor $\alpha$ against the achievable one-sigma limits on $r=0$ after CMB$\times$CMB (left), CMB$\times$CIB (center) and CMB$\times$LSS (right) delensing. For clarity, only the data the two experiments have in common (i.e., the overlapping \fsky\ and multipoles) is used to generate these plots. The plot is split into upper and lower panels, focussing on the pre- and post-2020 experiments respectively, and in all cases both dust and synchrotron have been removed (adopting the $n_p$ approach to spatially varying foreground spectral indices) prior to delensing. There are several observations to make. First, the limited impact of delensing on the pre-2020 experiments is clear: though around 40\% more of the lensing can be removed by, e.g., CMB$\times$CIB delensing than by iterative CMB$\times$CMB delensing, there is minimal improvement on the one-sigma limits on $r$. This is, as has been previously mentioned, because the post-component-separation $B$ modes are noise- rather than lensing-dominated. Regardless, and in particular in combination with \planck, the pre-2020 experiments are able to push the limits on $r=0$ down to $\sim 3 \times 10^{-3}$. The post-2020 experiments' increased sensitivity and frequency coverage, and resulting ability to delens, yield an order of magnitude improvement on the constraints one can place on $r=0$; indeed, the combinations of post-2020 experiments involving a space mission will be able to constrain $r=0$ with one-sigma errors of $10^{-4}$.

\begin{figure*}
\centering
\includegraphics[width=15cm]{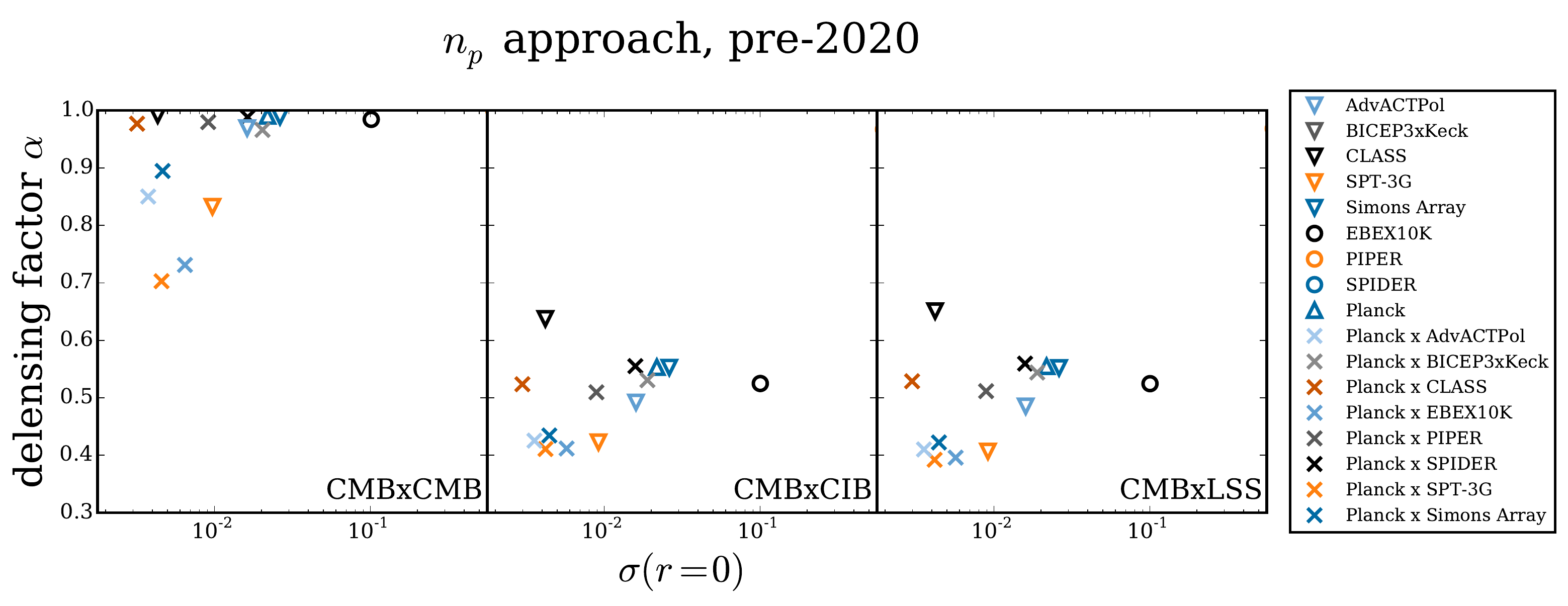}
\includegraphics[width=15cm]{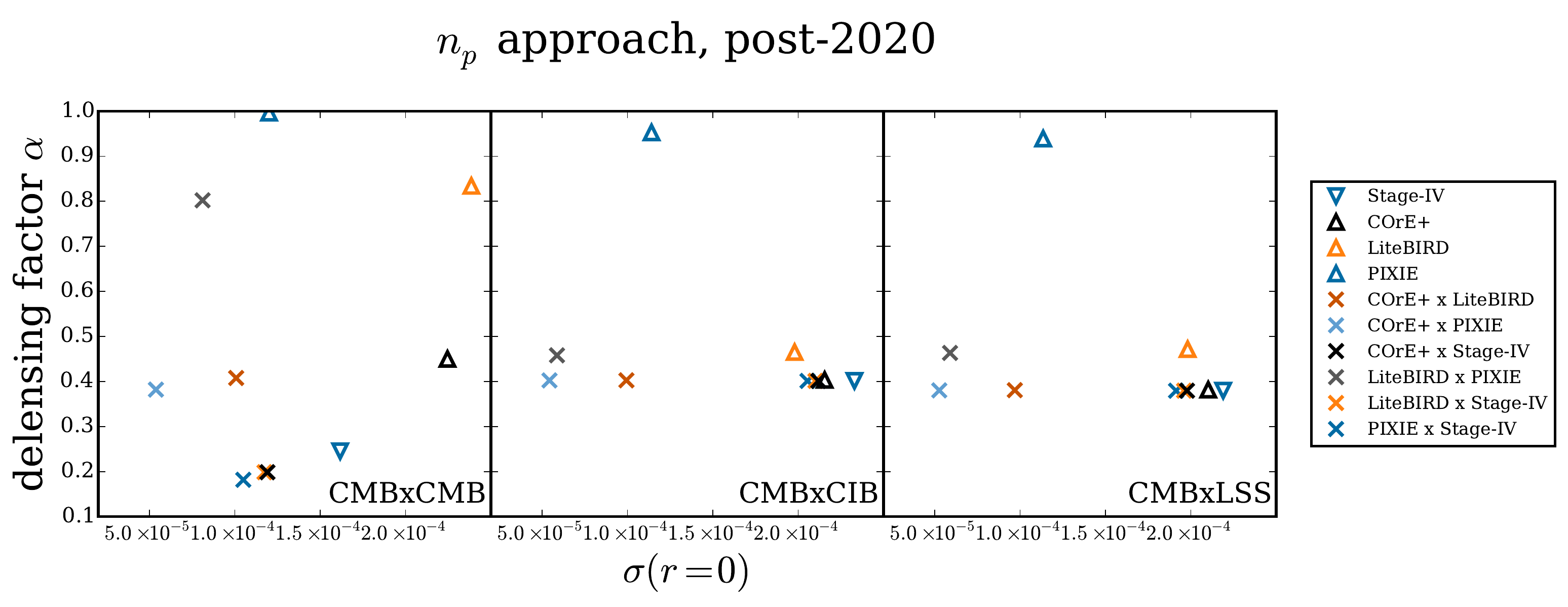}
\caption{\textit{Top panel:} One-sigma limits on $r$ for the pre-2020 experiments after iterative CMB$\times$CMB (left), CMB$\times$CIB (center) and CMB$\times$LSS (right) delensing. In all cases, dust and synchrotron have been removed --- assuming fixed spectral indices within $n_p$ independent patches of sky --- prior to delensing. \textit{Bottom panel:} as above, but for the post-2020 experiments.}
\label{fig:alpha_vs_sigma_r_results}
\end{figure*}

We also study the delensing performance and detectability limits on $r=0$ when adopting the more ambitious $\mathbf{A}$-expansion approach to fitting spatially varying foreground spectral indices. Examining the pre-2020 experiments (not plotted), we find that the presence of \planck\ and C-BASS is essential in any experimental configuration to improve the current limits of $\sigma(r=0) \simeq 0.1$, and that the addition of further experiments improves the \planck\ plus C-BASS limits of $\sim 4 \times 10^{-2}$ by only a few tens of percent. The post-2020 results, plotted in Fig.~\ref{fig:alpha_vs_sigma_r_results_stolyarov}, are more promising. When we assume the $n_p$-patch approach, the $\sigma(r = 0)$ limits for all post-2020 configurations are clustered in the range 1--2$\times 10^{-4}$. When we constrain the spatially varying spectral indices, this range increases markedly, with Stage-IV in particular struggling ($\sigma(r=0) \sim 10^{-2}$). The inclusion of C-BASS helps control the chaos, however: Stage-IV constraints improve by more than an order of magnitude, and all other combinations by a factor of at least two. In summary, with the addition of C-BASS, limits on $r=0$ of $\mathcal{O}(10^{-4})$ are achievable with post-2020 experiments, even with the most complex foreground composition considered.

\begin{figure*}
\centering
\includegraphics[width=15cm]{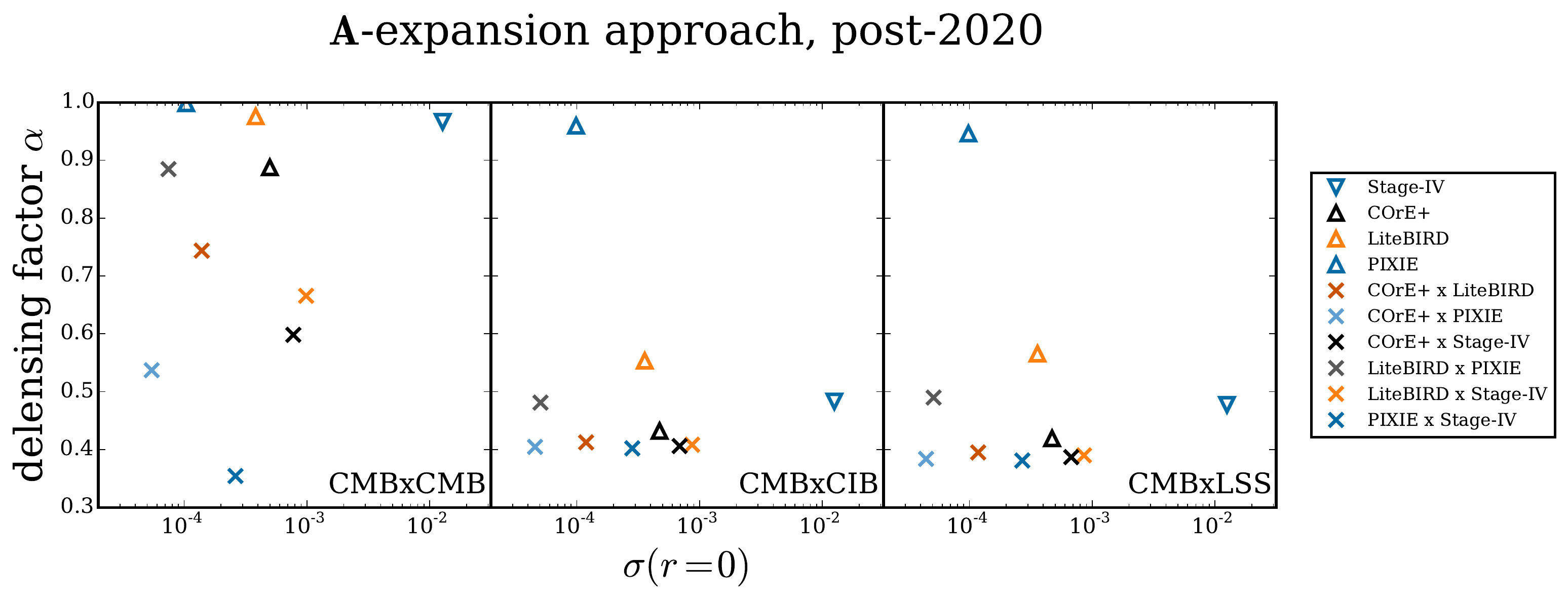}
\includegraphics[width=15cm]{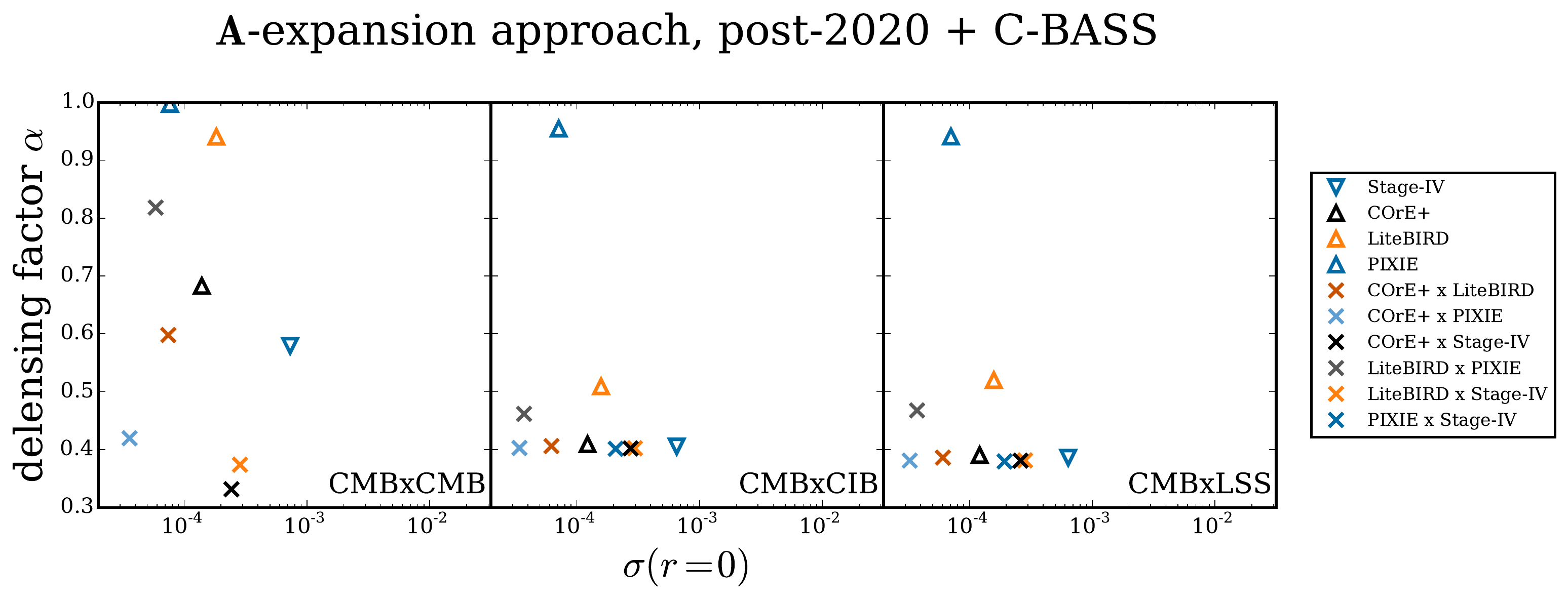}
\caption{\textit{Top panel:} One-sigma limits on $r$ for the post-2020 experiments after iterative CMB$\times$CMB (left), CMB$\times$CIB (center) and CMB$\times$LSS (right) delensing. In all cases, dust and synchrotron have been removed --- accounting for spatially varying spectral indices to first order --- prior to delensing. \textit{Bottom panel:} as above, but with the addition of C-BASS as a dedicated synchrotron monitor.}
\label{fig:alpha_vs_sigma_r_results_stolyarov}
\end{figure*}

\subsection{Full cosmological parameter constraints}
\label{ssec:params_constraints}

For the interested reader, a comprehensive collection of constraints on the parameters extending $\Lambda$CDM is presented in Appendix~\ref{app:fisher_tables}. Constraints are tabulated for all pre-2020 experiments in combination with \planck, as well as for all post-2020 experiments and their cross-correlations. When considering Stage IV in isolation, we add \planck\ to provide low-$\ell$ information. Within each experimental table the constraints are grouped by the foregrounds removed (none or synchrotron and dust -- the $n_p$ approach is adopted by default) and the delensing method employed (none, CMB$\times$CMB, CMB$\times$CIB or CMB$\times$LSS). For the post-2020 experiments, constraints are also provided for the $\mathbf{A}$-expansion approach to foreground removal, additionally combining the experiment(s) with C-BASS and, for the full-sky satellites, the Northern hemisphere-based QUIJOTE-CMB. As previously stated, the constraints are calculated using the minimal model containing the particular parameter of interest, as detailed in Table~\ref{table:models}.

In this section we highlight some particularly interesting findings selected from this full set of forecasts. Concentrating first on the pre-2020 instruments (in combination with \planck), the noise degradation due to component separation is, in the majority of cases, less than 6, with the lowest value ($\Delta = 1.2$) reached by PIPER $\times$ \planck. Foreground residuals are typically at the same level as the fiducial primordial $B$ modes, $\reff \sim 10^{-3}$, though $\reff \leq 10^{-5}$ is reached by EBEX10K $\times$ \planck\ and BICEP3 $\times$ Keck Array $\times$ \planck. Though all experiments are able to remove more than 50\% of the lensing signal by cross-correlating the CMB with tracers of the lensing potential, the constraints on $r$ do not improve significantly: they are limited by the noise and residuals after component separation, not by the lensing. None of the pre-2020 missions can measure primordial tensors with $r=0.001$ --- indeed, doing so would be beyond their reach even without foregrounds to clean --- but  AdvACTPol, CLASS, EBEX10K, Simons Array and SPT-3G (in combination with \planck) are able to rule out $r \gtrsim 10^{-2}$ at the two sigma level. All experiments are able to place one-sigma errors on $\nt$ of $\sim 0.2$ (assuming $r=0.1$). Though these constraints are not tight enough to test the inflationary consistency relation~\cite{1992PhLB..279..244L, 1993PhRvL..71..219C} --- assuming, of course, that tensors are detected at the appropriate level --- they will be able to test alternatives to inflation such as the ekpyrotic model~\cite{2008PhR...465..223L} and some models of loop quantum cosmology~\cite{2010PhRvD..81j4049M} at the 10-sigma level, as well as beginning to distinguish between standard slow-roll inflation and more general single-field models (see, e.g., Ref~\cite{2010PhRvL.105w1302K}). When allowing the scalar spectral index to run, we find $\sigma(\ns) \sim 2-3 \times 10^{-3}$, leading to a $10$-sigma measurement of the deviation from scale invariance, and $\sigma(\alphas)\sim 3-5 \times 10^{-3}$, allowing an important test of slow roll inflation~\cite{2006JCAP...09..010E}. Constraints on spatial curvature of $\sigma(\omk) \leq 2.5\times 10^{-3}$ are achievable by AdvACTPol and Simons Array (in combination with \planck).\footnote{Note that we restrict ourselves to considering CMB$\times$CMB and CMB$\times$CIB delensing when reporting the pre-2020 results here, as the observations required for CMB$\times$LSS delensing will not be available on this timescale. Results for all delensing options are tabulated in Appendix~\ref{app:fisher_tables}.}

While delensing does not help in the search for primordial $B$ modes, the improved measurement of the lensing deflection power spectrum enabled by the cross-correlation can significantly improve constraints on the sum of neutrino masses: AdvACTPol $\times$ Planck and Simons Array $\times$ Planck, for example, are able to place one-sigma limits of $\sigma(\mnu) = 55$ meV by exploiting \planck's CIB measurements, near the threshold for distinguishing the degenerate from the normal and inverted neutrino mass hierarchies~\cite{2015APh....63...66A}. We stress that these constraints are derived from CMB (and CIB) information alone: combination with data from galaxy surveys such as eBOSS~\cite{2015arXiv150804473D} or DESI~\cite{2013arXiv1308.0847L} will yield even greated discriminatory power~\cite{2014JCAP...05..023F,2015arXiv150607493P}. Considering the number of neutrino species (or indeed other relativistic degrees of freedom), $\sigma(\neff)$ approaches $\sim\,0.06$, a factor of $\sim 6$ improvement on current CMB-only limits~\cite{2015arXiv150201589P}. In the dark energy sector, assuming a redshift-independent equation of state yields errors on the equation of state parameter $\sigma(\w)$ of at least $\sim0.16$. Expanding the model to include redshift dependence understandably weakens the constraints, with optimal errors of $\sim0.4$ and $\sim0.7$ on $w_0$ and $w_a$, respectively. We would expect similar error-bar expansion for other models of dark energy with a time-varying equation of state.

Turning to the post-2020 instruments, the noise degradation due to component separation, assuming the $n_p$ approach, is less than 2.5 in all cases, and the foreground residuals at least an order of magnitude beneath the fiducial tensor signal (though not necessarily the noise: see Fig.~\ref{fig:comp_sep}); these experiments efficiently and thoroughly clean polarized foregrounds. Post-component-separation noise levels are low enough for delensing to make a clear impact in all cases, with any combination involving Stage-IV or COrE+ in particular benefitting from CMB$\times$CMB delensing. In the inflationary sector, one-sigma error bars on $r=10^{-3}$ of $1.8\times10^{-4}$ (a 5-sigma measurement) are achievable with Stage-IV $\times$ \planck, dropping to $1.3-1.4\times10^{-4}$ in combination with any satellite mission. Testing the single-field inflationary consistency relation is out of reach even with $r = 0.1$, with the minimum $\sigma(\nt) \simeq 0.03$; however, constraints at this level are of interest for testing modifications to the the standard consistency relation in multifield inflation~\cite{2015PhRvL.114c1301P}. Constraints on $\ns$ and $\alphas$ do not improve significantly between the pre- and post-2020 timeframes, with errors decreasing by a factor of approximately two in both cases. The minimum error on the energy density in curvature is $\sigma(\omk) = 1.5 \times 10^{-3}$ (for Stage-IV $\times$ COrE+), five times stronger than current constraints but falling short of the $10^{-4}$ level interesting for testing the framework of eternal inflation when combined to other observations~\cite{1980PhRvD..21.3305C,1982Natur.295..304G,1995PhRvD..52.3314B,2006JHEP...03..039F,2012JCAP...06..029K,2012PhRvD..86b3534G}.} When considering the more challenging case of fitting for the spatial variation in foreground spectral indices (in which case C-BASS or QUIJOTE-CMB are added as synchrotron monitors), $\Delta < 9.8$ in all cases apart from Stage-IV $\times$ \planck, with residuals remaining sub-dominant to the fiducial tensor signal. Consequently, a 3 -- 7-$\sigma$ measurement of $r=10^{-3}$ is still attainable (after delensing) for certain combinations of experiments, even under the most challenging foreground scenario considered here.

In the neutrino sector, the combination of Stage-IV with any space mission leads to constraints on the sum of the neutrino masses down to $\sigma(\mnu)\sim 30-40$ meV, and constraints on the effective number of relativistic species of $\sigma(\neff) < 0.046$, even under the $\mathbf{A}$-expansion approach. Such measurements will begin to discriminate between the normal and inverted neutrino-mass hierarchies, and could confirm or spectacularly contradict our understanding of the thermal physics and neutrino makeup of the early universe~\cite{2015APh....63...66A}. The constraints on even the simplest model for the dark energy equation of state are relatively weak, with $\sigma(\w)\geq 0.09$. While this is considerably larger than the $1\%$ constraints expected for Euclid~\cite{2010arXiv1001.0061R} and LSST~\cite{LSST_Science_Book}, adding a precision measurement of the Hubble parameter has the potential to significantly improve this constraint. Allowing a redshift-dependent dark energy equation of state yields errors on $w_0$ and $w_a$ of $\sim0.10$ and $\sim0.14$ at best, though these numbers rely on delensing using measurements of large-scale structure: with access to only CMB and CIB data, these errors expand to $\sim0.25$ and $\sim0.50$, respectively.

Overall, there is a clear synergy between the ground-based Stage-IV instruments (probing small angular scales, with deep integration at the CMB frequencies) and the space missions (probing large angular scales with broad frequency coverage).

\section{ Web Interface }
\label{sec:web_interface}

In order to aid the community's planning for future experimental designs, we make the toolkit used to obtain these results publicly available. The web interface is accessible at \url{turkey.lbl.gov}.
The user is asked to fill in the following instrumental specifications:
\begin{itemize}
	\item central frequencies in GHz,
	\item sensitivities in \ukarc\ per frequency channel,
	\item FWHM in arcmin per frequency channel,
	\item $\fsky$,
	\item fractional bandpass per frequency channel,
	\item $\lmin$ and $\lmax$,
	\item information channels among $T$, $E$, $B$ and $d$ -- this is used to construct the covariance matrix, Eq.~\eqref{eq:covariance_definition}.
\end{itemize}
In addition, the user can choose the foreground properties and analysis techniques, in particular:
\begin{itemize}
	\item dust temperature $\Td$ and spectral index $\betad$, synchrotron spectral index $\betas$,
	\item type of foreground separation among options described in Sect.~\ref{ssec:foregrounds},
	\item type of delensing among the different options described in Sect.~\ref{ssec:delensing}.
	\item prior on $\betad$ and $\betas$, Eq.~\eqref{eq:priors_def}
	\item the combination of the chosen specifications with \planck, following the procedure described in paragraph~\ref{ssec:missions}.
\end{itemize}
Finally, the user has to choose the set of cosmological parameters for forecasts to be produced. When these settings are chosen, the interface produces the following outputs:
\begin{itemize}
	\item the boosted noise in the reconstructed CMB map, $\sigma_{\rm CMB}$, Eq.~\eqref{eq:sigma_CMB_def},
	\item the level of the foreground residual power spectrum $\reff$, Eq.~\eqref{eq:reff_def},
	\item the delensing factor $\alpha$, Eq.~\eqref{eq:alpha_def},
	\item the forecasted marginalized constraints on the chosen set of cosmological parameters.
\end{itemize}
Wherever foregrounds are included in the input components, the constraints on cosmological parameters are also marginalized over the amplitude and tilt of the foreground residuals, cf. Sect.~\ref{sec:forecast_formalism}.

\section{Conclusions}
\label{sec:conclusions}

\begin{figure*}
\centering
\includegraphics[width=15cm]{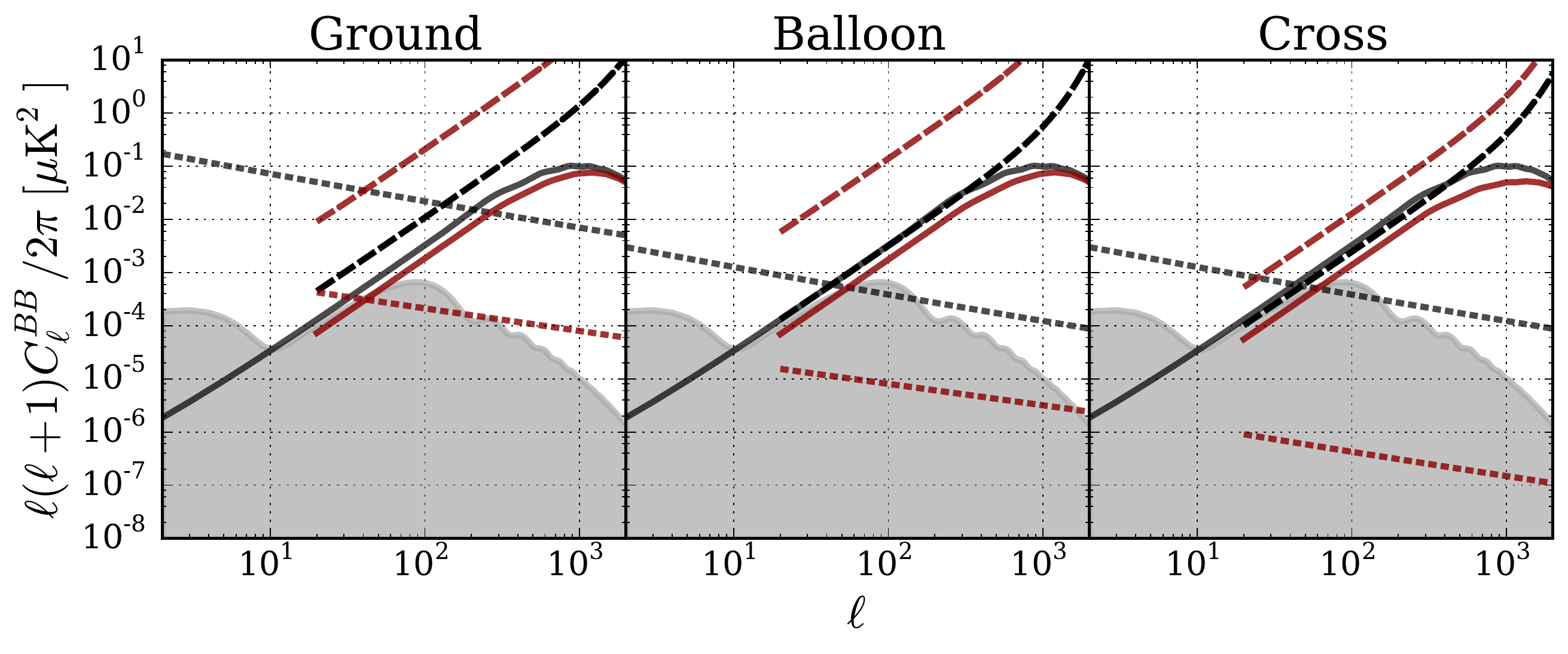}
\includegraphics[width=15cm]{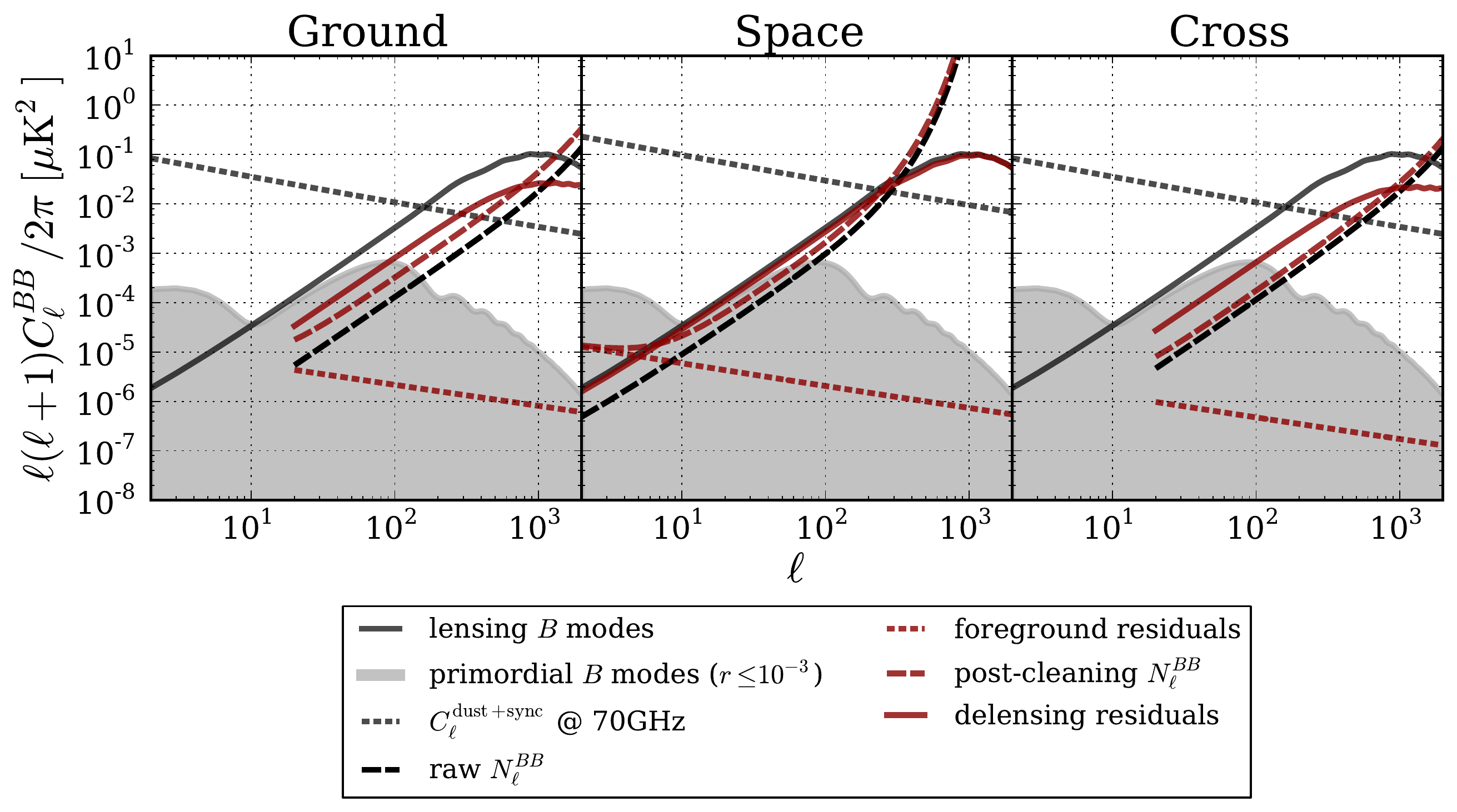}
\caption{\textit{Top panel:} Figure summarizing the performance of typical pre-2020 balloon and ground-based instruments, as well as their combination. The delensed $B$ modes are obtained using CMB $\times$ CIB delensing. \textit{Bottom panel:} Same as top panel but for post-2020 ground and space projects. Delensing is this time derived using the CMB $\times$ CMB estimator.}
\label{fig:conclusion_figure}
\end{figure*}

We have introduced a powerful and versatile framework for self-consistently evaluating the performance of next-generation CMB polarization experiments in the presence of astrophysical foregrounds and lensing. Our method, described in detail in Sect.~\ref{sec:methodology}, uses a parametric maximum-likelihood approach for component separation, CMB four-point estimators and cross-correlations with the cosmic infrared background and large-scale structure for delensing and the Fisher matrix formalism for parameter forecasts. The results of applying this framework to a suite of future ground-based, balloon and space-borne CMB polarization instruments were presented in Sect.~\ref{sec:results}.

Figure~\ref{fig:conclusion_figure} provides a summary of the capabilities of this framework, demonstrating the ability of different experimental combinations to remove polarized foregrounds and lensing to reveal the primordial CMB. The frequency coverage and sensitivity of pre-2020 experiments (top row) means they are noise-dominated in the search for $B$ modes once foregrounds have been removed. Consequently, delensing does not improve constraints on primordial tensors. The post-2020 projects considered (bottom row) have much lower post-component-separation noise, even in the case of complex foreground emission, and their constraints on tensors are hence (de)lensing-limited. However, they will potentially be biased on large angular scales by foreground residuals. It is therefore important to note the power of the Stage-IV $\times$ space mission combination, which blends small angular scales and broad frequency coverage.

\textbf{Component separation ---} The component separation technique employed in this framework, based on a parametric maximum-likelihood method~\cite{2009MNRAS.392..216S}, consists of two steps. The first step is the estimation of the mixing matrix governing the frequency behaviour of the signal components, or more precisely the spectral indices of the polarized foregrounds; the second step is the linear combination of all available frequency maps to recover the cleaned CMB. Imperfect estimation of the mixing matrix leads to residual foregrounds contaminating the cleaned CMB maps, and inverting the mixing matrix to extract the CMB boosts the noise in the resulting maps. These effects are passed on to later stages of the pipeline, and the amplitude and angular dependence of the foreground residuals are marginalized over in all cosmological parameter constraints reported.

Under the assumption that the foreground spectral indices are constant in independent patches of angular size $\sim\,15\,\deg$, we show that \planck\ data are required for pre-2020 projects to clean polarized dust and synchrotron, with the post-cleaning noise power spectra of the best experimental combinations approximately twice that expected from a na\"ive combination of input frequency channels. Typical foreground residual amplitudes for these combinations correspond to an effective tensor-to-scalar ratio of $\reff\,\leq\,5\times10^{-3}$. As a result, the search for primordial $B$ modes in this timeframe will be noise-dominated. Post-2020 instruments are more robust to the foreground composition, with noise degradations of $\Delta\,\sim\,1.2-2.5$ after component separation and foreground residual levels of $\reff\,\leq\,10^{-4}$. These residuals are comparable to the instrumental noise at the largest angular scales, and must therefore be carefully treated; the noise itself is sub-dominant to the lensing, and thus we expect delensing to be a powerful tool in the post-2020 timeframe.

In an alternative approach to handling spatially varying spectral indices, we expand the mixing matrix to first order in the indices to allow them to vary from pixel to pixel. This approach is much more challenging, with noise degradations of $\Delta\,\sim\,1.4-25.0$ for even the post-2020 experiments, and requires dedicated synchrotron monitors such as C-BASS and QUIJOTE-CMB in addition to frequency bands above $\sim\,300$\;GHz in order to maximize the scientific output of the experiments considered. As emphasized in the $\mathbf{A}$-expansion approach, C-BASS and QUIJOTE-CMB prove essential in disentangling the components in the case of an expansion of the mixing matrix, and the high-frequency coverage of PIXIE makes it particularly robust in this more challenging scenario. This result, arising from the addition of new degrees of freedom to the mixing matrix, may hold in the cases of more complex foreground emissions (i.e., more components or more complex parametrizations).

Exploration of the possible effects of mis-calibration or frequency band mismatch --- potentially important systematic effects for component separation~\cite{2009MNRAS.392..216S,PhysRevD.85.083006} --- is left for further work.

\textbf{Delensing ---} Our delensing method allows estimation of the lensing potential using the CMB polarization four-point function or the correlation between the CMB and two ``external'' tracers of the lensing kernel: the CIB and large-scale structure. Delensing through cross-correlation, which requires only high-precision $E$ mode measurements, outperforms ``internal'' CMB delensing for post-component-separation noise levels $\gtrsim 3$ \ukarc, and represents the most promising avenue for pre-2020 experiments. We find that delensing using existing CIB data from \planck's 545 GHz channel is just as effective as using the highly futuristic large-scale structure measurements considered here, with both methods able to remove a maximum of 60\% of the lensing using cosmic-variance-limited $E$-mode measurements (post-component-separation noise levels of $\lesssim10$ \ukarc).

For the post-2020 experiments whose post-component-separation noise levels and resolution allow precise measurements of the lensing $B$ mode, the iterability of pure-CMB delensing means it rapidly outperforms the cross-correlation methods. In the limit of noiseless polarization measurements perfect delensing should be possible~\cite{2004PhRvD..69d3005S}; in the cases considered here, up to around 80\% of the lensing $B$ modes can be removed by the combination of the ground-based Stage-IV and any satellite, assuming polarized dust and synchrotron have previously been cleaned from the map. Note that we do not explore in this work the possible impact of instrumental systematics which could limit lensing reconstruction, such as pointing mis-calibration, beam mis-characterization, optical systematics, etc. (see, e.g., Ref.~\cite{2009PhRvD..79f3008M}); the impact of such instrumental systematics will grow with ever-increasing instrumental sensitivity.

\textbf{Parameter constraints ---}  The Fisher estimate combines the two previous algorithms and produces constraints on any cosmological parameter. Our results are focused on $r$, \nt, \ns, \alphas, \mnu, \w, \neff\ and \omk. We show that in the case of CMB, synchrotron and dust, and after delensing and marginalization over foreground residuals, the best pre-2020 instruments in combination with \planck\ can reach  $\sigma(r)\,\sim\,3\times10^{-3}$, $\sigma(\nt)\,\sim\,0.2$, $\sigma( \ns )\,\sim\,2.2\times10^{-3}$, $\sigma(\alphas)\,\sim\,3\times10^{-3}$, $\sigma( \mnu )\,\sim\,55$ meV, $\sigma( \w )\,\sim\,0.16$, $\sigma( \wo )\,\sim\,0.36$, $\sigma( \wa )\,\sim\,0.71$, $\sigma( \neff )\,\sim\,0.05-0.06$ and $\sigma( \omk )\,\sim\,2.5\times10^{-3}$ when delensing using the CMB$\times$CIB method. Post-2020 instruments, in particular the combination of the ground-based Stage-IV and a space mission, could reach constraints $\sigma(r)\,\sim1.3\times10^{-4}$, $\sigma(\nt)\,\sim\,0.03$, $\sigma( \ns )\,\sim\,1.8\times10^{-3}$, $\sigma(\alphas)\,\sim\,1.7\times10^{-3}$, $\sigma( \mnu )\,\sim\,31$ meV, $\sigma( \w )\,\sim\,0.09$, $\sigma( \wo )\,\sim\,0.25$, $\sigma( \wa )\,\sim\,0.50$, $\sigma( \neff )\,\sim\,0.024$ and $\sigma( \omk )\,\sim\,1.5\times10^{-3}$. These constraints are robust under both approaches to spatially-varying foreground spectral indices.
The combination of CMB measurements with {\it Euclid}, DESI, LSST and supernovae data will significantly improve the constraints on the dark sector. For example, constraints on \w\ derived from CMB  observatories seem to be around 10\%, compared with the target constraints from {\it Euclid} or LSST ($\sigma(\w)\,\sim\,1\%$). Finally, none of the CMB experiments studied is able to verify the inflationary consistency condition $\nt = -r/8$ after disentangling dust, synchrotron and CMB, even assuming a level $r=0.1$ of primordial $B$ modes (confirming the results of Ref.~\cite{2015arXiv150902676H}, which are derived in a more optimistic setting). Again, external priors like $H(z)$ given by e.g., DESI could significantly improve all the quoted marginalized constraints.

Finally, we have made our forecasting toolkit publicly available through a web interface at \url{turkey.lbl.gov}. We hope that this tool and its extensions will serve the community by enabling the optimal design of future CMB observatories for a broad range of scientific goals, thus unleashing the full potential of the polarized CMB sky to constrain fundamental physics. 

\acknowledgments 

The authors would like to thank Vamsi Vytla\footnote{The Center for X-Ray Optics, Lawrence Berkeley National Lab, Berkeley CA 94720; \href{mailto:vkvytla@lbl.gov}{vkvytla@lbl.gov}.} who helped create the web interface described in Sect.~\ref{sec:web_interface}. We would also like to thank Radek Stompor for very useful discussions and for his critical insights on the parametric maximum-likelihood component separation algorithm, Carlo Baccigalupi for sharing his expertise on foreground properties, Jacques Delabrouille for his pertinent comments on an early draft, and Raphael Flauger for very helpful discussions about curvature derivatives.

This work was partially completed within the Labex ILP (reference ANR-10-LABX-63) part of the Idex SUPER, and received financial state aid managed by the Agence Nationale de la Recherche, as part of the programme Investissements d'avenir under the reference ANR-11-IDEX-0004-02. SMF is supported by the Science and Technology Facilities Council in the UK. This work was also partially supported by a New Frontiers in Astronomy and Cosmology grant \#37426, as well as by National Science Foundation Grant No. PHYS-1066293 and the hospitality of the Aspen Center for Physics. HVP was partially supported by the European Research Council under the European Community's Seventh Framework Programme (FP7/2007- 2013) / ERC grant agreement no 306478-CosmicDawn.


\appendix
\section{Instrument specifications}
\label{app:instruments_specifications}

\Cref{table:quijote_cbass_planck_specs,table:pre2020_specs,table:post2020_specs} give the specifications of each instrument considered in this work. The maximum multipole tabulated for each instrument is indicative: as in Fig.~\ref{fig:freq_ell_coverage}, it is taken to be the minimum of 4000 and the multipole at which the beam-deconvolved noise reaches $10^4 \, \mu{\rm K}^2$, where it dominates over even the temperature power spectrum. Fig.~\ref{fig:post-2020-sensitivities} shows polarized sensitivities of all post-2020 instruments as a function of frequency.

\begin{table}[htb!]
\centering
\caption{\planck, C-BASS, QUIJOTE-CMB}
\label{table:quijote_cbass_planck_specs}
\resizebox{12cm}{!}{
\hspace{-2cm}
\begin{tabular}{|c|c|c|c|c|c|c|} 
\hline
\multicolumn{7}{|c|}{\planck\ specifications, Ref.~\cite{2015arXiv150201588P}}\\
\hline
frequencies [GHz] & fractional bandpass [\%] & sensitivities [\ukarc] & $\fsky$ [\%] & FWHM [arcmin] & $\lmin$ & $\lmax$ \\
\hline
30.0 & \multirow{7}{*}{30.0} & 300.0 & \multirow{7}{*}{50.0} 	&  33.2   & \multirow{7}{*}{ 3 } & \multirow{7}{*}{2800} \\
44.0 & & 300.0  &    			      	&  28.1  &   &  \\
70.0 & &192.0 & 					& 13.1  &   &   \\
100.0 & &52.0 & 					& 9.7 &   &   \\
143.0 & &44.0 & 					& 7.3  &   &   \\
217.0 & &64.0 & 					& 5.0  &   &   \\
353.0 & &276.0 & 				& 4.9  &   &   \\
\hline
\hline
\multicolumn{7}{|c|}{C-BASS specifications, \url{http://www.astro.caltech.edu/cbass}}\\
\hline
frequencies [GHz] & fractional bandpass [\%] & sensitivities [\ukarc] & $\fsky$ [\%] & FWHM [arcmin] & $\lmin$ & $\lmax$ \\
\hline
5.0 & 20.0 & 4500.0 & 80.0 	&  45.0   & 20 & 500 \\
\hline
\hline
\multicolumn{7}{|c|}{QUIJOTE-CMB specifications, Ref.~\cite{2012SPIE.8444E..2YR}}\\
\hline
frequencies [GHz] & fractional bandpass [\%] & sensitivities [\ukarc] & $\fsky$ [\%] & FWHM [arcmin] & $\lmin$ & $\lmax$ \\
\hline
11.0 & 18.0 & 840.0 & \multirow{6}{*}{44.0} 	&  55.2   & \multirow{6}{*}{ 20 } & \multirow{6}{*}{500} \\
13.0 & 15.0 & 840.0  &    			     	&  55.2  &   &  \\
17.0 & 12.0 & 840.0 & 				& 36.0  &   &   \\
19.0 & 11.0 & 840.0 & 				& 36.0 &   &   \\
30.0 & 27.0 & 66.6 & 				&  22.2  &   &   \\
42.0 & 24.0 & 50.4 & 				& 16.8  &   &   \\
\hline
\end{tabular}}
\end{table}


\begin{table}[htb!]
\caption{Pre-2020 instruments}
\centering
\label{table:pre2020_specs}
\resizebox{12cm}{!}{
\hspace{-2cm}
\begin{tabular}{|c|c|c|c|c|c|c|} 
\hline
\multicolumn{7}{|c|}{Advanced ACTPol specifications, Ref.~\cite{2014JCAP...08..010C}}\\
\hline
frequencies [GHz] & fractional bandpass [\%] & sensitivities [\ukarc] & $\fsky$ [\%] & FWHM [arcmin] & $\lmin$ & $\lmax$ \\
\hline
90.0 &  \multirow{3}{*}{30.0} & 11.0 & \multirow{3}{*}{50.0} 	&  2.2   & \multirow{3}{*}{ 20 } & \multirow{3}{*}{4000} \\
150.0 & &9.8  &    			      	&  1.3  &   &  \\
230.0 & & 35.4 & 					& 0.9  &   &   \\
\hline
\hline
\multicolumn{7}{|c|}{BICEP3 + Keck specifications}\\
\hline
frequencies [GHz] & fractional bandpass [\%] & sensitivities [\ukarc] & $\fsky$ [\%] & FWHM [arcmin] & $\lmin$ & $\lmax$ \\
\hline
95.0 &  \multirow{2}{*}{30.0}  & 1.7  & \multirow{2}{*}{1.0} 	&  25.0   & \multirow{2}{*}{ 20 } & \multirow{2}{*}{1300} \\
150.0 &  & 3.4  &    			      	&  30.0  &   &  \\
\hline
\hline
\multicolumn{7}{|c|}{CLASS specifications, Ref.~\cite{2014SPIE.9153E..1IE}}\\
\hline
frequencies [GHz] & fractional bandpass [\%] & sensitivities [\ukarc] & $\fsky$ [\%] & FWHM [arcmin] & $\lmin$ & $\lmax$ \\
\hline
38.0 & \multirow{4}{*}{30.0} & 39.0 & \multirow{4}{*}{70.0} 	&  90.0   & \multirow{4}{*}{ 20 } & \multirow{4}{*}{1100} \\
93.0 & & 10.0  &    			      	&  40.0  &   &  \\
148.0 & & 15.0 & 					& 24.0  &   &   \\
217.0 & & 43.0 & 					& 18.0  &   &   \\
\hline
\hline
\multicolumn{7}{|c|}{EBEX10K specifications, proposal to NASA in 2015}\\
\hline
frequencies [GHz] & fractional bandpass [\%] & sensitivities [\ukarc] & $\fsky$ [\%] & FWHM [arcmin] & $\lmin$ & $\lmax$ \\
\hline
150.0 & \multirow{4}{*}{30.0} &5.5 & \multirow{4}{*}{2.5} 	&  6.6   & \multirow{4}{*}{20} & \multirow{4}{*}{4000} \\
220.0 & &11.0  &    		      		&  4.7  &   &   \\
280.0 & &25.4 & 					& 3.9 &   &   \\
350.0 & &53.0 &					& 3.3 &   &    \\
\hline
\hline
\multicolumn{7}{|c|}{PIPER specifications, Ref.~\cite{2014SPIE.9153E..1LL}}\\
\hline
frequencies [GHz] & fractional bandpass [\%] & sensitivities [\ukarc] & $\fsky$ [\%] & FWHM [arcmin] & $\lmin$ & $\lmax$ \\
\hline
200.0 & 30.0 & 31.4 & \multirow{4}{*}{85.0} 	&  21.0   & \multirow{4}{*}{ 20 } & \multirow{4}{*}{1000} \\
270.0 & 30.0 & 45.9 &    			      	&  21.0  &   &  \\
350.0 & 16.0 & 162.0 & 				&  21.0  &   &   \\
600.0 & 10.0 & 2659.2 & 				&  21.0 &   &   \\
\hline
\multicolumn{7}{|c|}{Simons Array specifications, Ref.~\cite{2014SPIE.9153E..1FA}}\\
\hline
frequencies [GHz] & fractional bandpass [\%] & sensitivities [\ukarc] & $\fsky$ [\%] & FWHM [arcmin] & $\lmin$ & $\lmax$ \\
\hline
90.0 &  \multirow{3}{*}{30.0} & 14.4 & \multirow{3}{*}{65.0} 	&  5.2   & \multirow{3}{*}{ 20 } & \multirow{3}{*}{4000} \\
150.0 & & 11.8  &    			      	&  3.5  &   &  \\
220.0 & & 40.3 & 					& 2.7 &   &   \\
\hline
\hline
\multicolumn{7}{|c|}{SPIDER specifications, Ref.~\cite{2008SPIE.7010E..2PC}}\\
\hline
frequencies [GHz] & fractional bandpass [\%] & sensitivities [\ukarc] & $\fsky$ [\%] & FWHM [arcmin] & $\lmin$ & $\lmax$ \\
\hline
90.0 & 24.0 & 21.2 & \multirow{2}{*}{8.0} 	&  45.0   & \multirow{2}{*}{ 20 } & \multirow{2}{*}{800} \\
150.0 & 24.0 & 17.7  &    			      	&  30.0  &   &  \\
\hline
\hline
\multicolumn{7}{|c|}{SPT-3G specifications, Ref.~\cite{2014SPIE.9153E..1PB}}\\
\hline
frequencies [GHz] & fractional bandpass [\%] & sensitivities [\ukarc] & $\fsky$ [\%] & FWHM [arcmin] & $\lmin$ & $\lmax$ \\
\hline
95.0 & 27.0 & 7.0 & \multirow{3}{*}{6.0} &  1.6   & \multirow{3}{*}{ 20 } & \multirow{3}{*}{4000} \\
148.0 & 26.0 & 4.5  &    			&  1.1  &   &  \\
223.0 & 23.0 & 7.5  &    			&  1.0  &   &  \\
\hline
\end{tabular}}
\end{table}


\begin{table}[htb!]
\caption{Post-2020 instruments}
\centering
\label{table:post2020_specs}
\resizebox{12cm}{!}{
\hspace{-2cm}
\begin{tabular}{|c|c|c|c|c|c|c|} 
\hline
\multicolumn{7}{|c|}{\core\ specifications, \url{http://conservancy.umn.edu/handle/11299/169642}}\\
\hline
frequencies [GHz] & fractional bandpass [\%] & sensitivities [\ukarc] & $\fsky$ [\%] & FWHM [arcmin] & $\lmin$ & $\lmax$ \\
\hline 
60.0 & \multirow{16}{*}{ 30.0 } &  16.0 & \multirow{16}{*}{70.0} 	&  14.0   & \multirow{16}{*}{ 2 } & \multirow{19}{*}{4000} \\
70.0 & & 14.9 &    			      	&  12.0  &   &  \\
80.0 & & 12.9 & 					& 10.5  &   &   \\
90.0 & & 9.2 & 					& 9.3  &   &   \\
100.0 & & 8.5 & 				& 8.4  &   &   \\
115.0 & & 7.0 & 				& 7.3  &   &   \\
130.0 & & 5.9 & 				& 6.5  &   &   \\
145.0 & & 5.0 & 				& 5.8  &   &   \\
160.0 & & 5.4 & 				& 5.3  &   &   \\
175.0 & & 5.3 & 				& 4.8  &   &   \\
195.0 & & 5.3 & 				& 4.3  &   &   \\
220.0 & & 8.1 & 				& 3.8  &   &   \\
255.0 & & 12.6 & 				& 3.3  &   &   \\
295.0 & & 27.4 & 				& 2.9  &   &   \\
340.0 & & 43.7 & 				& 2.5  &   &   \\
390.0 & & 77.8 & 				& 2.2  &   &   \\
450.0 & & 164.8 & 				& 1.9  &   &   \\
520.0 & & 418.2 & 				& 1.6  &   &   \\
600.0 & & 1272.4 & 				& 1.4  &   &   \\
\hline
\hline
\multicolumn{7}{|c|}{\lb\ specifications \url{http://ltd16.grenoble.cnrs.fr/IMG/UserFiles/Images/09\_TMatsumura\_20150720\_LTD\_v18.pdf}}\\
\hline
frequencies [GHz] & fractional bandpass [\%] & sensitivities [\ukarc] & $\fsky$ [\%] & FWHM [arcmin] & $\lmin$ & $\lmax$ \\
\hline
40.0 & \multirow{15}{*}{30.0} & 42.5 & \multirow{15}{*}{70.0} &  108   & \multirow{15}{*}{ 2 } & \multirow{15}{*}{1350}  \\
50.0 &   & 26.0  &    			       &  86  &   &   \\
60.0 &  &20.0 &    		  	       & 72 &   &  \\
68.4 &  &15.5 &    			       & 63 &   &    \\
78.0 &  &12.5  &    			       &  55  &   &  \\
88.5 &  &10.0  &    			       &  49  &   &    \\
100.0 &  &12.0   &    			       &  43  &   &  \\
118.9 & & 9.5  &    			       &  36  &   & \\
140.0 & &7.5  &    			       &  31  &   &    \\
166.0 &  &7.0  &    			       &  26  &   &   \\
195.0 &  &5.0  &    			       &  22  &   &  \\
234.9 &  &6.5  &    			       &  18  &   &   \\
280.0 & &10.0  &    			       &  37  &   &   \\
337.4 & &10.0  &    			       &  31  &   & \\
402.1 & &19.0   &    			       &  26  &   &  \\
\hline
\hline
\multicolumn{7}{|c|}{Stage-IV specifications, derived so that the noise after component separation, $\sigma_{CMB}$, is $\sim1$ \ukarc, Refs.~\cite{2014ApJ...788..138W,2015APh....63...66A}}\\
\hline
frequencies [GHz] & fractional bandpass [\%] & sensitivities [\ukarc] & $\fsky$ [\%] & FWHM [arcmin] & $\lmin$ & $\lmax$ \\
\hline
40.0 & \multirow{5}{*}{30.0} &3.0 & \multirow{5}{*}{50.0} 	&  11.0	& \multirow{5}{*}{20} & \multirow{5}{*}{4000} \\
90.0 & &1.5  &    			      		&  5.0		&   &    \\
150.0 & &1.5 & 					& 3.0		&   &  \\
220.0 & &5.0 &					& 2.0		&   &   \\
280.0 & & 9.0  &					& 1.5	&  &  \\
\hline
\hline
\multicolumn{7}{|c|}{PIXIE specifications, Ref.~\cite{2011JCAP...07..025K}}\\
\hline
frequencies [GHz] & fractional bandpass [\%] & sensitivities [\ukarc] & $\fsky$ [\%] & FWHM [arcmin] & $\lmin$ & $\lmax$ \\
\hline
\multicolumn{3}{|c|}{see Fig.~\ref{fig:pixie_sensitivity}}  & 70.0  &  96.0   &  2 & 500  \\
\hline
\end{tabular}}
\end{table}

\begin{figure}[htb!]
	\centering
		\includegraphics[width=10cm]{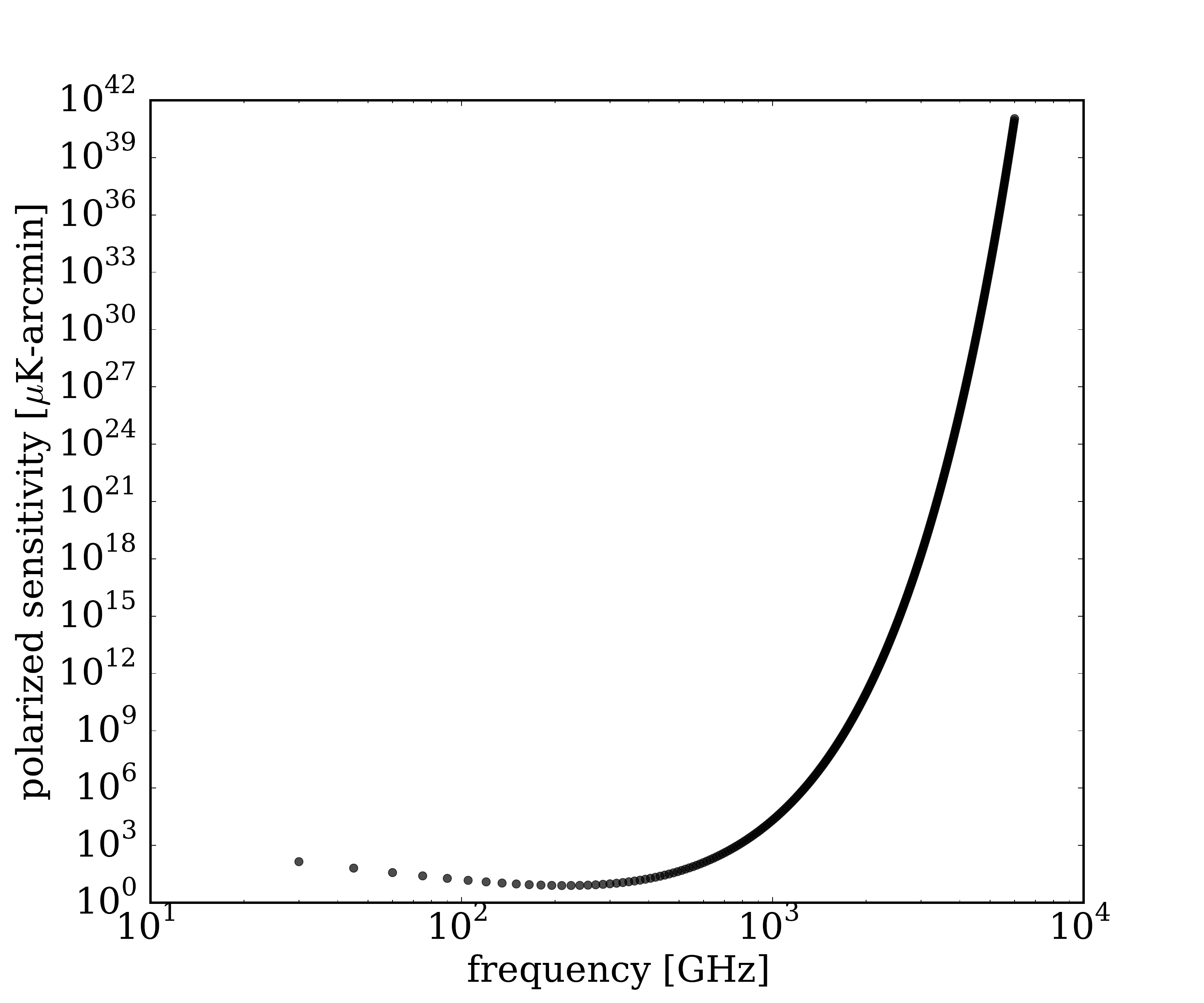}
	\caption{Polarized sensitivities for PIXIE. Bandpasses are $15$ GHz wide across all frequency channels.}
	\label{fig:pixie_sensitivity}
\end{figure}

\begin{figure}[htb!]
	\centering
		\includegraphics[width=12cm]{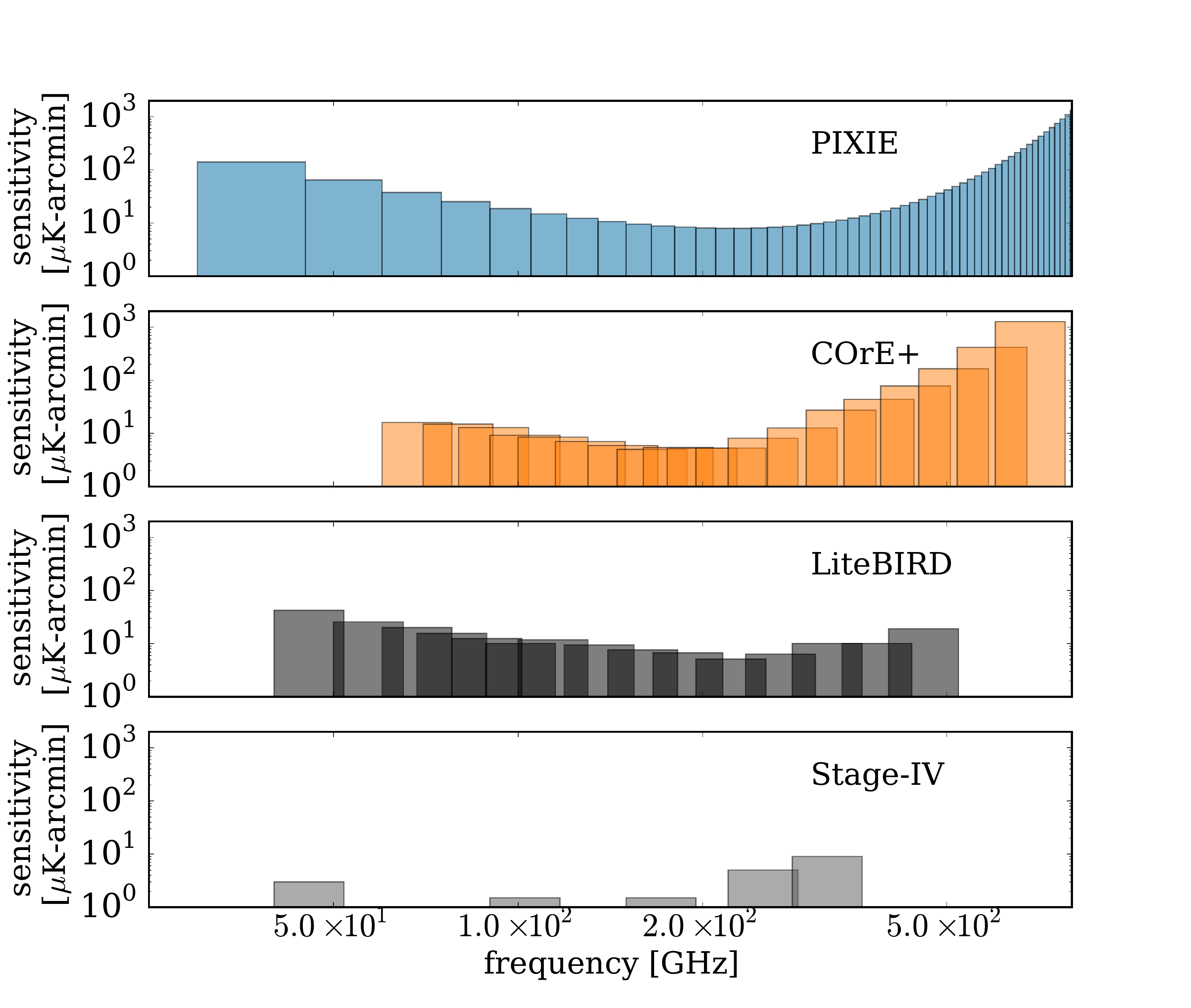}
		\caption{Comparison of polarized sensitivities for post-2020 instruments between 30 and 800 GHz. }
	\label{fig:post-2020-sensitivities}
\end{figure}
\clearpage
\section{Error bars on cosmological parameters for each project}
\label{app:fisher_tables}

We detail in this section the error bars on several relevant cosmological parameters as estimated by our algorithm, for the cross-correlation of all pre-2020 instruments with \planck, as well as for the post-2020 observatories. In addition, we summarize the delensing factor, the effective level of foreground residuals and the noise degradation introduced in Sect.~\ref{sec:methodology}.
Each constraint is derived assuming a particular cosmology, as detailed in Table~\ref{table:models}.

In the cases considering combinations of datasets, $\Delta$ and $\reff$ are the values obtained over the overlapping sky, i.e., when all the frequency maps and sensitivities from the two instruments are combined.

\subsection{Pre-2020 instruments, cross-correlated with \planck}

\begin{table}[htb!]
\label{table:bicep3Xplanck}
\resizebox{18cm}{!}{
\hspace{-3cm}
\begin{tabular}{|l|l||l|l||l|l||l|l||l|l|}		
\hline
\multicolumn{10}{|c|}{BICEP3 $\times$ Keck Array $\times$ \planck} \\
\hline
\hline
\multicolumn{2}{|l||}{Delensing option $\rightarrow$} &  \multicolumn{2}{c||}{no} & \multicolumn{2}{c||}{CMB x} &  \multicolumn{2}{c||}{CMB x} &  \multicolumn{2}{c|}{CMB x} \\
\multicolumn{2}{|l||}{$\downarrow$ comp. sep. option} &  \multicolumn{2}{c||}{delensing} & \multicolumn{2}{c||}{CMB} & \multicolumn{2}{c||}{CIB} & \multicolumn{2}{c|}{LSS} \\
\hline
 \multirow{6}{*}{\parbox{2.3cm}{cmb-only}} & \multirow{3}{*}{$\Delta=1.0$} & \multicolumn{2}{c||}{$\alpha=1.0$} & \multicolumn{2}{c||}{$\alpha=0.66$} & \multicolumn{2}{c||}{$\alpha=0.43$} & \multicolumn{2}{c|}{$\alpha=0.43$}\\ & & $\sigma(r)=2.5\times10^{-3}$ & $\sigma(\mnu)=1.7\times10^{2}$ & $\sigma(r)=2.1\times10^{-3}$ & $\sigma(\mnu)=1.7\times10^{2}$ & $\sigma(r)=1.6\times10^{-3}$ & $\sigma(\mnu)=69$ & $\sigma(r)=1.6\times10^{-3}$ & $\sigma(\mnu)=50$ \\ & & $\sigma(\nt)=0.10$ & $\sigma(\neff)=0.15$ & $\sigma(\nt)=0.098$ & $\sigma(\neff)=0.15$ & $\sigma(\nt)=0.094$ & $\sigma(\neff)=0.10$ & $\sigma(\nt)=0.094$ & $\sigma(\neff)=0.069$ \\ & \multirow{3}{*}{$\reff=0.0$} & $\sigma(\ns)=3.0\times10^{-3}$ & $\sigma(\w)=0.41$ & $\sigma(\ns)=3.0\times10^{-3}$ & $\sigma(\w)=0.41$ & $\sigma(\ns)=3.0\times10^{-3}$ & $\sigma(\w)=0.18$ & $\sigma(\ns)=2.9\times10^{-3}$ & $\sigma(\w)=0.13$ \\ & & $\sigma(\alphas)=4.8\times10^{-3}$ & $\sigma(\wo)=1.3$ & $\sigma(\alphas)=4.8\times10^{-3}$ & $\sigma(\wo)=1.3$ & $\sigma(\alphas)=4.4\times10^{-3}$ & $\sigma(\wo)=0.45$ & $\sigma(\alphas)=2.6\times10^{-3}$ & $\sigma(\wo)=0.14$ \\
\hline
\end{tabular}}
\end{table}

\begin{table}[htb!]
\label{table:ebex10kXplanck}
\resizebox{18cm}{!}{
\hspace{-3cm}
\begin{tabular}{|l|l||l|l||l|l||l|l||l|l|}		
\hline
\multicolumn{10}{|c|}{EBEX10K $\times$ \planck} \\
\hline
\hline
\multicolumn{2}{|l||}{Delensing option $\rightarrow$} &  \multicolumn{2}{c||}{no} & \multicolumn{2}{c||}{CMB x} &  \multicolumn{2}{c||}{CMB x} &  \multicolumn{2}{c|}{CMB x} \\
\multicolumn{2}{|l||}{$\downarrow$ comp. sep. option} &  \multicolumn{2}{c||}{delensing} & \multicolumn{2}{c||}{CMB} & \multicolumn{2}{c||}{CIB} & \multicolumn{2}{c|}{LSS} \\
\hline
 \multirow{6}{*}{\parbox{2.3cm}{cmb-only}} & \multirow{3}{*}{$\Delta=1.0$} & \multicolumn{2}{c||}{$\alpha=1.0$} & \multicolumn{2}{c||}{$\alpha=0.61$} & \multicolumn{2}{c||}{$\alpha=0.41$} & \multicolumn{2}{c|}{$\alpha=0.39$}\\ & & $\sigma(r)=2.7\times10^{-3}$ & $\sigma(\mnu)=1.2\times10^{2}$ & $\sigma(r)=2.5\times10^{-3}$ & $\sigma(\mnu)=1.2\times10^{2}$ & $\sigma(r)=2.3\times10^{-3}$ & $\sigma(\mnu)=66$ & $\sigma(r)=2.3\times10^{-3}$ & $\sigma(\mnu)=48$ \\ & & $\sigma(\nt)=0.10$ & $\sigma(\neff)=0.13$ & $\sigma(\nt)=0.099$ & $\sigma(\neff)=0.13$ & $\sigma(\nt)=0.097$ & $\sigma(\neff)=0.099$ & $\sigma(\nt)=0.097$ & $\sigma(\neff)=0.067$ \\ & \multirow{3}{*}{$\reff=0.0$} & $\sigma(\ns)=2.9\times10^{-3}$ & $\sigma(\w)=0.33$ & $\sigma(\ns)=2.9\times10^{-3}$ & $\sigma(\w)=0.33$ & $\sigma(\ns)=2.9\times10^{-3}$ & $\sigma(\w)=0.17$ & $\sigma(\ns)=2.8\times10^{-3}$ & $\sigma(\w)=0.13$ \\ & & $\sigma(\alphas)=4.6\times10^{-3}$ & $\sigma(\wo)=1.2$ & $\sigma(\alphas)=4.6\times10^{-3}$ & $\sigma(\wo)=1.2$ & $\sigma(\alphas)=4.3\times10^{-3}$ & $\sigma(\wo)=0.45$ & $\sigma(\alphas)=2.6\times10^{-3}$ & $\sigma(\wo)=0.14$ \\
\hline
\end{tabular}}
\end{table}

\begin{table}[htb!]
\label{table:advactpolXplanck}
\resizebox{18cm}{!}{
\hspace{-3cm}
\begin{tabular}{|l|l||l|l||l|l||l|l||l|l|}		
\hline
\multicolumn{10}{|c|}{AdvACTPol $\times$ \planck} \\
\hline
\hline
\multicolumn{2}{|l||}{Delensing option $\rightarrow$} &  \multicolumn{2}{c||}{no} & \multicolumn{2}{c||}{CMB x} &  \multicolumn{2}{c||}{CMB x} &  \multicolumn{2}{c|}{CMB x} \\
\multicolumn{2}{|l||}{$\downarrow$ comp. sep. option} &  \multicolumn{2}{c||}{delensing} & \multicolumn{2}{c||}{CMB} & \multicolumn{2}{c||}{CIB} & \multicolumn{2}{c|}{LSS} \\
\hline
 \multirow{6}{*}{\parbox{2.3cm}{cmb-only}} & \multirow{3}{*}{$\Delta=1.0$} & \multicolumn{2}{c||}{$\alpha=1.0$} & \multicolumn{2}{c||}{$\alpha=0.66$} & \multicolumn{2}{c||}{$\alpha=0.41$} & \multicolumn{2}{c|}{$\alpha=0.39$}\\ & & $\sigma(r)=1.3\times10^{-3}$ & $\sigma(\mnu)=53$ & $\sigma(r)=1.2\times10^{-3}$ & $\sigma(\mnu)=53$ & $\sigma(r)=1.1\times10^{-3}$ & $\sigma(\mnu)=49$ & $\sigma(r)=1.1\times10^{-3}$ & $\sigma(\mnu)=39$ \\ & & $\sigma(\nt)=0.087$ & $\sigma(\neff)=0.044$ & $\sigma(\nt)=0.084$ & $\sigma(\neff)=0.044$ & $\sigma(\nt)=0.082$ & $\sigma(\neff)=0.042$ & $\sigma(\nt)=0.081$ & $\sigma(\neff)=0.034$ \\ & \multirow{3}{*}{$\reff=0.0$} & $\sigma(\ns)=2.2\times10^{-3}$ & $\sigma(\w)=0.14$ & $\sigma(\ns)=2.2\times10^{-3}$ & $\sigma(\w)=0.14$ & $\sigma(\ns)=2.2\times10^{-3}$ & $\sigma(\w)=0.14$ & $\sigma(\ns)=2.2\times10^{-3}$ & $\sigma(\w)=0.11$ \\ & & $\sigma(\alphas)=2.6\times10^{-3}$ & $\sigma(\wo)=0.51$ & $\sigma(\alphas)=2.6\times10^{-3}$ & $\sigma(\wo)=0.51$ & $\sigma(\alphas)=2.6\times10^{-3}$ & $\sigma(\wo)=0.36$ & $\sigma(\alphas)=2.1\times10^{-3}$ & $\sigma(\wo)=0.12$ \\
\hline
\end{tabular}}
\end{table}

\begin{table}[htb!]
\label{table:piperXplanck}
\resizebox{18cm}{!}{
\hspace{-3cm}
\begin{tabular}{|l|l||l|l||l|l||l|l||l|l|}		
\hline
\multicolumn{10}{|c|}{PIPER $\times$ \planck} \\
\hline
\hline
\multicolumn{2}{|l||}{Delensing option $\rightarrow$} &  \multicolumn{2}{c||}{no} & \multicolumn{2}{c||}{CMB x} &  \multicolumn{2}{c||}{CMB x} &  \multicolumn{2}{c|}{CMB x} \\
\multicolumn{2}{|l||}{$\downarrow$ comp. sep. option} &  \multicolumn{2}{c||}{delensing} & \multicolumn{2}{c||}{CMB} & \multicolumn{2}{c||}{CIB} & \multicolumn{2}{c|}{LSS} \\
\hline
 \multirow{6}{*}{\parbox{2.3cm}{cmb-only}} & \multirow{3}{*}{$\Delta=1.0$} & \multicolumn{2}{c||}{$\alpha=1.0$} & \multicolumn{2}{c||}{$\alpha=0.97$} & \multicolumn{2}{c||}{$\alpha=0.50$} & \multicolumn{2}{c|}{$\alpha=0.50$}\\ & & $\sigma(r)=3.8\times10^{-3}$ & $\sigma(\mnu)=2.3\times10^{2}$ & $\sigma(r)=3.8\times10^{-3}$ & $\sigma(\mnu)=2.3\times10^{2}$ & $\sigma(r)=3.7\times10^{-3}$ & $\sigma(\mnu)=59$ & $\sigma(r)=3.7\times10^{-3}$ & $\sigma(\mnu)=40$ \\ & & $\sigma(\nt)=0.12$ & $\sigma(\neff)=0.13$ & $\sigma(\nt)=0.12$ & $\sigma(\neff)=0.13$ & $\sigma(\nt)=0.12$ & $\sigma(\neff)=0.088$ & $\sigma(\nt)=0.12$ & $\sigma(\neff)=0.058$ \\ & \multirow{3}{*}{$\reff=0.0$} & $\sigma(\ns)=2.4\times10^{-3}$ & $\sigma(\w)=0.51$ & $\sigma(\ns)=2.4\times10^{-3}$ & $\sigma(\w)=0.51$ & $\sigma(\ns)=2.4\times10^{-3}$ & $\sigma(\w)=0.15$ & $\sigma(\ns)=2.4\times10^{-3}$ & $\sigma(\w)=0.11$ \\ & & $\sigma(\alphas)=4.1\times10^{-3}$ & $\sigma(\wo)=1.2$ & $\sigma(\alphas)=4.1\times10^{-3}$ & $\sigma(\wo)=1.2$ & $\sigma(\alphas)=3.7\times10^{-3}$ & $\sigma(\wo)=0.36$ & $\sigma(\alphas)=2.2\times10^{-3}$ & $\sigma(\wo)=0.11$ \\
\hline
\end{tabular}}
\end{table}

\begin{table}[htb!]
\label{table:spiderXplanck}
\resizebox{18cm}{!}{
\hspace{-3cm}
\begin{tabular}{|l|l||l|l||l|l||l|l||l|l|}		
\hline
\multicolumn{10}{|c|}{SPIDER $\times$ \planck} \\
\hline
\hline
\multicolumn{2}{|l||}{Delensing option $\rightarrow$} &  \multicolumn{2}{c||}{no} & \multicolumn{2}{c||}{CMB x} &  \multicolumn{2}{c||}{CMB x} &  \multicolumn{2}{c|}{CMB x} \\
\multicolumn{2}{|l||}{$\downarrow$ comp. sep. option} &  \multicolumn{2}{c||}{delensing} & \multicolumn{2}{c||}{CMB} & \multicolumn{2}{c||}{CIB} & \multicolumn{2}{c|}{LSS} \\
\hline
 \multirow{6}{*}{\parbox{2.3cm}{cmb-only}} & \multirow{3}{*}{$\Delta=1.0$} & \multicolumn{2}{c||}{$\alpha=1.0$} & \multicolumn{2}{c||}{$\alpha=0.97$} & \multicolumn{2}{c||}{$\alpha=0.49$} & \multicolumn{2}{c|}{$\alpha=0.50$}\\ & & $\sigma(r)=3.1\times10^{-3}$ & $\sigma(\mnu)=2.6\times10^{2}$ & $\sigma(r)=3.1\times10^{-3}$ & $\sigma(\mnu)=2.6\times10^{2}$ & $\sigma(r)=3.1\times10^{-3}$ & $\sigma(\mnu)=69$ & $\sigma(r)=3.1\times10^{-3}$ & $\sigma(\mnu)=50$ \\ & & $\sigma(\nt)=0.11$ & $\sigma(\neff)=0.15$ & $\sigma(\nt)=0.11$ & $\sigma(\neff)=0.15$ & $\sigma(\nt)=0.11$ & $\sigma(\neff)=0.10$ & $\sigma(\nt)=0.11$ & $\sigma(\neff)=0.069$ \\ & \multirow{3}{*}{$\reff=0.0$} & $\sigma(\ns)=3.0\times10^{-3}$ & $\sigma(\w)=0.52$ & $\sigma(\ns)=3.0\times10^{-3}$ & $\sigma(\w)=0.52$ & $\sigma(\ns)=2.9\times10^{-3}$ & $\sigma(\w)=0.18$ & $\sigma(\ns)=2.9\times10^{-3}$ & $\sigma(\w)=0.13$ \\ & & $\sigma(\alphas)=4.9\times10^{-3}$ & $\sigma(\wo)=1.3$ & $\sigma(\alphas)=4.9\times10^{-3}$ & $\sigma(\wo)=1.3$ & $\sigma(\alphas)=4.4\times10^{-3}$ & $\sigma(\wo)=0.45$ & $\sigma(\alphas)=2.6\times10^{-3}$ & $\sigma(\wo)=0.14$ \\
\hline
\end{tabular}}
\end{table}

\begin{table}[htb!]
\label{table:simonsarrayXplanck}
\resizebox{18cm}{!}{
\hspace{-3cm}
\begin{tabular}{|l|l||l|l||l|l||l|l||l|l|}		
\hline
\multicolumn{10}{|c|}{Simons Array $\times$ \planck} \\
\hline
\hline
\multicolumn{2}{|l||}{Delensing option $\rightarrow$} &  \multicolumn{2}{c||}{no} & \multicolumn{2}{c||}{CMB x} &  \multicolumn{2}{c||}{CMB x} &  \multicolumn{2}{c|}{CMB x} \\
\multicolumn{2}{|l||}{$\downarrow$ comp. sep. option} &  \multicolumn{2}{c||}{delensing} & \multicolumn{2}{c||}{CMB} & \multicolumn{2}{c||}{CIB} & \multicolumn{2}{c|}{LSS} \\
\hline
 \multirow{6}{*}{\parbox{2.3cm}{cmb-only}} & \multirow{3}{*}{$\Delta=1.0$} & \multicolumn{2}{c||}{$\alpha=1.0$} & \multicolumn{2}{c||}{$\alpha=0.76$} & \multicolumn{2}{c||}{$\alpha=0.41$} & \multicolumn{2}{c|}{$\alpha=0.40$}\\ & & $\sigma(r)=1.6\times10^{-3}$ & $\sigma(\mnu)=56$ & $\sigma(r)=1.5\times10^{-3}$ & $\sigma(\mnu)=56$ & $\sigma(r)=1.4\times10^{-3}$ & $\sigma(\mnu)=47$ & $\sigma(r)=1.4\times10^{-3}$ & $\sigma(\mnu)=37$ \\ & & $\sigma(\nt)=0.089$ & $\sigma(\neff)=0.055$ & $\sigma(\nt)=0.088$ & $\sigma(\neff)=0.055$ & $\sigma(\nt)=0.085$ & $\sigma(\neff)=0.049$ & $\sigma(\nt)=0.085$ & $\sigma(\neff)=0.038$ \\ & \multirow{3}{*}{$\reff=0.0$} & $\sigma(\ns)=2.0\times10^{-3}$ & $\sigma(\w)=0.14$ & $\sigma(\ns)=2.0\times10^{-3}$ & $\sigma(\w)=0.14$ & $\sigma(\ns)=2.0\times10^{-3}$ & $\sigma(\w)=0.14$ & $\sigma(\ns)=2.0\times10^{-3}$ & $\sigma(\w)=0.11$ \\ & & $\sigma(\alphas)=2.7\times10^{-3}$ & $\sigma(\wo)=0.63$ & $\sigma(\alphas)=2.7\times10^{-3}$ & $\sigma(\wo)=0.63$ & $\sigma(\alphas)=2.7\times10^{-3}$ & $\sigma(\wo)=0.36$ & $\sigma(\alphas)=2.0\times10^{-3}$ & $\sigma(\wo)=0.11$ \\
\hline
\end{tabular}}
\end{table}

\begin{table}[htb!]
\label{table:classXplanck}
\resizebox{18cm}{!}{
\hspace{-3cm}
\begin{tabular}{|l|l||l|l||l|l||l|l||l|l|}		
\hline
\multicolumn{10}{|c|}{CLASS $\times$ \planck} \\
\hline
\hline
\multicolumn{2}{|l||}{Delensing option $\rightarrow$} &  \multicolumn{2}{c||}{no} & \multicolumn{2}{c||}{CMB x} &  \multicolumn{2}{c||}{CMB x} &  \multicolumn{2}{c|}{CMB x} \\
\multicolumn{2}{|l||}{$\downarrow$ comp. sep. option} &  \multicolumn{2}{c||}{delensing} & \multicolumn{2}{c||}{CMB} & \multicolumn{2}{c||}{CIB} & \multicolumn{2}{c|}{LSS} \\
\hline
 \multirow{6}{*}{\parbox{2.3cm}{cmb-only}} & \multirow{3}{*}{$\Delta=1.0$} & \multicolumn{2}{c||}{$\alpha=1.0$} & \multicolumn{2}{c||}{$\alpha=0.95$} & \multicolumn{2}{c||}{$\alpha=0.48$} & \multicolumn{2}{c|}{$\alpha=0.48$}\\ & & $\sigma(r)=1.4\times10^{-3}$ & $\sigma(\mnu)=1.6\times10^{2}$ & $\sigma(r)=1.4\times10^{-3}$ & $\sigma(\mnu)=1.6\times10^{2}$ & $\sigma(r)=1.2\times10^{-3}$ & $\sigma(\mnu)=60$ & $\sigma(r)=1.2\times10^{-3}$ & $\sigma(\mnu)=42$ \\ & & $\sigma(\nt)=0.092$ & $\sigma(\neff)=0.13$ & $\sigma(\nt)=0.092$ & $\sigma(\neff)=0.13$ & $\sigma(\nt)=0.089$ & $\sigma(\neff)=0.092$ & $\sigma(\nt)=0.089$ & $\sigma(\neff)=0.061$ \\ & \multirow{3}{*}{$\reff=0.0$} & $\sigma(\ns)=2.4\times10^{-3}$ & $\sigma(\w)=0.35$ & $\sigma(\ns)=2.4\times10^{-3}$ & $\sigma(\w)=0.35$ & $\sigma(\ns)=2.4\times10^{-3}$ & $\sigma(\w)=0.16$ & $\sigma(\ns)=2.4\times10^{-3}$ & $\sigma(\w)=0.11$ \\ & & $\sigma(\alphas)=4.2\times10^{-3}$ & $\sigma(\wo)=1.2$ & $\sigma(\alphas)=4.2\times10^{-3}$ & $\sigma(\wo)=1.2$ & $\sigma(\alphas)=3.8\times10^{-3}$ & $\sigma(\wo)=0.39$ & $\sigma(\alphas)=2.3\times10^{-3}$ & $\sigma(\wo)=0.12$ \\
\hline
\end{tabular}}
\end{table}

\begin{table}[htb!]
\label{table:spt3gXplanck}
\resizebox{18cm}{!}{
\hspace{-3cm}
\begin{tabular}{|l|l||l|l||l|l||l|l||l|l|}		
\hline
\multicolumn{10}{|c|}{SPT-3G $\times$ \planck} \\
\hline
\hline
\multicolumn{2}{|l||}{Delensing option $\rightarrow$} &  \multicolumn{2}{c||}{no} & \multicolumn{2}{c||}{CMB x} &  \multicolumn{2}{c||}{CMB x} &  \multicolumn{2}{c|}{CMB x} \\
\multicolumn{2}{|l||}{$\downarrow$ comp. sep. option} &  \multicolumn{2}{c||}{delensing} & \multicolumn{2}{c||}{CMB} & \multicolumn{2}{c||}{CIB} & \multicolumn{2}{c|}{LSS} \\
\hline
 \multirow{6}{*}{\parbox{2.3cm}{cmb-only}} & \multirow{3}{*}{$\Delta=1.0$} & \multicolumn{2}{c||}{$\alpha=1.0$} & \multicolumn{2}{c||}{$\alpha=0.40$} & \multicolumn{2}{c||}{$\alpha=0.40$} & \multicolumn{2}{c|}{$\alpha=0.38$}\\ & & $\sigma(r)=1.7\times10^{-3}$ & $\sigma(\mnu)=66$ & $\sigma(r)=1.1\times10^{-3}$ & $\sigma(\mnu)=66$ & $\sigma(r)=1.1\times10^{-3}$ & $\sigma(\mnu)=56$ & $\sigma(r)=1.1\times10^{-3}$ & $\sigma(\mnu)=44$ \\ & & $\sigma(\nt)=0.090$ & $\sigma(\neff)=0.073$ & $\sigma(\nt)=0.083$ & $\sigma(\neff)=0.073$ & $\sigma(\nt)=0.083$ & $\sigma(\neff)=0.065$ & $\sigma(\nt)=0.083$ & $\sigma(\neff)=0.050$ \\ & \multirow{3}{*}{$\reff=0.0$} & $\sigma(\ns)=2.6\times10^{-3}$ & $\sigma(\w)=0.20$ & $\sigma(\ns)=2.6\times10^{-3}$ & $\sigma(\w)=0.20$ & $\sigma(\ns)=2.6\times10^{-3}$ & $\sigma(\w)=0.16$ & $\sigma(\ns)=2.6\times10^{-3}$ & $\sigma(\w)=0.12$ \\ & & $\sigma(\alphas)=3.7\times10^{-3}$ & $\sigma(\wo)=0.76$ & $\sigma(\alphas)=3.7\times10^{-3}$ & $\sigma(\wo)=0.76$ & $\sigma(\alphas)=3.5\times10^{-3}$ & $\sigma(\wo)=0.41$ & $\sigma(\alphas)=2.4\times10^{-3}$ & $\sigma(\wo)=0.13$ \\
\hline
\end{tabular}}
\end{table}

\clearpage
\subsection{Post-2020 instruments}

In addition to the $n_p$ approach used for the sync+dust cases shown in previous tables, the following results include the $\mathbf{A}$-expansion approach after combining the considered instruments with either C-BASS or QUIJOTE-CMB.
Independently of their specific sky coverage (80\% for C-BASS and 44\% for QUIJOTE-CMB), we assume these two observatories help characterize synchrotron emission over the sky observed by any post-2020 instrument. 
Moreover, QUIJOTE-CMB is only combined with space missions as it is a project observing the Northern hemisphere; as with all the ground-based Stage-III instruments we consider throughout this work, we assume that the Stage-IV project will likely be observing the Southern hemisphere. We add \planck\ when considering Stage-IV in isolation in order to include information about the largest angular scales.

\begin{table}[htb!]
\label{table:stage4}
\resizebox{18cm}{!}{
\hspace{-3cm}
\begin{tabular}{|l|l||l|l||l|l||l|l||l|l|}		
\hline
\multicolumn{10}{|c|}{Stage-IV $\times$ \planck} \\
\hline
\hline
\multicolumn{2}{|l||}{Delensing option $\rightarrow$} &  \multicolumn{2}{c||}{no} & \multicolumn{2}{c||}{CMB x} &  \multicolumn{2}{c||}{CMB x} &  \multicolumn{2}{c|}{CMB x} \\
\multicolumn{2}{|l||}{$\downarrow$ comp. sep. option} &  \multicolumn{2}{c||}{delensing} & \multicolumn{2}{c||}{CMB} & \multicolumn{2}{c||}{CIB} & \multicolumn{2}{c|}{LSS} \\
\hline
 \multirow{6}{*}{\parbox{2.3cm}{cmb-only}} & \multirow{3}{*}{$\Delta=1.0$} & \multicolumn{2}{c||}{$\alpha=1.0$} & \multicolumn{2}{c||}{$\alpha=0.16$} & \multicolumn{2}{c||}{$\alpha=0.40$} & \multicolumn{2}{c|}{$\alpha=0.38$}\\ & & $\sigma(r)=4.9\times10^{-4}$ & $\sigma(\mnu)=44$ & $\sigma(r)=1.2\times10^{-4}$ & $\sigma(\mnu)=44$ & $\sigma(r)=2.2\times10^{-4}$ & $\sigma(\mnu)=46$ & $\sigma(r)=2.1\times10^{-4}$ & $\sigma(\mnu)=36$ \\ & & $\sigma(\nt)=0.060$ & $\sigma(\neff)=0.029$ & $\sigma(\nt)=0.027$ & $\sigma(\neff)=0.029$ & $\sigma(\nt)=0.040$ & $\sigma(\neff)=0.029$ & $\sigma(\nt)=0.040$ & $\sigma(\neff)=0.025$ \\ & \multirow{3}{*}{$\reff=0.0$} & $\sigma(\ns)=2.1\times10^{-3}$ & $\sigma(\w)=0.11$ & $\sigma(\ns)=2.1\times10^{-3}$ & $\sigma(\w)=0.11$ & $\sigma(\ns)=2.2\times10^{-3}$ & $\sigma(\w)=0.12$ & $\sigma(\ns)=2.1\times10^{-3}$ & $\sigma(\w)=0.098$ \\ & & $\sigma(\alphas)=2.1\times10^{-3}$ & $\sigma(\wo)=0.27$ & $\sigma(\alphas)=2.1\times10^{-3}$ & $\sigma(\wo)=0.27$ & $\sigma(\alphas)=2.2\times10^{-3}$ & $\sigma(\wo)=0.26$ & $\sigma(\alphas)=1.9\times10^{-3}$ & $\sigma(\wo)=0.11$ \\
\hline
\end{tabular}}
\end{table}

\begin{table}[htb!]
\label{table:litebird}
\resizebox{18cm}{!}{
\hspace{-3cm}
\begin{tabular}{|l|l||l|l||l|l||l|l||l|l|}		
\hline
\multicolumn{10}{|c|}{\lb} \\
\hline
\hline
\multicolumn{2}{|l||}{Delensing option $\rightarrow$} &  \multicolumn{2}{c||}{no} & \multicolumn{2}{c||}{CMB x} &  \multicolumn{2}{c||}{CMB x} &  \multicolumn{2}{c|}{CMB x} \\
\multicolumn{2}{|l||}{$\downarrow$ comp. sep. option} &  \multicolumn{2}{c||}{delensing} & \multicolumn{2}{c||}{CMB} & \multicolumn{2}{c||}{CIB} & \multicolumn{2}{c|}{LSS} \\
\hline
 \multirow{6}{*}{\parbox{2.3cm}{cmb-only}} & \multirow{3}{*}{$\Delta=1.0$} & \multicolumn{2}{c||}{$\alpha=1.0$} & \multicolumn{2}{c||}{$\alpha=0.79$} & \multicolumn{2}{c||}{$\alpha=0.45$} & \multicolumn{2}{c|}{$\alpha=0.46$}\\ & & $\sigma(r)=3.0\times10^{-4}$ & $\sigma(\mnu)=72$ & $\sigma(r)=2.7\times10^{-4}$ & $\sigma(\mnu)=72$ & $\sigma(r)=2.1\times10^{-4}$ & $\sigma(\mnu)=61$ & $\sigma(r)=2.1\times10^{-4}$ & $\sigma(\mnu)=45$ \\ & & $\sigma(\nt)=0.052$ & $\sigma(\neff)=0.15$ & $\sigma(\nt)=0.049$ & $\sigma(\neff)=0.15$ & $\sigma(\nt)=0.044$ & $\sigma(\neff)=0.11$ & $\sigma(\nt)=0.045$ & $\sigma(\neff)=0.074$ \\ & \multirow{3}{*}{$\reff=0.0$} & $\sigma(\ns)=3.7\times10^{-3}$ & $\sigma(\w)=0.16$ & $\sigma(\ns)=3.7\times10^{-3}$ & $\sigma(\w)=0.16$ & $\sigma(\ns)=3.4\times10^{-3}$ & $\sigma(\w)=0.15$ & $\sigma(\ns)=2.9\times10^{-3}$ & $\sigma(\w)=0.11$ \\ & & $\sigma(\alphas)=5.6\times10^{-3}$ & $\sigma(\wo)=0.95$ & $\sigma(\alphas)=5.6\times10^{-3}$ & $\sigma(\wo)=0.95$ & $\sigma(\alphas)=4.6\times10^{-3}$ & $\sigma(\wo)=0.39$ & $\sigma(\alphas)=2.3\times10^{-3}$ & $\sigma(\wo)=0.12$ \\
\hline
\end{tabular}}
\end{table}

\begin{table}[htb!]
\label{table:core+}
\resizebox{18cm}{!}{
\hspace{-3cm}
\begin{tabular}{|l|l||l|l||l|l||l|l||l|l|}		
\hline
\multicolumn{10}{|c|}{COrE+} \\
\hline
\hline
\multicolumn{2}{|l||}{Delensing option $\rightarrow$} &  \multicolumn{2}{c||}{no} & \multicolumn{2}{c||}{CMB x} &  \multicolumn{2}{c||}{CMB x} &  \multicolumn{2}{c|}{CMB x} \\
\multicolumn{2}{|l||}{$\downarrow$ comp. sep. option} &  \multicolumn{2}{c||}{delensing} & \multicolumn{2}{c||}{CMB} & \multicolumn{2}{c||}{CIB} & \multicolumn{2}{c|}{LSS} \\
\hline
 \multirow{6}{*}{\parbox{2.3cm}{cmb-only}} & \multirow{3}{*}{$\Delta=1.0$} & \multicolumn{2}{c||}{$\alpha=1.0$} & \multicolumn{2}{c||}{$\alpha=0.33$} & \multicolumn{2}{c||}{$\alpha=0.40$} & \multicolumn{2}{c|}{$\alpha=0.38$}\\ & & $\sigma(r)=2.9\times10^{-4}$ & $\sigma(\mnu)=38$ & $\sigma(r)=1.7\times10^{-4}$ & $\sigma(\mnu)=38$ & $\sigma(r)=1.8\times10^{-4}$ & $\sigma(\mnu)=39$ & $\sigma(r)=1.8\times10^{-4}$ & $\sigma(\mnu)=32$ \\ & & $\sigma(\nt)=0.048$ & $\sigma(\neff)=0.036$ & $\sigma(\nt)=0.034$ & $\sigma(\neff)=0.036$ & $\sigma(\nt)=0.036$ & $\sigma(\neff)=0.035$ & $\sigma(\nt)=0.036$ & $\sigma(\neff)=0.029$ \\ & \multirow{3}{*}{$\reff=0.0$} & $\sigma(\ns)=1.9\times10^{-3}$ & $\sigma(\w)=0.100$ & $\sigma(\ns)=1.9\times10^{-3}$ & $\sigma(\w)=0.100$ & $\sigma(\ns)=1.9\times10^{-3}$ & $\sigma(\w)=0.12$ & $\sigma(\ns)=1.8\times10^{-3}$ & $\sigma(\w)=0.094$ \\ & & $\sigma(\alphas)=2.1\times10^{-3}$ & $\sigma(\wo)=0.39$ & $\sigma(\alphas)=2.1\times10^{-3}$ & $\sigma(\wo)=0.39$ & $\sigma(\alphas)=2.2\times10^{-3}$ & $\sigma(\wo)=0.31$ & $\sigma(\alphas)=1.8\times10^{-3}$ & $\sigma(\wo)=0.10$ \\
\hline
\end{tabular}}
\end{table}

\begin{table}[htb!]
\label{table:pixie}
\resizebox{18cm}{!}{
\hspace{-3cm}
\begin{tabular}{|l|l||l|l||l|l||l|l||l|l|}		
\hline
\multicolumn{10}{|c|}{PIXIE} \\
\hline
\hline
\multicolumn{2}{|l||}{Delensing option $\rightarrow$} &  \multicolumn{2}{c||}{no} & \multicolumn{2}{c||}{CMB x} &  \multicolumn{2}{c||}{CMB x} &  \multicolumn{2}{c|}{CMB x} \\
\multicolumn{2}{|l||}{$\downarrow$ comp. sep. option} &  \multicolumn{2}{c||}{delensing} & \multicolumn{2}{c||}{CMB} & \multicolumn{2}{c||}{CIB} & \multicolumn{2}{c|}{LSS} \\
\hline
 \multirow{6}{*}{\parbox{2.3cm}{cmb-only}} & \multirow{3}{*}{$\Delta=1.0$} & \multicolumn{2}{c||}{$\alpha=1.0$} & \multicolumn{2}{c||}{$\alpha=1.00$} & \multicolumn{2}{c||}{$\alpha=0.95$} & \multicolumn{2}{c|}{$\alpha=0.94$}\\ & & $\sigma(r)=3.1\times10^{-4}$ & $\sigma(\mnu)=4.6\times10^{2}$ & $\sigma(r)=3.0\times10^{-4}$ & $\sigma(\mnu)=4.6\times10^{2}$ & $\sigma(r)=2.9\times10^{-4}$ & $\sigma(\mnu)=86$ & $\sigma(r)=2.9\times10^{-4}$ & $\sigma(\mnu)=64$ \\ & & $\sigma(\nt)=0.070$ & $\sigma(\neff)=1.3$ & $\sigma(\nt)=0.070$ & $\sigma(\neff)=1.3$ & $\sigma(\nt)=0.069$ & $\sigma(\neff)=0.72$ & $\sigma(\nt)=0.067$ & $\sigma(\neff)=0.68$ \\ & \multirow{3}{*}{$\reff=0.0$} & $\sigma(\ns)=0.070$ & $\sigma(\w)=0.61$ & $\sigma(\ns)=0.070$ & $\sigma(\w)=0.61$ & $\sigma(\ns)=6.3\times10^{-3}$ & $\sigma(\w)=0.18$ & $\sigma(\ns)=4.8\times10^{-3}$ & $\sigma(\w)=0.14$ \\ & & $\sigma(\alphas)=0.034$ & $\sigma(\wo)=2.5$ & $\sigma(\alphas)=0.034$ & $\sigma(\wo)=2.5$ & $\sigma(\alphas)=7.6\times10^{-3}$ & $\sigma(\wo)=0.63$ & $\sigma(\alphas)=4.0\times10^{-3}$ & $\sigma(\wo)=0.21$ \\
\hline
\end{tabular}}
\end{table}

\begin{table}[htb!]
\label{table:litebirdXcore+}
\resizebox{18cm}{!}{
\hspace{-3cm}
\begin{tabular}{|l|l||l|l||l|l||l|l||l|l|}		
\hline
\multicolumn{10}{|c|}{\lb\ $\times$ COrE+} \\
\hline
\hline
\multicolumn{2}{|l||}{Delensing option $\rightarrow$} &  \multicolumn{2}{c||}{no} & \multicolumn{2}{c||}{CMB x} &  \multicolumn{2}{c||}{CMB x} &  \multicolumn{2}{c|}{CMB x} \\
\multicolumn{2}{|l||}{$\downarrow$ comp. sep. option} &  \multicolumn{2}{c||}{delensing} & \multicolumn{2}{c||}{CMB} & \multicolumn{2}{c||}{CIB} & \multicolumn{2}{c|}{LSS} \\
\hline
 \multirow{6}{*}{\parbox{2.3cm}{cmb-only}} & \multirow{3}{*}{$\Delta=1.0$} & \multicolumn{2}{c||}{$\alpha=1.0$} & \multicolumn{2}{c||}{$\alpha=0.33$} & \multicolumn{2}{c||}{$\alpha=0.40$} & \multicolumn{2}{c|}{$\alpha=0.38$}\\ & & $\sigma(r)=2.8\times10^{-4}$ & $\sigma(\mnu)=38$ & $\sigma(r)=1.5\times10^{-4}$ & $\sigma(\mnu)=38$ & $\sigma(r)=1.7\times10^{-4}$ & $\sigma(\mnu)=39$ & $\sigma(r)=1.6\times10^{-4}$ & $\sigma(\mnu)=32$ \\ & & $\sigma(\nt)=0.047$ & $\sigma(\neff)=0.036$ & $\sigma(\nt)=0.032$ & $\sigma(\neff)=0.036$ & $\sigma(\nt)=0.035$ & $\sigma(\neff)=0.035$ & $\sigma(\nt)=0.035$ & $\sigma(\neff)=0.029$ \\ & \multirow{3}{*}{$\reff=0.0$} & $\sigma(\ns)=1.9\times10^{-3}$ & $\sigma(\w)=0.100$ & $\sigma(\ns)=1.9\times10^{-3}$ & $\sigma(\w)=0.100$ & $\sigma(\ns)=1.9\times10^{-3}$ & $\sigma(\w)=0.12$ & $\sigma(\ns)=1.8\times10^{-3}$ & $\sigma(\w)=0.094$ \\ & & $\sigma(\alphas)=2.1\times10^{-3}$ & $\sigma(\wo)=0.39$ & $\sigma(\alphas)=2.1\times10^{-3}$ & $\sigma(\wo)=0.39$ & $\sigma(\alphas)=2.2\times10^{-3}$ & $\sigma(\wo)=0.31$ & $\sigma(\alphas)=1.8\times10^{-3}$ & $\sigma(\wo)=0.10$ \\
\hline
\end{tabular}}
\end{table}

\begin{table}[htb!]
\label{table:pixieXlitebird}
\resizebox{18cm}{!}{
\hspace{-3cm}
\begin{tabular}{|l|l||l|l||l|l||l|l||l|l|}		
\hline
\multicolumn{10}{|c|}{PIXIE $\times$ \lb} \\
\hline
\hline
\multicolumn{2}{|l||}{Delensing option $\rightarrow$} &  \multicolumn{2}{c||}{no} & \multicolumn{2}{c||}{CMB x} &  \multicolumn{2}{c||}{CMB x} &  \multicolumn{2}{c|}{CMB x} \\
\multicolumn{2}{|l||}{$\downarrow$ comp. sep. option} &  \multicolumn{2}{c||}{delensing} & \multicolumn{2}{c||}{CMB} & \multicolumn{2}{c||}{CIB} & \multicolumn{2}{c|}{LSS} \\
\hline
\multirow{6}{*}{\parbox{2.3cm}{cmb-only}} & \multirow{3}{*}{$\Delta=1.0$} & \multicolumn{2}{c||}{$\alpha=1.0$} & \multicolumn{2}{c||}{$\alpha=0.78$} & \multicolumn{2}{c||}{$\alpha=0.45$} & \multicolumn{2}{c|}{$\alpha=0.46$}\\ & & $\sigma(r)=2.8\times10^{-4}$ & $\sigma(\mnu)=72$ & $\sigma(r)=2.5\times10^{-4}$ & $\sigma(\mnu)=72$ & $\sigma(r)=1.9\times10^{-4}$ & $\sigma(\mnu)=61$ & $\sigma(r)=1.9\times10^{-4}$ & $\sigma(\mnu)=45$ \\ & & $\sigma(\nt)=0.052$ & $\sigma(\neff)=0.15$ & $\sigma(\nt)=0.049$ & $\sigma(\neff)=0.15$ & $\sigma(\nt)=0.044$ & $\sigma(\neff)=0.11$ & $\sigma(\nt)=0.044$ & $\sigma(\neff)=0.074$ \\ & \multirow{3}{*}{$\reff=0.0$} & $\sigma(\ns)=3.7\times10^{-3}$ & $\sigma(\w)=0.16$ & $\sigma(\ns)=3.7\times10^{-3}$ & $\sigma(\w)=0.16$ & $\sigma(\ns)=3.4\times10^{-3}$ & $\sigma(\w)=0.15$ & $\sigma(\ns)=2.9\times10^{-3}$ & $\sigma(\w)=0.11$ \\ & & $\sigma(\alphas)=5.6\times10^{-3}$ & $\sigma(\wo)=0.94$ & $\sigma(\alphas)=5.5\times10^{-3}$ & $\sigma(\wo)=0.94$ & $\sigma(\alphas)=4.6\times10^{-3}$ & $\sigma(\wo)=0.39$ & $\sigma(\alphas)=2.3\times10^{-3}$ & $\sigma(\wo)=0.12$ \\
\hline
\end{tabular}}
\end{table}

\begin{table}[htb!]
\label{table:pixieXcore+}
\resizebox{18cm}{!}{
\hspace{-3cm}
\begin{tabular}{|l|l||l|l||l|l||l|l||l|l|}		
\hline
\multicolumn{10}{|c|}{PIXIE $\times$ COrE+} \\
\hline
\hline
\multicolumn{2}{|l||}{Delensing option $\rightarrow$} &  \multicolumn{2}{c||}{no} & \multicolumn{2}{c||}{CMB x} &  \multicolumn{2}{c||}{CMB x} &  \multicolumn{2}{c|}{CMB x} \\
\multicolumn{2}{|l||}{$\downarrow$ comp. sep. option} &  \multicolumn{2}{c||}{delensing} & \multicolumn{2}{c||}{CMB} & \multicolumn{2}{c||}{CIB} & \multicolumn{2}{c|}{LSS} \\
\hline
\multirow{6}{*}{\parbox{2.3cm}{cmb-only}} & \multirow{3}{*}{$\Delta=1.0$} & \multicolumn{2}{c||}{$\alpha=1.0$} & \multicolumn{2}{c||}{$\alpha=0.33$} & \multicolumn{2}{c||}{$\alpha=0.40$} & \multicolumn{2}{c|}{$\alpha=0.38$}\\ & & $\sigma(r)=2.8\times10^{-4}$ & $\sigma(\mnu)=38$ & $\sigma(r)=1.5\times10^{-4}$ & $\sigma(\mnu)=38$ & $\sigma(r)=1.7\times10^{-4}$ & $\sigma(\mnu)=39$ & $\sigma(r)=1.6\times10^{-4}$ & $\sigma(\mnu)=32$ \\ & & $\sigma(\nt)=0.048$ & $\sigma(\neff)=0.036$ & $\sigma(\nt)=0.034$ & $\sigma(\neff)=0.036$ & $\sigma(\nt)=0.036$ & $\sigma(\neff)=0.035$ & $\sigma(\nt)=0.035$ & $\sigma(\neff)=0.029$ \\ & \multirow{3}{*}{$\reff=0.0$} & $\sigma(\ns)=1.9\times10^{-3}$ & $\sigma(\w)=0.100$ & $\sigma(\ns)=1.9\times10^{-3}$ & $\sigma(\w)=0.100$ & $\sigma(\ns)=1.9\times10^{-3}$ & $\sigma(\w)=0.12$ & $\sigma(\ns)=1.8\times10^{-3}$ & $\sigma(\w)=0.094$ \\ & & $\sigma(\alphas)=2.1\times10^{-3}$ & $\sigma(\wo)=0.39$ & $\sigma(\alphas)=2.1\times10^{-3}$ & $\sigma(\wo)=0.39$ & $\sigma(\alphas)=2.2\times10^{-3}$ & $\sigma(\wo)=0.31$ & $\sigma(\alphas)=1.8\times10^{-3}$ & $\sigma(\wo)=0.10$ \\
\hline
\end{tabular}}
\end{table}

\begin{table}[htb!]
\label{table:stage4Xcore+}
\resizebox{18cm}{!}{
\hspace{-3cm}
\begin{tabular}{|l|l||l|l||l|l||l|l||l|l|}		
\hline
\multicolumn{10}{|c|}{Stage-IV $\times$ COrE+} \\
\hline
\hline
\multicolumn{2}{|l||}{Delensing option $\rightarrow$} &  \multicolumn{2}{c||}{no} & \multicolumn{2}{c||}{CMB x} &  \multicolumn{2}{c||}{CMB x} &  \multicolumn{2}{c|}{CMB x} \\
\multicolumn{2}{|l||}{$\downarrow$ comp. sep. option} &  \multicolumn{2}{c||}{delensing} & \multicolumn{2}{c||}{CMB} & \multicolumn{2}{c||}{CIB} & \multicolumn{2}{c|}{LSS} \\
\hline
\multirow{6}{*}{\parbox{2.3cm}{cmb-only}} & \multirow{3}{*}{$\Delta=1.0$} & \multicolumn{2}{c||}{$\alpha=1.0$} & \multicolumn{2}{c||}{$\alpha=0.15$} & \multicolumn{2}{c||}{$\alpha=0.40$} & \multicolumn{2}{c|}{$\alpha=0.38$}\\ & & $\sigma(r)=2.6\times10^{-4}$ & $\sigma(\mnu)=35$ & $\sigma(r)=9.5\times10^{-5}$ & $\sigma(\mnu)=35$ & $\sigma(r)=1.6\times10^{-4}$ & $\sigma(\mnu)=37$ & $\sigma(r)=1.5\times10^{-4}$ & $\sigma(\mnu)=30$ \\ & & $\sigma(\nt)=0.044$ & $\sigma(\neff)=0.026$ & $\sigma(\nt)=0.023$ & $\sigma(\neff)=0.026$ & $\sigma(\nt)=0.032$ & $\sigma(\neff)=0.026$ & $\sigma(\nt)=0.031$ & $\sigma(\neff)=0.022$ \\ & \multirow{3}{*}{$\reff=0.0$} & $\sigma(\ns)=1.8\times10^{-3}$ & $\sigma(\w)=0.087$ & $\sigma(\ns)=1.8\times10^{-3}$ & $\sigma(\w)=0.087$ & $\sigma(\ns)=1.8\times10^{-3}$ & $\sigma(\w)=0.10$ & $\sigma(\ns)=1.8\times10^{-3}$ & $\sigma(\w)=0.083$ \\ & & $\sigma(\alphas)=1.8\times10^{-3}$ & $\sigma(\wo)=0.24$ & $\sigma(\alphas)=1.8\times10^{-3}$ & $\sigma(\wo)=0.24$ & $\sigma(\alphas)=1.9\times10^{-3}$ & $\sigma(\wo)=0.23$ & $\sigma(\alphas)=1.6\times10^{-3}$ & $\sigma(\wo)=0.095$ \\
\hline
\end{tabular}}
\end{table}

\begin{table}[htb!]
\label{table:stage4Xpixie}
\resizebox{18cm}{!}{
\hspace{-3cm}
\begin{tabular}{|l|l||l|l||l|l||l|l||l|l|}		
\hline
\multicolumn{10}{|c|}{Stage-IV $\times$ PIXIE} \\
\hline
\hline
\multicolumn{2}{|l||}{Delensing option $\rightarrow$} &  \multicolumn{2}{c||}{no} & \multicolumn{2}{c||}{CMB x} &  \multicolumn{2}{c||}{CMB x} &  \multicolumn{2}{c|}{CMB x} \\
\multicolumn{2}{|l||}{$\downarrow$ comp. sep. option} &  \multicolumn{2}{c||}{delensing} & \multicolumn{2}{c||}{CMB} & \multicolumn{2}{c||}{CIB} & \multicolumn{2}{c|}{LSS} \\
\hline
 \multirow{6}{*}{\parbox{2.3cm}{cmb-only}} & \multirow{3}{*}{$\Delta=1.0$} & \multicolumn{2}{c||}{$\alpha=1.0$} & \multicolumn{2}{c||}{$\alpha=0.16$} & \multicolumn{2}{c||}{$\alpha=0.40$} & \multicolumn{2}{c|}{$\alpha=0.38$}\\ & & $\sigma(r)=2.6\times10^{-4}$ & $\sigma(\mnu)=40$ & $\sigma(r)=1.1\times10^{-4}$ & $\sigma(\mnu)=40$ & $\sigma(r)=1.8\times10^{-4}$ & $\sigma(\mnu)=41$ & $\sigma(r)=1.7\times10^{-4}$ & $\sigma(\mnu)=32$ \\ & & $\sigma(\nt)=0.047$ & $\sigma(\neff)=0.028$ & $\sigma(\nt)=0.024$ & $\sigma(\neff)=0.028$ & $\sigma(\nt)=0.034$ & $\sigma(\neff)=0.028$ & $\sigma(\nt)=0.034$ & $\sigma(\neff)=0.023$ \\ & \multirow{3}{*}{$\reff=0.0$} & $\sigma(\ns)=1.9\times10^{-3}$ & $\sigma(\w)=0.099$ & $\sigma(\ns)=1.9\times10^{-3}$ & $\sigma(\w)=0.099$ & $\sigma(\ns)=1.9\times10^{-3}$ & $\sigma(\w)=0.11$ & $\sigma(\ns)=1.9\times10^{-3}$ & $\sigma(\w)=0.087$ \\ & & $\sigma(\alphas)=2.0\times10^{-3}$ & $\sigma(\wo)=0.26$ & $\sigma(\alphas)=2.0\times10^{-3}$ & $\sigma(\wo)=0.26$ & $\sigma(\alphas)=2.0\times10^{-3}$ & $\sigma(\wo)=0.25$ & $\sigma(\alphas)=1.7\times10^{-3}$ & $\sigma(\wo)=0.098$ \\
\hline
\end{tabular}}
\end{table}

\begin{table}[htb!]
\label{table:stage4Xlitebird}
\resizebox{18cm}{!}{
\hspace{-3cm}
\begin{tabular}{|l|l||l|l||l|l||l|l||l|l|}		
\hline
\multicolumn{10}{|c|}{Stage-IV $\times$ \lb} \\
\hline
\hline
\multicolumn{2}{|l||}{Delensing option $\rightarrow$} &  \multicolumn{2}{c||}{no} & \multicolumn{2}{c||}{CMB x} &  \multicolumn{2}{c||}{CMB x} &  \multicolumn{2}{c|}{CMB x} \\
\multicolumn{2}{|l||}{$\downarrow$ comp. sep. option} &  \multicolumn{2}{c||}{delensing} & \multicolumn{2}{c||}{CMB} & \multicolumn{2}{c||}{CIB} & \multicolumn{2}{c|}{LSS} \\
\hline
\multirow{6}{*}{\parbox{2.3cm}{cmb-only}} & \multirow{3}{*}{$\Delta=1.0$} & \multicolumn{2}{c||}{$\alpha=1.0$} & \multicolumn{2}{c||}{$\alpha=0.16$} & \multicolumn{2}{c||}{$\alpha=0.40$} & \multicolumn{2}{c|}{$\alpha=0.38$}\\ & & $\sigma(r)=2.6\times10^{-4}$ & $\sigma(\mnu)=37$ & $\sigma(r)=1.1\times10^{-4}$ & $\sigma(\mnu)=37$ & $\sigma(r)=1.6\times10^{-4}$ & $\sigma(\mnu)=38$ & $\sigma(r)=1.6\times10^{-4}$ & $\sigma(\mnu)=31$ \\ & & $\sigma(\nt)=0.045$ & $\sigma(\neff)=0.028$ & $\sigma(\nt)=0.024$ & $\sigma(\neff)=0.028$ & $\sigma(\nt)=0.033$ & $\sigma(\neff)=0.027$ & $\sigma(\nt)=0.033$ & $\sigma(\neff)=0.023$ \\ & \multirow{3}{*}{$\reff=0.0$} & $\sigma(\ns)=1.8\times10^{-3}$ & $\sigma(\w)=0.093$ & $\sigma(\ns)=1.8\times10^{-3}$ & $\sigma(\w)=0.093$ & $\sigma(\ns)=1.9\times10^{-3}$ & $\sigma(\w)=0.10$ & $\sigma(\ns)=1.8\times10^{-3}$ & $\sigma(\w)=0.085$ \\ & & $\sigma(\alphas)=1.9\times10^{-3}$ & $\sigma(\wo)=0.26$ & $\sigma(\alphas)=1.9\times10^{-3}$ & $\sigma(\wo)=0.26$ & $\sigma(\alphas)=2.0\times10^{-3}$ & $\sigma(\wo)=0.24$ & $\sigma(\alphas)=1.7\times10^{-3}$ & $\sigma(\wo)=0.096$ \\
\hline
\end{tabular}}
\end{table}

\bibliographystyle{JHEP}
\renewcommand{\bibname}{References} 
\bibliography{references}

\end{document}